\pdfoutput=1
\documentclass[usenatbib]{mn2e}
\usepackage{amsmath}
\usepackage{amssymb}
\usepackage{amsfonts}
\usepackage{bm}
\usepackage{graphicx}
\usepackage{placeins}
\usepackage{aas_macros}
\usepackage{array}
\usepackage{hhline}

\title{Tidal interactions of a Maclaurin spheroid. I: Properties of free oscillation modes}
\author[Harry J. Braviner and Gordon I. Ogilvie]
{Harry J. Braviner and Gordon I. Ogilvie \\
Department of Applied Mathematics and Theoretical Physics, University of Cambridge, Centre for Mathematical Sciences,\\
Wilberforce Road, Cambridge, CB3 0WA}

\begin{document}

\maketitle

\begin{abstract}
We review the work of \citet{Bryan1889} on the normal modes of a Maclaurin spheroid, carrying out numerical calculations of the frequencies and spatial forms of these modes that have not been previously published.
We study all modes of degree $l \le 4$, which includes both inertial modes and surface gravity modes, with the aim of better understanding the effect of rapid rotation on tidal interactions.
The inclusion of these higher degree modes greatly increases the number of frequencies at which tidal resonances may occur.
We derive an expression for the decay rates of these modes to first order in viscosity and explicitly plot these for modes.
We see that the equatorial bulge of the spheroid has a significant effect on the decay rates (changing some of these by a factor of $2$ between an eccentricity of $e=0$ and $0.5$), and a more modest effect on the mode frequencies.
This suggests that models of tidal interaction between rapidly rotating stars and giant planets that model the Coriolis force while neglecting the centrifugal distortion of the body may be in error by an order unity factor.
In a subsequent paper we shall examine the case of a forced flow in this spheroid, and complete the model by considering how the tides raised by the orbiting companion change the orbital elements.
\end{abstract}

\section{Introduction}
\label{sec:introduction}

At the time of writing over 750 confirmed exoplanets have been discovered, with over 3400 unconfirmed candidates from the Kepler satellite, according to the \texttt{exoplanets.org} database (\citet{exoplanets}).
Of these confirmed planets, over 200 have orbital periods shorter than 5 days.
Tidal interaction should be expected to have played, and to continue to play, a significant role in the evolution of the orbits of such systems.

\citet{JGB2008} considered the 36 then-known planets with semi-major axes less than $0.2\,\mathrm{AU}$ and well-measured eccentricities.
The equations for the tidal evolution of the orbital elements were integrated backwards over the age of the system for various tidal quality factors in the star and planet, with the aim of producing an initial eccentricity distribution matching that of the population of planets with semi-major axes greater than $0.2\,\mathrm{AU}$.
The latter population were assumed to have not been significantly affected by tidal evolution.
The authors concluded that the observations were best explained by a quality factor of $Q^{\prime}_{\mathrm{p}} = 10^{6.5}$ in the planets and $Q^{\prime}_* = 10^{5.5}$ in the host stars.
The planetary quality factor is roughly compatible with that of Jupiter inferred from observations of the Galilean satellites (specifically, it lies within the range found by \cite{YP1981} by considering the evolution of these satellites over the age of the Solar System, but is higher than that found by \cite{Lainey2009} by considering astrometric observations since 1891) and a quality factor of $10^{5.5}$ for stars is similar to that inferred from observations of binaries (\citet{OgilvieLin2007} and \citet{MeibomMathieu2005}).
However, \citet{OgilvieLin2007} find that models of solar type stars can only explain $Q^{\prime}_* \gtrsim 10^7$.
The results of \cite{JGB2008} are much more sensitive to $Q^{\prime}_{\mathrm{p}}$ than $Q^{\prime}_*$, and applying $Q^{\prime}_* = 10^{5.5}$ to future evolution of the orbit would predict very short future lifetimes for many of the planets observed.
The two preceding facts suggest the study of \citeauthor{JGB2008} provides compelling evidence only for significant tidal dissipation within the giant planets themselves, not their host stars.

\citet{Hansen2010} also examined the eccentricity distribution of extra-solar planets, but based on the assumption that all stars shared a common tidal dissipation constant, rather than a common quality factor (as \citet{JGB2008} did).
This is equivalent to assuming that $Q^{\prime} \propto 1 /(\omega R^5)$, where $\omega$ is the orbital frequency and $R$ is the radius of the body on which the tide is raised.
\citeauthor{Hansen2010} also assumes that the planets share a common dissipation constant, which may differ from that of the stars.
The study includes the then-known planets orbiting stars with masses between $0.7 M_{\sun}$ and $1.5 M_{\sun}$, with planets of masses less than $0.3 M_J$ rejected from the study due to these not being fully convective.
At the time this was over 200 planets, which are used to produce an initial estimate of the planetary dissipation constant.
However, the dissipation constants are further constrained by considering the planets tracing the edge of the occupied region in a plot of period versus eccentricity, nine of which yield useful constraints.
\citeauthor{Hansen2010} found that the stellar dissipation constant inferred from the circularisation of binary stars is too great to be compatible with the eccentricities of planets more massive than $3 M_J$, but notes that the tidal interaction of binaries may be enhanced by their ability to excite inertial waves in one another, which is not included in the equilibrium tide model of the paper.
On the basis of this, the dissipation constant of the stars suggested by binary circularisation is discarded and replaced by such a constant calibrated purely on observations of planets.
Whilst \citeauthor{Hansen2010}'s model does not produce a constant $Q^{\prime}_{\mathrm{p}}$, the majority of the planets studied had $Q^{\prime}_{\mathrm{p}} \approx 3 \times 10^7$.
The spread of $Q^{\prime}_*$ values was somewhat greater, due to the strong dependence on the stellar radius, with most stars lying between $3 \times 10^7 \lesssim Q^{\prime}_* \lesssim 8 \times 10^8$.
This is a much weaker stellar tide than that suggested by \citet{JGB2008}, and seems more compatible with both the theoretical predictions of \citet{OgilvieLin2007} and the continued survival of extremely short period planets such as WASP-18b and WASP-103b.

More recently, \citet{Husnoo2012} considered the eccentricity distribution of a sample of 54 giant planets with orbital periods of up to 20 days and computed circularisation timescales based on the orbital period and the mass of the star and the planet.
Again, the relation between these timescales and the observed orbital eccentricities provides good evidence for the circularisation of these systems by the action of tidal dissipation within the planets, though again there is only very weak dependence of the result on the action of the stellar tide.
The authors did find evidence of excess rotation in the host stars of six systems, with rotational periods between $13\%$ and $25\%$ of the value expected for a star of such a type and age.
These may well be explained by tidal dissipation within the stars.
\citet{Albrecht2012} studied instead the spin-orbit misalignments of gas giant planets with short orbital periods and found that those orbiting stars cooler than $6250\,\mathrm{K}$ (which possess thick convective envelopes that may enhance tidal dissipation) exhibited lower misalignments than those orbiting stars hotter than this temperature;
this dichotomy was interpreted as evidence of such planets being formed with initially isotropically distributed orbits, which are then aligned with the equatorial plane of the star by the action of tides in those cases where efficient tidal dissipation may operate.

Whilst the previously mentioned papers studied the statistics of large populations of planets, it may be possible in the near future to observe tidal evolution directly.
\citet{Hellieretal2009} reported that the transiting planet WASP-18b had an orbital period of only $0.94$ days and that, if the quality factor within the host star is as low as that found by \citeauthor{JGB2008}, the timing of WASP-18b's transits should change by a measurable amount over a ten year period.
If such a shift is not observed, it would provide an upper bound on the dissipation rate.

In the observational studies described above, relatively simple tidal models were employed.
In \citet{JGB2008} a constant phase-lag model for the tides in both the star and the planet was used to compute the coupled evolution of the eccentricity and semi-major axis of the orbit;
in \citet{Husnoo2012}, a constant time-lag tidal model for the star and planet was used to compute circularisation timescales.
The model of \citet{Hansen2010} is equivalent in its dependence on orbital period to this constant time-lag model.
\citet{Albrecht2012} split systems into populations according to the effective temperature of the host star and applied a simplified version of the model of \citeauthor{Zahn1977} described in the following paragraph.
Whilst such models are sufficient for comparing to the statistics of a large sample of exoplanets, they unlikely to be accurate models for the detailed tidal evolution of any single system.
Indeed, the failure to explain both the eccentricity distribution of binary stars and that of extra-solar planets with a single calibration of either $Q^{\prime}_*$ or the stellar dissipation constant suggests that more sophisticated models with more realistic frequency dependences are required.
To build such models we require a theoretical understanding of the flow driven in the star by the gravitational forcing from the planet and of the mechanisms by which the tide can be dissipated.
If the dynamical tide (as opposed to the quasi-hydrostatic equilibrium tide) is significant, this involves understanding the excitation and dissipation of waves within the star.
All of these processes are likely to be dependent on the frequency of the tidal forcing, with dissipation rates (and hence evolution rates of orbital elements) being significantly enhanced close to the natural frequencies of such waves if they form global modes.

Theoretical work on tidal dissipation in stars predates the discovery of extra-solar planets, and was focused on understanding the synchronisation and circularisation of binary stars.
\citet{Zahn1977} considered two sources of dissipation.
In cooler stars, which possess an extensive convective envelope, the quasi-hydrostatic bulge (termed the equilibrium tide) raised by the companion induces some shear as it travels around the star.
The dissipation arising from the interaction of this shear with the convection was estimated using an eddy viscosity model.
In hotter stars, which possess a convective core, the most significant source of dissipation was estimated to be the radiative damping of internal gravity waves excited by tidal forcing within the radiative envelope.

With the discovery of extra-solar planets, interest in tidal dissipation in stars was renewed.
Even the dissipation rate of the equilibrium tide may not carry over from studies of binary stars, owing to the fact that binaries are usually already synchronised during the tidal circularisation of the orbit.
In planetary systems for which the period is short enough that tidal evolution will occur the majority of the angular momentum is usually contained in the spin of the star, and hence synchronisation of the star will not occur.
As mentioned above, the wave modes of the star that may be excited by tidal forcing (conventionally referred to as the dynamical tide) offer the possibility of resonances.
In the radiative region of a star, where the stratification is stable, internal gravity waves may be excited.
In a typical star the buoyancy frequency varies with depth and the maximum buoyancy frequency is much higher than the tidal forcing frequency.
However, close to the convective region the buoyancy frequency is much lower and the tidal forcing may be locally resonant, resulting in waves being launched from near the radiative-convective interface.
\citet{Zahn1977} considered such waves in the context of forcing of a high-mass star by a binary companion.
\citet{GN1989} further studied this model, finding that such waves transported angular momentum outwards and that the star could be synchronised from the outside in.
\citet{Terquem1998} and \cite{GD1998} studied such waves in a solar-type star (with a radiative core and convective envelope).

Since a star is a rotating body it might be expected to support some form of inertial waves.
These are modes for which the restoring force is the Coriolis force, hence no stratification is required to support them; indeed, the fact that the buoyancy frequency of the radiative zone is much higher than the Coriolis frequency results in waves in those regions being strongly dominated by buoyancy, and pure inertial waves can be considered to be confined to the convection zone.
However, global inertial modes for inviscid fluids do not exist in arbitrary containers.
A spherical shell with impermeable boundaries, which is a reasonable model for inertial waves in a solar-type star, is an example of such a container lacking global modes.
Such shells instead possess wave attractors, as discussed in \citet{RV1997} and \citet{Ogilvie2005}.
The model we consider herein (a spheroid of homogeneous, incompressible fluid with no core) does possess global modes, and we expect that many properties of the weakly viscous problem will carry over to closely related models, such as a neutrally stratified fluid with a small core.
Moreover, the frequencies of inertial waves are comparable to the rotation frequency of the body, as is the tidal forcing frequency in many giant planets (as shown by \citet{Goldreich1963}, the tide in the less massive body is usually the more significant tide for circularising the orbit).
A great deal of work has been undertaken on this problem, mainly concentrating on the properties of inertial waves in a variety of neutrally stable density profiles.
\citet{Papaloizou1981} numerically investigated particular low order modes in an $n=1.5$ polytrope and \citet{Wu2005a, Wu2005b} developed a analytic model for inertial waves in a body with a power-law density variation and applied the results to estimate the tidal dissipation rate for Jupiter.

In all of the cases described above, the star was assumed to be slowly rotating and spherical.
Whilst this considerably simplifies the analysis, real stars do not possess spherical symmetry; the centrifugal force deforms both the stellar surface and the internal surfaces of constant density by raising an equatorial bulge.
In principle this might change the natural frequencies and dissipation rates of the inertial modes, and how strongly such modes couple to a forcing potential.
Models for rotating stars are not simple, and to determine analytically even a background density gradient for this problem would be difficult.

To perform a first investigation of whether the equatorial bulge does significantly modify the tidal response of the star, we entirely neglect density variations and differential rotation within the star and consider it to be composed a homogeneous, incompressible fluid.
The steady, axisymmetric configurations of such a fluid in solid body rotation are the Maclaurin spheroids.
In principle, the normal modes of a Maclaurin spheroid of inviscid fluid have long been known, having first been derived by \citet{Bryan1889}, and can be expressed in closed form.
However, the coordinate system employed therein is cumbersome and does not lend itself well to visualising the modes.
In addition, the equation for the mode frequencies is a polynomial that cannot generally be solved analytically for modes of high degree.
For these reasons \citeauthor{Bryan1889} was unable to make many explicit numerical statements about these normal modes beyond the slow rotation limit.
Since the availability of modern computing power several applications of this work have concentrated on the (mathematically very similar) case of a spheroid with a rigid surface, with the aim of understanding waves in planetary cores.
See \citet{Zhang2004} for one such example.
\citet{LI1999} reformulated Bryan's original work in an effort to understand gravitational wave emission from neutron stars.
In the present paper we shall review the theory of the inviscid normal modes of Maclaurin spheroids with free surfaces, and derive the decay rates due to the introduction of a small viscosity.
In a subsequent paper we shall apply these results to understand the response of a tidally forced spheroid, illustrating how the response is modified significantly by the oblateness of the body.
This will be the first study of dynamical tides in a rapidly rotating dissipative body.

In section \ref{sec:maclaurin_spheroid_base} we briefly describe the Maclaurin spheroids.
In section \ref{sec:general_Bryan_mode} we review the derivation of the normal modes, largely following the notation and method of \citet{LI1999}, but emphasising that the derivation works for the high frequency (surface gravity) modes in addition to the low frequency (inertial) modes that were the focus of \citeauthor{LI1999}.
The necessary coordinate systems to carry out this derivation are described in appendices \ref{sec:maclaurin_spheroid_obl_coords} and \ref{sec:maclaurin_spheroid_bi_spheroidal_coords}, and the special cases of zero frequency modes and modes having a frequency of twice the rotation rate are described in appendices \ref{sec:zeromodes} and \ref{sec:kappa4modes} respectively.
We shall find that the modes may be grouped by their \emph{degree}, $l$, and that we should expect the modes of lower degree to be of more importance for tidal interactions.
In section \ref{sec:lowdegreemodes} we describe the modes of degree $l \le 2$, since many of these are special cases.
In particular, there are no inertial modes of degree $l \le 2$.
In section \ref{sec:sectoral_modes} we describe the \emph{sectoral modes}.
These are modes for which the azimuthal wavenumber, $m$, matches the degree, $l$, and we will see that these possess particularly simple properties.
In sections \ref{sec:degree3modes} and \ref{sec:degree4modes} we describe the modes of degree three and four respectively.
In these sections we will see the appearance of inertial modes, which we expect to be strongly excited by tidal forcing.
In section \ref{sec:eigenmode_problem} we describe the energy equation and the decay rate of the modes (detailed justification of this result is given in appendix \ref{sec:decay_rate_derivation}).
We graph the decay rates of the modes of degree four and lower.
We make concluding remarks in section \ref{sec:conclusion}.

\section{The Maclaurin Spheroid}
\label{sec:maclaurin_spheroid_base}

Consider a homogeneous, incompressible fluid occupying the oblate spheroidal volume $V$ given by
\begin{equation}\label{spheroid_cylindricals} \frac{x^2 + y^2}{R_e^2} + \frac{z^2}{R_p^2}
= \frac{\varpi^2}{R_e^2} + \frac{z^2}{R_p^2}
\le 1 \,,
\end{equation}
where $R_e > R_p$ and $\varpi$ is cylindrical polar radius measured from the $z$ axis.
We will specify the shape of the spheroid by its eccentricity, $e = \left(1 - (R_p/R_e)^2\right)^{1/2}$,
and its overall scale by the mean radius, $R$.
These are related to the equatorial and polar radii by $R = (1-e^2)^{1/6} R_e = (1-e^2)^{-1/3} R_p$.
In many calculations it will be more concise to work with the focal radius, $c = e R_e$.
Note that the volume of the spheroid is $\frac{4}{3} \pi R^3$.

Up to an additive constant, the gravitational potential within the fluid is
\begin{equation}\label{MacSpherPhi}
\Phi = \pi G \rho \left( A_1 \varpi^2 + 2\left(1-A_1\right) z^2 \right) \,,
\end{equation}
where
\begin{equation}
A_1 = \frac{(1-e^2)^{1/2}}{e^2}\left(\frac{\sin^{-1}e}{e} - \left(1-e^2\right)^{1/2}\right) \,.
\end{equation}
It is easy to show that this satisfies $\nabla^2 \Phi = 4\pi G \rho$, but considerably harder to show that it connects to an appropriate exterior potential.
A derivation of this may be found in \citet{binney2008galactic}.

\begin{figure}
\centering
\input{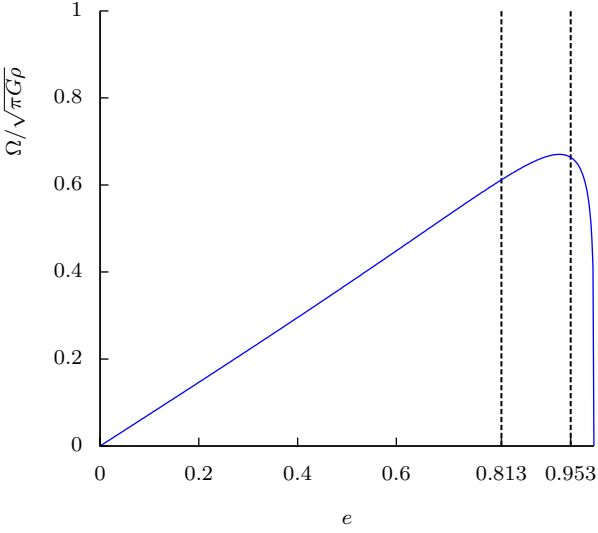}
\caption{Non-dimensionalised angular velocity ($\Omega$) for a Maclaurin spheroid against eccentricity ($e$).
For $e \gtrsim 0.8127$ the spheroid has a secular instability and for $e \gtrsim 0.9529$ it has a dynamical instability.
}
\label{fig:maclaurin_spheroid_angular}
\end{figure}

At what rate must the spheroid rotate to support an equatorial bulge of a given eccentricity?
In order for the spheroid to satisfy the static Navier-Stokes equations we require the surface of the spheroid to be an equipotential of the combined gravitational and centrifugal potentials, $\Phi - \frac{1}{2}\Omega^2 \varpi^2$.
This fixes the relation between eccentricity and angular velocity to be
\begin{equation}\label{MacSpherOmega}
\frac{\Omega^2}{\pi G \rho} = \frac{2}{e^2} \left(\left(3-2e^2\right)\sqrt{1-e^2} \frac{\sin^{-1}e}{e} - 3\left(1-e^2\right)\right) \,,
\end{equation}
as shown in figure \ref{fig:maclaurin_spheroid_angular}.
There is a maximum rotation rate, $\Omega/\sqrt{\pi G \rho} \approx 0.6703$, at which $e \approx 0.9300$, and for each slower rotation rate there exist two spheroids, with different eccentricities.
However, the high eccentricity branch is unstable:
spheroids with $e \gtrsim 0.8127$ are secularly unstable and those with $e \gtrsim 0.9529$ are dynamically unstable.
Derivations of these instabilities may be found in \citet{Chandrasekhar1987}.

The normal modes of a such a spheroid of inviscid fluid with a free boundary were derived by \citet{Bryan1889}.
We shall briefly review the derivation of these modes, largely following the notation of \cite{LI1999}, but without restricting our attention solely to the inertial modes.
In order to do this we will need to introduce two new coordinate systems.
The first of these, the \emph{oblate spheroidal coordinates}, cover the whole of space and will be used to solve the Poisson equation for the perturbation to the gravitational potential.
The second of these, the so-called \emph{bi-spheroidal coordinates}, cover only the interior of the spheroid and will be used to solve the Poincar\'{e} equation for the perturbation to the hydrodynamic potential (to be introduced in section \ref{sec:general_Bryan_mode}).
These coordinate systems are described in appendices \ref{sec:maclaurin_spheroid_obl_coords} and \ref{sec:maclaurin_spheroid_bi_spheroidal_coords} respectively.

At this stage it is convenient to introduce an alternative parametrisation for the oblateness of the spheroid, $\zeta_0^2 = (1-e^2) / e^2$.
The spherical limit, $e\rightarrow 0$, corresponds to $\zeta_0 \rightarrow \infty$ and the highly eccentric limit, $e \rightarrow 1$, to $\zeta_0 \rightarrow 0$.
This parameter arises as the location of the surface of the spheroid in the oblate coordinate system described in appendix \ref{sec:maclaurin_spheroid_obl_coords}, and many formulae will be simpler when expressed in terms of $\zeta_0$ rather than $e$.
For this reason we shall give formulae in terms of $\zeta_0$ rather than $e$ from now on.
However, we shall continue to plot graphs against $e$ since this has a finite range and is more familiar%
\footnote{For comparison it may be helpful to note that the Sun has $e=0.004$, Jupiter $e=0.354$, and Saturn $e=0.432$.}%
.
It is useful to note that
\begin{align} \label{MacSpherOmegaZ0}
\frac{\Omega^2}{\pi G \rho} &= 2 \zeta_0 \left((1+3\zeta_0^2)\cot^{-1} \zeta_0 - 3\zeta_0\right)
\end{align}
and that the pressure, $p$, is given by
\begin{align} \label{hydrostaticP}
\frac{p}{\pi G \rho^2} &= 2 \zeta_0^2 (1+\zeta_0^2)\left(1- \zeta_0 \cot^{-1} \zeta_0\right)\left(c^2 - \frac{\varpi^2}{1 + \zeta_0^2} - \frac{z^2}{\zeta_0^2}\right) \,.
\end{align}
The latter equation may be derived using the conditions that the interior of the spheroid must be in hydrostatic equilibrium and the pressure must vanish at the surface, and equations \eqref{MacSpherPhi} and \eqref{MacSpherOmegaZ0}.

\section{The General Modes of Bryan}
\label{sec:general_Bryan_mode}

Small oscillations of the fluid spheroid obey the linearised incompressible Euler equations
\begin{align}\label{inviscidNS}
\partial_t \bm{u} + 2 \bm{\Omega} \times \bm{u} = - \bm{\nabla} W \;\;
\mathrm{and} \;\; \bm{\nabla}\cdot \bm{u} = 0 \;\; \mathrm{inside} \; V \,,
\end{align}
where $W = p^{\prime}/\rho + \Phi^{\prime}$ is the \emph{hydrodynamic potential}.
We will find it easier to work in terms of the displacement, $\bm{\xi}$, related to the velocity by $\bm{u} = \partial_t \bm{\xi}$.
We use coordinates such that $\bm{\Omega} = \Omega \hat{\bm{z}}$, $\Omega > 0$, and seek modes with harmonic time-dependence $e^{-i\kappa \Omega t}$.
Note that the dimensionless mode frequency $\kappa$ differs by a sign from the convention of \citeauthor{LI1999}, and by a factor of $-2$ from Bryan's original paper.
We may eliminate the velocity field by taking both the divergence and curl of \eqref{inviscidNS} to obtain the Poincar\'{e} equation,
\begin{align} \label{W_Poincare}
\partial_x^2 W + \partial_y^2 W + \left(1 - \frac{4}{\kappa^2}\right) \partial_z^2 W = 0 \,,
\end{align}
which must be satisfied throughout the interior of the spheroid.
Since there is vacuum outside of the spheroid, the Lagrangian pressure perturbation at the surface must vanish, that is
\begin{align} \label{pbc}
p^{\prime} + \bm{\xi} \cdot \bm{\nabla} p = 0 \; \mathrm{on} \; \partial V \,.
\end{align}

The perturbation to the gravitational field, $\Phi^{\prime}$, obeys the Poisson equation sourced by the density perturbation, $\nabla^2 \Phi^{\prime} = 4\pi G \rho^{\prime}$.
Note that our $\Phi^{\prime}$ also differs in sign from that of \citeauthor{LI1999}.
Since our fluid density is constant and we are making a linear approximation in the displacement, this reduces to Laplace's equation
\begin{align} \label{Poisson_eqn}
\nabla^2 \Phi^{\prime} = 0 \; \mathrm{away}\;\mathrm{from}\; \partial V \end{align}
and a boundary condition due to a surface mass density,
\begin{align}\label{Phibc}
\left[\bm{n} \cdot \bm{\nabla} \Phi^{\prime}\right]^{\partial V^+}_{\partial V^-} &= 4 \pi G \rho \bm{n} \cdot \bm{\xi} \;\; \mathrm{across} \; \partial V.
\end{align}

Note that $\bm{u}$, $\bm{\xi}$, $W$ and all primed quantities are small Eulerian perturbations about the basic state of steady solid body rotation.
Owing to the fact that we are working to linear order in the displacement, it is consistent to impose the boundary conditions at $\partial V$, the unperturbed surface of the spheroid.

We will first solve \eqref{Poisson_eqn} for the gravitational field.
To do so we must find a coordinate system covering the whole of space in which the equation is separable and the surface of the spheroid takes a simple form.
The oblate spheroidal coordinates $(\zeta, \mu, \varphi)$ are such a coordinate system, and are described in detail in appendix \ref{sec:maclaurin_spheroid_obl_coords}.
Letting the azimuthal dependence of the mode be $e^{im\varphi}$, we may use \eqref{laplacianObl} to write
\begin{align} \label{Laplace_sep}
\frac{\partial}{\partial \zeta} \left( (1+\zeta^2) \frac{\partial \Phi^{\prime}}{\partial \zeta} \right)
+\frac{ m^2 \Phi^{\prime}}{1 + \zeta^2} & \nonumber \\
+ \frac{\partial}{\partial \mu} \left( (1-\mu^2) \frac{\partial \Phi^{\prime}}{\partial \mu} \right)
-\frac{ m^2 \Phi^{\prime}}{1 - \mu^2} & = 0 \,.
\end{align}
We may express the solutions of \eqref{Laplace_sep} in terms of associated Legendre functions.
These are the solutions of $\partial_x \left((1-x^2) \partial_x f\right) + \left( l(l+1) -m^2/(1-x^2) \right) f = 0$, conventionally denoted $P_l^m(x)$ and $Q_l^m(x)$.

Inspection of \eqref{Laplace_sep} suggests that we have four possible solutions: $P_l^m (i \zeta) P_l^m (\mu) e^{im\varphi}$, $Q_l^m (i \zeta) P_l^m(\mu) e^{im\varphi}$, $P_l^m(i\zeta) Q_l^m(\mu) e^{im\varphi}$ and $Q_l^m(i\zeta) Q_l^m(\mu) e^{im\varphi}$.
However, the third and fourth of these are ruled out by the requirement that the solution be regular on the north and south polar axes ($\mu = \pm 1$) where $Q_l^m(\mu)$ diverges.
The requirement that $P_l^m(\mu)$ be non-singular on $-1 \le \mu \le +1$ fixes $l$ and $m$ to be integers with $\left|m\right| \le l$.
Within the spheroid, only the first solution is acceptable, since $Q_l^m(i \zeta)$ diverges as $\zeta \rightarrow 0$.
Outside the spheroid, only the second solution is acceptable, since $P_l^m(i \zeta)$ diverges as $\zeta \rightarrow \infty$, whereas $Q_l^m(i\zeta) = \mathcal{O}\left(\zeta^{-(l+1)}\right)$.%
\footnote{Note that some computer algebra systems define $Q_l^m$ for imaginary arguments in such a way that it diverges as $\zeta \rightarrow \infty$.
Ambiguity can be removed by using the definition from \citet{AbramowitzStegun}
\begin{align} \label{ASQdef}
Q_l^m(x) &= e^{l\pi i} \frac{\pi^{1/2} \Gamma\left(l+m+1\right) \left(x^2 - 1\right)^{m/2}}{\Gamma\left(l+\frac{3}{2}\right)2^{l+1}x^{l+m+1}} \nonumber \\
& \times {_2}F_1\left(\frac{1}{2}l + \frac{1}{2}m + 1, \frac{1}{2}l + \frac{1}{2}m + \frac{1}{2}; l + \frac{3}{2}; \frac{1}{x^2}\right)
\end{align}
for $m\ge0$, and $Q_l^{-m}(x)=(-1)^m\frac{(l-m)!}{(l+m)!}Q_l^m(x)$.
}
Continuity of $\Phi^{\prime}$ at the spheroid's surface then fixes the solution up to a constant, $A_{l, m, \kappa}$, as
\begin{align}\label{gravsoln}
\Phi^{\prime} &= \left\{ \begin{matrix}A_{l,m,\kappa} \frac{P_l^m\left(i\zeta\right)}{P_l^m\left(i\zeta_0\right)} P_l^m\left(\mu\right) e^{-i\kappa \Omega t + i m \varphi} \;\mathrm{for}\;\; \zeta \le \zeta_0 \\
\,\,A_{l,m,\kappa} \frac{Q_l^m\left(i\zeta\right)}{Q_l^m\left(i\zeta_0\right)} P_l^m\left(\mu\right) e^{-i\kappa \Omega t + i m \varphi} \;\mathrm{for}\;\; \zeta \ge \zeta_0 . \end{matrix} \right.
\end{align}

Note from \eqref{zeta_large_r} and \eqref{mu_large_r} that, for $r \gg c$, we have $\zeta \sim r/c$ and $\mu \sim \cos \theta$.
Therefore at large distances from the spheroid we have the approximate relation $Q_l^m(i\zeta) P_l^m(\mu) e^{im\varphi} \propto r^{-l-1} P_l^m(\cos \theta) e^{im\varphi}$, the usual spherical harmonic solution of Laplace's equation.
This justifies, in an approximate sense, our assertion that tidal forcing due to a distant body will most strongly excite the small $l$ modes. 
We intend to make this more precise in a subsequent publication.

Having found the solutions for the gravitational potential we will now solve \eqref{W_Poincare} for the hydrodynamic potential.
In the bi-spheroidal coordinates $\left( \xi, \tilde{\mu}, \varphi \right)$ introduced by appendix \ref{sec:maclaurin_spheroid_bi_spheroidal_coords} we may rewrite the Poincar\'{e} operator as shown in equation \eqref{PoinBiSpher}, allowing us to rewrite \eqref{W_Poincare} as
\begin{align} \label{Poincare_sep}
\frac{\partial}{\partial \xi} \left((1-\xi^2) \frac{\partial W}{\partial \xi}\right)
- \frac{m^2 W}{1-\xi^2} & \nonumber \\
- \frac{\partial}{\partial \tilde{\mu}} \left((1-\tilde{\mu}^2) \frac{\partial W}{\partial \tilde{\mu}}\right)
+ \frac{m^2 W}{1-\tilde{\mu}^2} & = 0 \,.
\end{align}
Once again the associated Legendre functions provide four potential candidates for separable solutions, $Q_l^m (\xi) P_l^m (\tilde{\mu}) e^{im\varphi}$, $Q_l^m (\xi) Q_l^m(\tilde{\mu}) e^{im\varphi}$, $P_l^m(\xi) P_l^m(\tilde{\mu}) e^{im\varphi}$ and $P_l^m(\xi) Q_l^m(\tilde{\mu}) e^{im\varphi}$.
For $\kappa^2 < 4\left(1 + \zeta_0^2\right)$ the surface $\xi = 1$ lies within the spheroid, ruling out the first two solutions.
For $\kappa^2 > 4\left(1 + \zeta_0^2\right)$ continuity across the plane $\xi = 0$ of either $W$ or its first derivative (depending on whether $l+m$ is respectively odd or even) again rules out the first two solutions.
For $\kappa^2 > 4$ the solution $P_l^m(\xi) Q_l^m(\tilde{\mu}) e^{im\varphi}$ would diverge at the points within the spheroid where $\tilde{\mu} = \pm 1$, and for $\kappa^2 < 4$ this solution will turn out to be incapable of satisfying the boundary conditions (as described in \cite{LI1999}).
Therefore, for all values of $\kappa$, the only acceptable solution for the hydrodynamic potential is
\begin{equation}\label{Wsoln} W = D_{l,m,\kappa} \frac{P_l^m\left(\xi\right)}{P_l^m\left(\xi_0\right)} P_l^m\left(\tilde{\mu}\right) e^{-i\kappa\Omega t + im\varphi} \end{equation}
for some constant $D_{l,m,\kappa}$.

For the remainder of this paper we shall be interested in the fluid dynamics within the spheroid rather than the gravitational excitation of these modes, and will focus on the modes of degree $l\le 4$.
For this purpose it will be easier to work with explicit expressions for the modes in cylindrical polar coordinates.
These can be obtained from the hydrodynamic potential by using Rodrigues' formula
\begin{align} P_l^m(x) = \frac{(-1)^m}{2^l l!} (1-x^2)^{m/2} \frac{d^{l+m}}{dx^{l+m}} (x^2-1)^l \end{align}
and the definition \eqref{bi_spher_def} for $\xi$ and $\tilde{\mu}$ to write
\begin{align} \label{W_cylind}
W &\propto \frac{P_l^m(\xi)}{P_l^m(\xi_0)} P_l^m(\widetilde{\mu}) e^{i(m\varphi -\kappa \Omega t)} \nonumber \\
&= \frac{e^{i(m\varphi - \kappa\Omega t)}}{(2^l l!)^2 P_l^m(\xi_0)} \left(\frac{\varpi}{b}\right)^m \left(\frac{\partial^2}{\partial\xi \partial\widetilde{\mu}}\right)^{l+m} \left(\frac{\varpi}{b}\right)^{2l} \nonumber \\
&= \frac{e^{i(m\varphi - \kappa\Omega t)}}{(2^l l!)^2 P_l^m(\xi_0)} \left(\frac{\varpi}{b}\right)^m
\Bigg[b \frac{\sqrt{4-\kappa^2}}{\kappa}\left(z\frac{\partial^2}{\partial z^2} + \frac{\partial}{\partial z}\right) \nonumber \\
&\hspace{0.6cm} + b \frac{\kappa}{\sqrt{4-\kappa^2}} z \left(\frac{\partial^2}{\partial\varpi^2} + \frac{1}{\varpi}\frac{\partial}{\partial\varpi}\right) \nonumber \\
&\hspace{0.6cm} + b^3 \frac{\sqrt{4-\kappa^2}}{\kappa} \frac{1}{\varpi} \bigg(\frac{\varpi^2}{b^2} \nonumber \\
& \;\;\;\;\;\;\;\;\;\;\;\;\;\;
+ \frac{\kappa^2}{4-\kappa^2} \frac{z^2}{b^2} - 1\bigg) \frac{\partial^2}{\partial z \partial \varpi}
\Bigg]^{l+m}
\left(\frac{\varpi}{b}\right)^{2l} \,.
\end{align}
where \eqref{cylind_bi_spher_derivs} has been used to express the $\partial^2 / \partial \xi \partial \tilde{\mu}$ operator in cylindrical polars.
$b$ is a length scale that occurs in the definition of the bi-spheroidal coordinates, and is defined in \eqref{bdef}.
It is apparent from this that, for any mode, the hydrodynamic potential will be a polynomial in $\varpi$ and $z$, a fact that was already known due to \citet{Kudlick1966}.
However, the expression \eqref{W_cylind} does not appear to have been previously published.
We shall make use of it to calculate $W$, and hence $\bm{u}$, for the modes of degree $l \le 4$ in sections \ref{sec:lowdegreemodes} to \ref{sec:degree4modes}.
We have also made use of it in our calculation of decay rates in section \ref{sec:eigenmode_problem}.
Expressing these quantities in cylindrical polars will greatly simplify the calculation of the norm of the rate-of-strain tensor, $e_{ij}$, which we will need when we calculate dissipation rates.


Before we can investigate even the spatial form of the normal modes, we must apply the boundary conditions \eqref{Phibc} and \eqref{pbc} to determine the allowed values of $\kappa$.
To obtain the hydrostatic pressure gradient, $\bm{n} \cdot \bm{\nabla} p$, at the surface we use \eqref{hydrostaticP} and note that $\left(\bm{n}\cdot \bm{\nabla} p\right)^2 = \bm{\nabla} p \cdot \bm{\nabla} p$ at the surface of the spheroid.
Using \eqref{pbc} to rewrite $\bm{n} \cdot \bm{\xi}$ as
\begin{equation} \label{p_prime_over_gradient}
\frac{-p^{\prime}}{\bm{n} \cdot \bm{\nabla} p}
= \frac{\left(W - \Phi^{\prime}\right)}{4 \pi c G \rho \zeta_0 \sqrt{1+\zeta_0^2} \left(1 - \zeta_0 \cot^{-1}\zeta_0\right) \sqrt{\zeta_0^2 + \mu^2}} \,,
\end{equation}
we are able to reduce the gravitational boundary condition \eqref{Phibc} to a ($\zeta_0$ dependent) relation between the $A_{l, m, \kappa}$ and $D_{l, m, \kappa}$ coefficients
\begin{align} \label{Phibc_A_D}
B_l^m(\zeta_0) A_{l,m,\kappa} = \frac{\left(D_{l,m,\kappa} - A_{l,m,\kappa}\right)}{\zeta_0 \left(1- \zeta_0 \cot^{-1}\zeta_0\right)} \,,
\end{align}
where
\begin{align} \label{Bdef}
B_l^m(\zeta_0) = \left(1+\zeta_0^2\right) \bigg( & \frac{1}{Q_l^m(i \zeta_0)} \frac{d Q_l^m(i\zeta)}{d \zeta} \nonumber \\
& - \left. \frac{1}{P_l^m(i \zeta_0)} \frac{d P_l^m(i\zeta)}{d \zeta} \bigg) \right|_{\zeta=\zeta_0} \,.
\end{align}
Whilst \eqref{Bdef} could be rewritten using the Wronskian of $P_l^m$  and $Q_l^m$, the above form is insensitive to the normalisation and argument of these two functions, so we recommend use of this form.

To apply the pressure boundary condition \eqref{pbc} we must evaluate $\bm{\xi} \cdot \bm{n}$ explicitly and equate it to the expression on the right hand side of \eqref{p_prime_over_gradient}.
Rearranging \eqref{inviscidNS} gives
\begin{align}
\label{xiw}
\xi_{\varpi} &= \frac{1}{\Omega^2 (4 - \kappa^2)} \left(-\partial_{\varpi} W + \frac{2m}{\kappa \varpi} W\right) \,, \\
\label{xiz}
\xi_z &= \frac{1}{\Omega^2 \kappa^2} \partial_z W \,,
\end{align}
and the normal vector may be written as
\begin{align} \label{surfacenormal}
\bm{n} = \zeta_0 \sqrt{\frac{1-\mu^2}{\zeta_0^2 + \mu^2}} \bm{e}_{\varpi} + \mu \sqrt{\frac{1+\zeta_0^2}{\zeta_0^2 + \mu^2}} \bm{e}_z \,.
\end{align}
The function $W$ is a function of $\xi$ and $\tilde{\mu}$ rather than $\zeta$ and $\mu$.
However, we need only evaluate it on the surface, where either $\xi = \xi_0$ and $\tilde{\mu} = \mu$, or (on some portion of the surface when $\kappa^2 < 4$) $\xi = \mu$, $\tilde{\mu} = \pm \xi_0$.
The latter case is dealt with in detail in \citet{LI1999}, so we simply note here that exactly the same working goes through in the case that $\kappa^2 > 4$ to give
\begin{align} \label{pbc_A_D}
A(\kappa, \zeta_0) D_{l, m, \kappa} = & \frac{\left((1+3\zeta_0^2) \cot^{-1}\zeta_0 - 3\zeta_0\right)}{2\left(1- \zeta_0 \cot^{-1}\zeta_0\right)} \nonumber \\
& \times \kappa \left(D_{l,m,\kappa} - A_{l,m,\kappa}\right) \,,
\end{align}
where we define
\begin{align}
A(\kappa, \zeta_0) = \frac{1+\zeta_0^2}{\sqrt{4\left(1+\zeta_0^2\right) - \kappa^2}}\frac{1}{P_l^m(\xi_0)}\frac{dP_l^m(\xi_0)}{d\xi} + \frac{2m\zeta_0}{4-\kappa^2} \,.
\end{align}
$\xi_0$ is a function of both $\zeta_0$ and $\kappa$, as defined in equation \eqref{xi0def}.
Note that these formulae differ slightly from those of \citet{LI1999} due to the difference in our sign conventions for $\kappa$ and $\Phi^{\prime}$.

Requiring that there exist non-zero solutions $(A_{l,m,\kappa}, D_{l,m,\kappa})$ that simultaneously satisfy \eqref{Phibc_A_D} and \eqref{pbc_A_D} gives us our condition for $\kappa$ to be an allowed frequency
\begin{align} \label{freqEqn}
&\frac{2A(\kappa, \zeta_0) - \kappa \zeta_0 B_l^m(\zeta_0) \left(\left(1+3\zeta_0^2\right) \cot^{-1} \zeta_0 - 3 \zeta_0\right)}{2\zeta_0\left(1 - \zeta_0\cot^{-1}\zeta_0\right)} \nonumber \\
& + A(\kappa, \zeta_0) B_l^m(\zeta_0) = 0 \, .
\end{align}
In general this is a polynomial for $\kappa$ that we cannot solve analytically.
There are two subtleties of which the reader should be aware if attempting to solve this numerically.
Firstly, as mentioned above, some computer algebra systems define $Q_l^m$ in such a way that is does not vanish for large imaginary arguments.
In such cases the definition \eqref{ASQdef} from the footnote on page \pageref{ASQdef} may be used.
Secondly, in the case that $\kappa^2 > 4\left(1 + \zeta_0^2\right)$ we must be careful with the arguments of imaginary quantities.
It is consistent to take $\arg \sqrt{4(1+\zeta_0^2) - \kappa^2} = - \arg \xi_0 = \pi/2$.
We do not consider complex $\kappa$ since we are not interested in the dynamically unstable spheroid.
We should also point out that the reasoning above is not valid for modes with either $\kappa = 0$ or $\kappa^2 = 4$, since the oblate spheroidal coordinates no longer cover the spheroid.
These cases are considered in appendices \ref{sec:zeromodes} and \ref{sec:kappa4modes} respectively.
Solutions to \eqref{freqEqn} for $l\le4$ are plotted in figures \ref{fig:l2freqplot}, \ref{fig:l3freqplot}, \ref{fig:l3kappaplot}, \ref{fig:l4freqplot} and \ref{fig:l4kappaplot}, and numerical values at $e=0$, $0.1$, $0.2$ and $0.5$ are given in appendix \ref{sec:kappa_tables}.

Throughout this paper we shall retain the complex notation used in this section.
Note that, up to a real multiplicative constant, the complex conjugate of a mode is obtained by replacing $(m,\kappa)$ by $(-m,-\kappa)$, and that inspection of \eqref{freqEqn} shows that this is indeed an allowed frequency.
If the reader wishes to use the real part of these modes instead, they must be aware that all quadratic integrals in this paper obey
$\int_V ff^* \mathrm{d} V = 2 \int_V \mathrm{Re}(f)^2 \mathrm{d} V$
for the $m \ne 0$ modes and
$ \int_V ff^* \mathrm{d} V = \frac{\Omega \kappa}{\pi} \int_0^{2\pi/\Omega \kappa} \int_V \mathrm{Re} (f)^2 \mathrm{d} V \mathrm{d} t$
for the axisymmetric modes.
There is of course a redundancy in our formalism, in that the $(l,m,\kappa)$ mode is the conjugate of the $(l,-m,-\kappa)$ mode and contains exactly the same physical information.
We remove such redundancy by considering all values of $\kappa$, but only $m \ge 0$, in the following sections.
Note that, for $m>0$, modes with $\kappa$ greater than or less than zero have respectively prograde or retrograde pattern speeds in the rotating frame, whereas those with $\kappa$ greater than or less than $-1/m$ are respectively prograde or retrograde in the inertial frame.

\section{Modes of degree $l\le 2$}
\label{sec:lowdegreemodes}

Many of the modes of degree $l\le2$ are special cases, so we consider them first.

The $l=m=0$ mode is just the addition of a constant to the hydrodynamic potential and has no physical significance.
For $l=1$, $m=0$ there is only the zero frequency mode describing the vertical translation of the whole spheroid.
For $l=1$, $m=1$
\begin{align} W &= \Omega \varpi e^{i\varphi-i\kappa\Omega t} \,, \\
\bm{u} &= \frac{(\bm{e}_{\varphi} - i\bm{e}_{\varpi})}{2+\kappa} e^{i\varphi -i \kappa\Omega t}
= -i\frac{\bm{e}_x + i\bm{e}_y}{2+\kappa} e^{-i\kappa\Omega t} \,,
\end{align}
and \eqref{freqEqn} is satisfied only by $\kappa=-1$.
This mode has zero frequency in the inertial frame; it is the steady horizontal translation of the spheroid.

\begin{figure}
\centering
\includegraphics[scale=0.35]{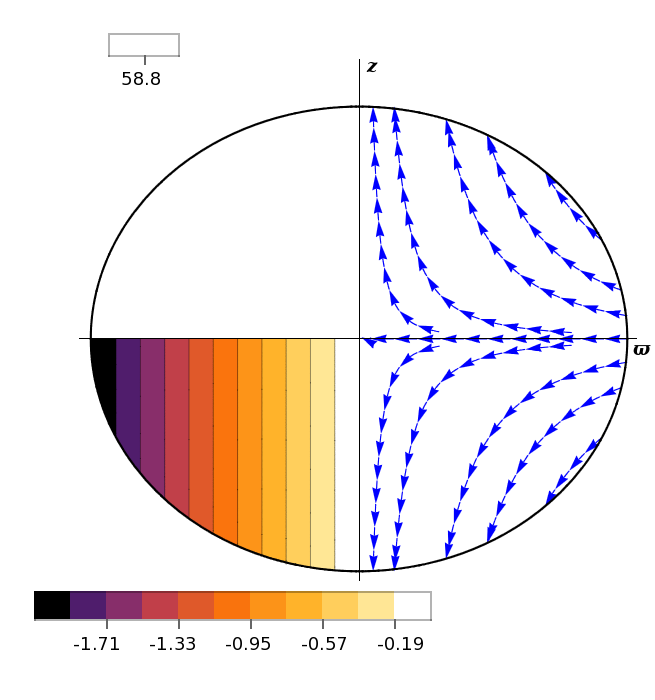}
\caption{The $l=2$, $m=0$ mode of non-zero frequency, describing the periodic spin-up / spin-down of the outer layers of the spheroid accompanied by an appropriate increase / decrease in the equatorial bulge.
The right-hand side shows the streamlines of the real meridional velocity at $t=\pi / 2 \kappa \Omega$.
The bottom-left quadrant is shaded by the real azimuthal velocity at $t=0$.
The top-left quadrant of this figure shows $e^*_{ij} e_{ij}$ (constant for this mode), which is physically the time-averaged squared magnitude of the strain rate.
These three regions are shaded according to the keys at the top, bottom and left of the figure respectively, and the numbers on these scales are correct for a spheroid of mean radius $R=1$.
The figure shown is for a spheroid of eccentricity $e=0.5$.}
\label{fig:l2m0spatial}
\end{figure}

Turning to the second degree modes, the $l=2$, $m=0$ potential and velocity may be written as
\begin{align} W &= \Omega\left((4-\kappa^2)\left(\frac{\varpi^2}{2} - \frac{b^2}{3}\right) + \kappa^2 z^2 \right) e^{-i\kappa \Omega t} \,, \\
\label{l2m0vel}
\bm{u} &= ( i\kappa(\varpi \bm{e}_{\varpi} - 2z\bm{e}_z) + 2\varpi \bm{e}_{\varphi} ) e^{-i\kappa \Omega t} \,.
\end{align}
The $\kappa=0$ solution, with $\bm{u} = 2 \varpi \bm{e}_{\varphi}$, corresponds to increasing the angular velocity of the spheroid with an appropriate increase in the eccentricity.
This is the \emph{spin-up mode}.
Note that the expression above gives the correct spatial dependence of $W$ for this zero mode, but not the correct constant - the spin-up mode properly is a superposition of this mode and the $l=m=0$ mode.
There is a non-zero solution to \eqref{freqEqn} for $l=2$, $m=0$, which may be found analytically as
\begin{align}
\kappa^2 = \frac{1+\zeta_0^2}{1 + 3\zeta_0^2}\left((1-9\zeta_0^2) + \frac{12\zeta_0\left(1 - \zeta_0\cot^{-1}\zeta_0 \right)}{(1+3\zeta_0^2)\cot^{-1}\zeta_0 - 3 \zeta_0} \right) \,.
\end{align}
In the spherical limit this describes an axisymmetric surface gravity mode in which the motion is purely meridional ($\kappa \propto 1/\Omega$ in this limit, giving a finite frequency and rendering the $\bm{e}_{\varphi}$ term in \eqref{l2m0vel} negligible).
For non-zero rotation rates, this mode is modified.
The mode now possesses a non-zero azimuthal velocity due to the Coriolis force acting on fluid elements moving towards and away from the rotation axis, and the gravitational restoring force is partially counter-acted by the centrifugal force.
We illustrate the velocity field of this mode, for a spheroid of eccentricity $e=0.5$, and the associated (constant) squared norm of the rate-of-strain tensor, $e_{ij}^* e_{ij} = 6\kappa^2$, on a meridional slice in figure \ref{fig:l2m0spatial}.
The rate-of-strain tensor is defined to be
$e_{ij} = \frac{1}{2} \left( \partial u_i / \partial x_j + \partial u_j / \partial u_i \right)$,
and we shall make use of this quantity in section \ref{sec:eigenmode_problem} when we compute the decay rate of a mode.

The general expressions for the $l=2$, $m=1$ modes are
\begin{align}
W &= \kappa (2+\kappa) \Omega \varpi z e^{i \varphi - i\kappa\Omega t} \,, \\
\bm{u} &= \left(\kappa z (\bm{e}_{\varphi} - i\bm{e}_{\varpi}) -i(2+\kappa)\varpi \bm{e}_z \right) e^{i\varphi - i \kappa\Omega t} \,.
\end{align}
Here also there is a $\kappa = -1$ solution to \eqref{freqEqn} which is steady in the inertial frame.
The velocity field is
$\bm{u} = i\left(z(\bm{e}_x + i\bm{e}_y) - \varpi\bm{e}_z e^{i \varphi}\right)e^{- i\Omega t}$,
corresponding to a perturbation of $\delta \bm{u} = \left(\bm{e}_x + i \bm{e}_y\right) \times \bm{x}$ in the inertial frame, which is an addition of $\delta \bm{\Omega} = \bm{e}_x + i \bm{e}_y$ to the angular velocity of the spheroid.
Taking the real part, we see that physically this is an addition of only $\bm{e}_x$ to the angular velocity;
real modes corresponding to arbitrary orientations of $\delta \bm{\Omega}$ in the $x$-$y$ plane may be achieved by combining this mode and its complex conjugate with different phases.
This is the \emph{spin-over mode}, and is equivalent to tilting the rotation axis through an infinitesimal angle.

The $l=2$, $m=2$ mode is a member of the family of sectoral modes, which we describe in section \ref{sec:sectoral_modes}, so we shall not comment on it further here.

\begin{figure}
\centering
\input{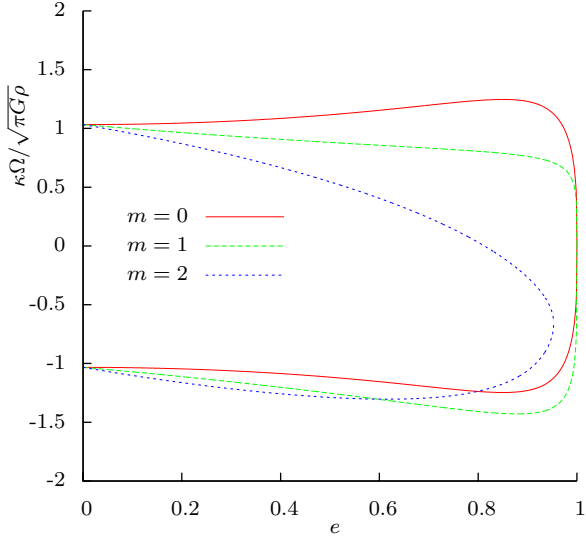}
\caption{Frequencies in the rotating frame of the degree $l=2$ modes, non-dimensionalised by the dynamical frequency of the spheroid, ($\kappa \Omega / \sqrt{\pi G \rho}$) against the eccentricity of the spheroid ($e$).
Note that the order $m=2$ modes become dynamically unstable at $e \approx 0.9529$, and hence no longer possess real frequencies.
}
\label{fig:l2freqplot}
\end{figure}

For the modes for which $\kappa + m \ne 0$ (that is, those that might be said to represent genuine waves) the frequencies are plotted in figure \ref{fig:l2freqplot}.
Note that in the low eccentricity limit these agree with the expression of \cite{Thomson1863} for the frequency of a surface gravity wave on a non-rotating sphere of fluid
\begin{align} \label{sphereSurfGrav}
\frac{\omega^2}{\pi G \rho} = \frac{8 l (l-1)}{3(2l+1)} \,.
\end{align}
All these modes have vanishing frequency as $e\rightarrow 1$.
Physically this can be understood by the lack of a restoring force;
the spheroid becomes increasingly thin and extended causing the surface gravity to become very weak, whilst the fact that $e\rightarrow 1$ is a non-rotating limit implies that the Coriolis force vanishes.
However, since this limit is dynamically unstable it is of very little astrophysical interest.

\section{Sectoral modes}
\label{sec:sectoral_modes}

\begin{figure*}
\centering
\includegraphics[scale=0.5]{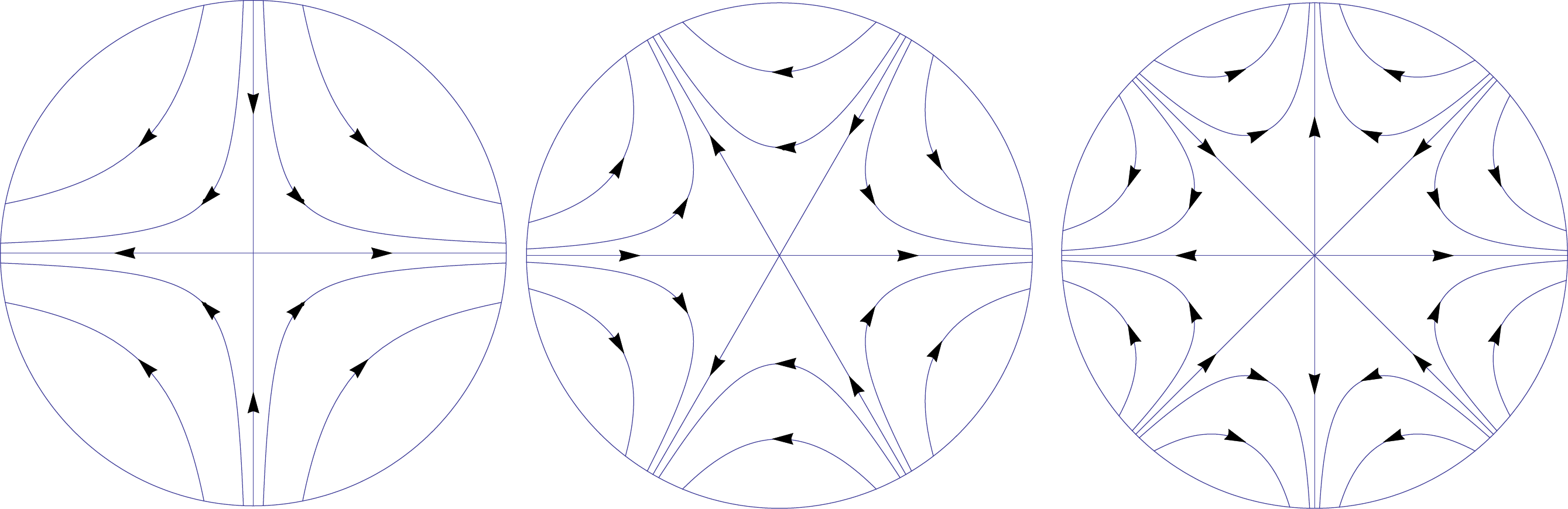}
\caption{Streamlines in the meridional plane of the spheroid for the sectoral ($l=m$) modes.
Shown from left to right are the $l=m=2$, $3$ and $4$ modes.
Since the velocity field of such a flow is independent of $z$ these are sufficient to illustrate the flow throughout the spheroid.
}
\label{fig:sectoral_streamlines}
\end{figure*}

We call a mode with $l=m$ a \emph{sectoral} mode of the spheroid.
For such modes the hydrodynamic potential may be written as
\begin{equation}
W = \left(2+\kappa\right)\Omega\varpi^l i e^{il\varphi - i\kappa\Omega t} \,.
\end{equation}
The velocity field is purely horizontal and may be written using a stream function
\begin{align}
\mathbf{u} = l \varpi^{l-1} \left(\mathbf{e}_{\varpi} + i\mathbf{e}_{\varphi}\right) e^{il\varphi - i\kappa \Omega t} \nonumber \\
= \bm{\nabla} \times \left( -i \varpi^l e^{il\varphi - i\kappa \Omega t} \bm{e}_z \right) \,.
\end{align}
Note that the fluid moves as rigid columns, and to illustrate this motion it is sufficient to plot the streamlines of the flow in the equatorial plane, as we have done in figure \ref{fig:sectoral_streamlines}.
For a general sectoral mode the squared norm of the rate-of-strain tensor is $e_{ij}^* e_{ij} = 4 l^2 \left(l-1\right)^2 \varpi^{2l-4}$, which is increasingly concentrated near the surface and the equator for increasing $l$.
However, we do not expect these modes to play a significant role in the tidal evolution of the system (or the spin-up of the outer layers of the spheroid), since their resonant frequencies are well above typical forcing frequencies.

For the sectoral modes the frequency equation \eqref{freqEqn} may be rewritten as the quadratic
\begin{equation} \label{sectoralFreqEqn}
\kappa^2 + 2 \kappa - 2 l \left( \frac{1 + \zeta_0\left(1 - \zeta_0 \cot^{-1} \zeta_0\right)B_l^m(\zeta_0)}{B_l^m(\zeta_0)\left((1+3\zeta_0^2)\cot^{-1}\zeta_0 - 3\zeta_0\right)} \right) = 0 \,.
\end{equation}
However, this obscures the origin of the terms in this equation.
If we seek modes with no $z$ dependence, purely horizontal displacement, and harmonic $\varphi$ dependence, then incompressibility forces us to take $\bm{\xi} \propto \bm{e}_{\varpi} + i \bm{e}_{\varphi}$.
Substituting this into the Euler equation \eqref{inviscidNS} allows us to write
\begin{align}
\kappa^2 + 2 \kappa = \frac{\int_V \frac{1}{2} \rho \bm{\xi}^* \cdot \bm{\nabla} W \mathrm{d}V}{\int_V \frac{1}{2}\rho \left| \bm{\xi}\right|^2 \mathrm{d} V} \,.
\end{align}
It is shown in equation \eqref{potentialEnergyIntegral} that the numerator of the right hand side is the potential energy of the displacement, and it is now evident that the linear term in \eqref{sectoralFreqEqn} is due to the Coriolis force.
We are also now able to explain the stability results described in section \ref{sec:maclaurin_spheroid_base}.
For $e \gtrsim 0.8127$ the potential energy of the $l=m=2$ surface displacement is negative, and a spheroid that is able to dissipate its kinetic energy through viscosity may continue to deform itself by increasing the amplitude of this mode, at least until linear theory is no longer valid.
However, for $0.8127 \lesssim e \lesssim 0.9529$ the inviscid mode itself is stabilised by the Coriolis force, manifested by the roots of \eqref{sectoralFreqEqn} still being real due to the term linear in $\kappa$.

In \citet{LI1999} the authors remarked that, for an inertial mode having $l=m+1$ (termed a \emph{classical r-mode} in that paper), the Lagrangian pressure perturbation, $\Delta p = p^{\prime} + \bm{\xi}\cdot \bm{\nabla} p$, vanishes throughout the spheroid.
This is due to the fact that $\Phi^{\prime} \propto W \propto z \varpi^m$ for such a mode.
Since this function is just a product of two powers, we also have $\bm{\xi}\cdot\bm{\nabla}p \propto z \varpi^m$, as may be seen by inspection of equations \eqref{hydrostaticP}, \eqref{xiw} and \eqref{xiz}.
This yields $\Delta p \propto z \varpi^m$.
Since $\Delta p$ must vanish on the surface of the spheroid, the constant of proportionality must vanish, therefore $\Delta p = 0$ throughout the spheroid for a mode having $l=m+1$.
We note in addition the following facts:
$\Phi^{\prime} \propto W \propto z \varpi^m$ also holds for surface gravity modes having $l=m+1$, and hence $\Delta p = 0$ throughout the spheroid for such modes;
the sectoral modes described in this section have $\Phi^{\prime} \propto W \propto \varpi^l$, and therefore have $\Delta p =0$ throughout the spheroid by the same argument as above;
modes having $\left| l - m \right| > 1$ do not have the above properties for $W$ and $\Phi^{\prime}$, and thus there is no reason to expect such modes (either inertial or surface gravity) to have vanishing $\Delta p$ throughout the interior of the spheroid.
The computer algebra routines used to produce the velocity fields and frequencies shown in the following sections were used to examine the Lagrangian pressure perturbation at $e=0.1$, $0.2$ and $0.5$.
In each case we found that $\Delta p$ vanished throughout the interior of the spheroid for modes having $l=m$ or $m+1$, and that none of the other modes described in this paper had this property.

\section{Modes of degree $l=3$}
\label{sec:degree3modes}

\begin{figure}
\centering
\input{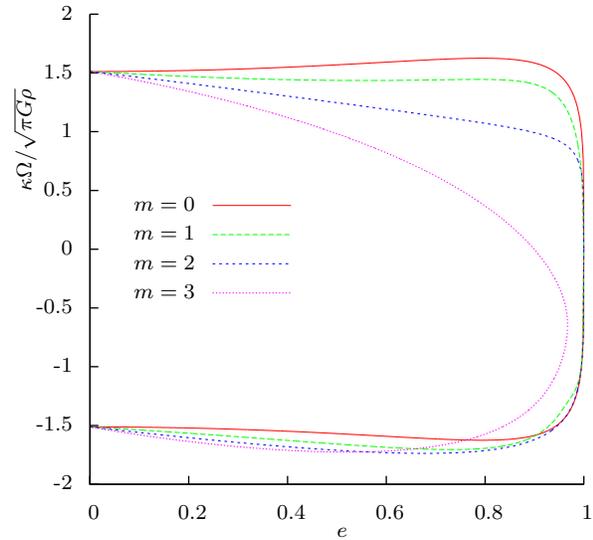}
\caption{Mode frequencies in the frame rotating with the spheroid, in units of the dynamical frequency of the spheroid ($\kappa \Omega / \sqrt{\pi G \rho}$) versus the eccentricity of the spheroid ($e$) for the surface gravity modes of degree $l=3$.
}
\label{fig:l3freqplot}
\end{figure}

Solving \eqref{freqEqn} for $l=3$ and $0 \le m \le 3$ we find two different behaviours as $e\rightarrow 0$.
Firstly, there are modes for which the frequency tends to that of a surface gravity wave on a sphere, as given by the expression \eqref{sphereSurfGrav}.
These frequencies are plotted in figure \ref{fig:l3freqplot} and we shall use the term \emph{surface gravity mode} to refer to these modes, even though they will actually be modified by the Coriolis and centrifugal forces for non-zero rotation rates.
Secondly, there are modes whose frequencies vanish linearly with the eccentricity (and hence the rotation rate) of the spheroid.
It is more helpful to plot the ratio of such frequencies to the rotation rate of the spheroid, and we do so in figure \ref{fig:l3kappaplot}.
The frequencies for pure inertial waves of degree three (that is, those that do not move the surface and which are restored only by the Coriolis force) in a slowly rotating sphere are shown in \cite{Ogilvie2013} to be
\begin{align}
\label{spherel3}
\kappa = \frac{2}{3}\left( -m \pm \left( \frac{9-m^2}{5} \right)^{1/2} \right)
\end{align}
for degrees $-2\le m \le 2$.
Our modes agree with this expression in the spherical limit, with the caveat that we do not believe that the $m=2$, $\kappa = -2$ mode exists.
Henceforth we use the term \emph{inertial mode} to refer to a mode with a linearly vanishing frequency in the $e \rightarrow 0$ limit.
We expect $-2 < \kappa < 2$ for inertial modes, the regime in which the Poincar\'{e} equation is hyperbolic.
However, surface gravity modes may also enter this $\kappa$ range for highly eccentric spheroids (inspection of figures \ref{fig:l2freqplot}, \ref{fig:l3freqplot} and \ref{fig:l4freqplot} shows that the sectoral modes do so while the spheroid is still secularly stable).
The reader should also be aware that, for non-zero eccentricities, the inertial modes do in fact move the surface of the spheroid, and so are restored by a combination of the Coriolis and gravitational forces.

\begin{figure}
\centering
\includegraphics[scale=0.35]{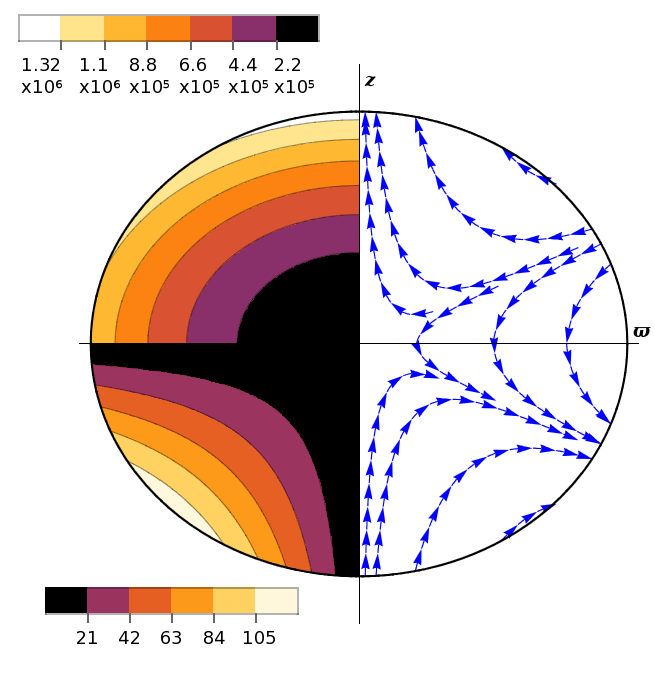}
\caption{
The $l=3$, $m=0$ surface gravity mode in a spheroid of unit mean radius and eccentricity $e=0.5$.
Quadrants and colour-keys are as described beneath figure \ref{fig:l2m0spatial}.
}
\label{fig:l3m0surfgrav}
\end{figure}

\begin{figure}
\centering
\includegraphics[scale=0.35]{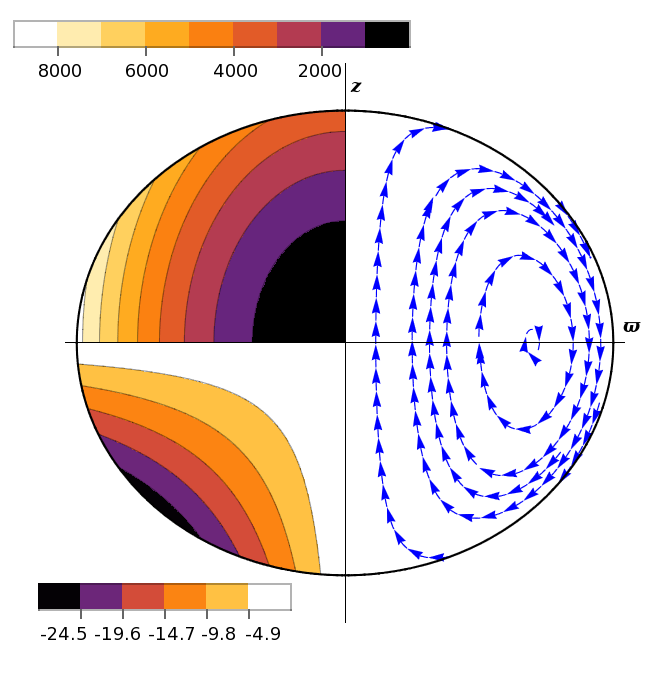}
\caption{
The $l=3$, $m=0$ inertial mode in a spheroid of unit mean radius and eccentricity $e=0.5$.
Quadrants and colour-keys are as described beneath figure \ref{fig:l2m0spatial}.
}
\label{fig:l3m0inert}
\end{figure}

The axisymmetric modes, $m=0$, may be written as
\begin{align} \label{l3m0W}
W =& \Omega \kappa z \left(3\left(\kappa^2 -4\right)\left(2b^2 - 5\varpi^2\right) + 10\kappa^2 z^2 \right) e^{-i\kappa \Omega t} \,, \\
\label{l3m0u}
\bm{u} =& 3 \Big( 10 \kappa \varpi z \left(2 \bm{e}_{\varphi} + i \kappa \bm{e}_{\varpi} \right) \nonumber \\
&-i \left((\kappa^2 - 4)(2b^2 - 5\varpi^2) + 10 \kappa^2 z^2 \right) \bm{e}_z \Big) e^{-i\kappa \Omega t} \,,
\end{align}
and the squared norm of the local rate-of-strain tensor is (after time-averaging over one period)
\begin{align} \label{l3m0e}
e_{ij}^* e_{ij} = 1800 \left(\left(\kappa^4 - 3 \kappa^2 + 4 \right) \varpi^2 + 3 \kappa^4 z^2 \right) \,.
\end{align}
Meridional sections of the $m=0$ surface gravity and inertial modes are shown in figures \ref{fig:l3m0surfgrav} and \ref{fig:l3m0inert} respectively;
one can see that the different frequencies produce remarkably different spatial dependence from the same expressions \eqref{l3m0W}, \eqref{l3m0u} and \eqref{l3m0e}.

\begin{figure}
\centering
\input{figures/kappaplots/l3kappaplot/l3kappaplot}
\caption{Mode frequencies in the frame rotating with the spheroid, in units of the angular velocity of the spheroid ($\kappa$) versus the eccentricity of the spheroid ($e$) for the inertial modes of degree $l=3$.}
\label{fig:l3kappaplot}
\end{figure}

\begin{figure}
\centering
\includegraphics[scale=0.35]{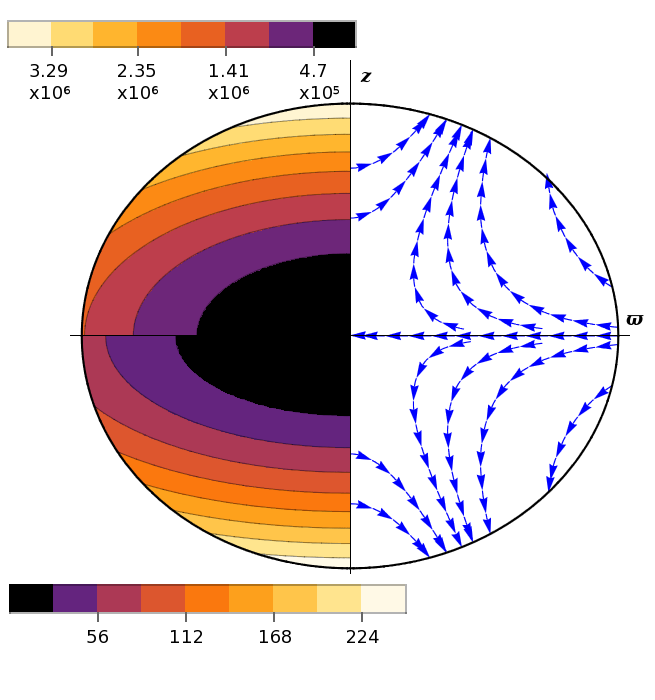}
\caption{
The $l=3$, $m=1$ prograde surface gravity mode in a spheroid of unit mean radius and eccentricity $e=0.5$.
Quadrants and colour-keys are as described beneath figure \ref{fig:l2m0spatial}.
}
\label{fig:l3m1prosurfgrav}
\end{figure}

\begin{figure}
\centering
\includegraphics[scale=0.35]{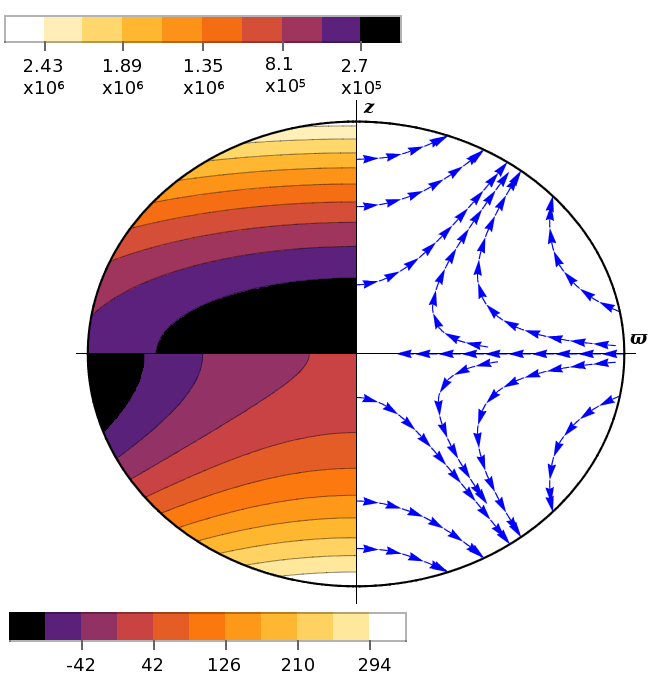}
\caption{
The $l=3$, $m=1$ retrograde surface gravity mode in a spheroid of unit mean radius and eccentricity $e=0.5$.
Quadrants and colour-keys are as described beneath figure \ref{fig:l2m0spatial}.
}
\label{fig:l3m1retsurfgrav}
\end{figure}

\begin{figure}
\centering
\includegraphics[scale=0.35]{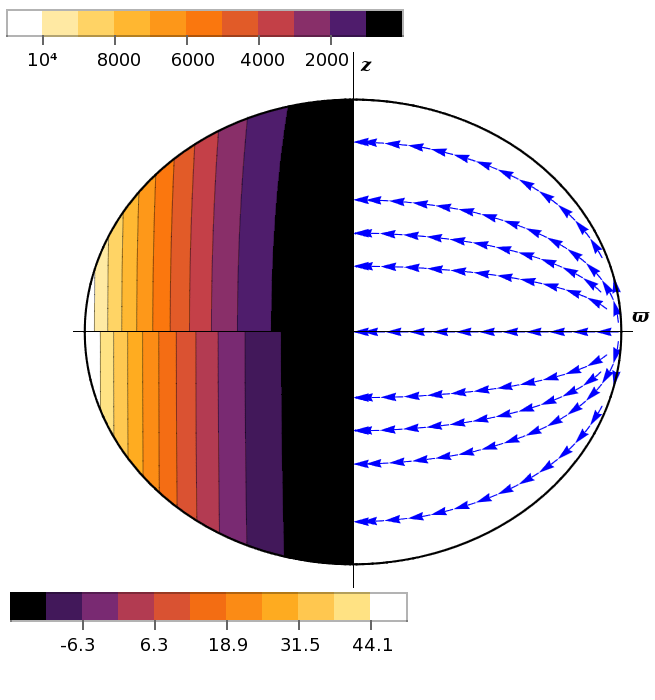}
\caption{
The $l=3$, $m=1$ prograde inertial mode in a spheroid of unit mean radius and eccentricity $e=0.5$.
Quadrants and colour-keys are as described beneath figure \ref{fig:l2m0spatial}.
}
\label{fig:l3m1proinert}
\end{figure}

\begin{figure}
\centering
\includegraphics[scale=0.35]{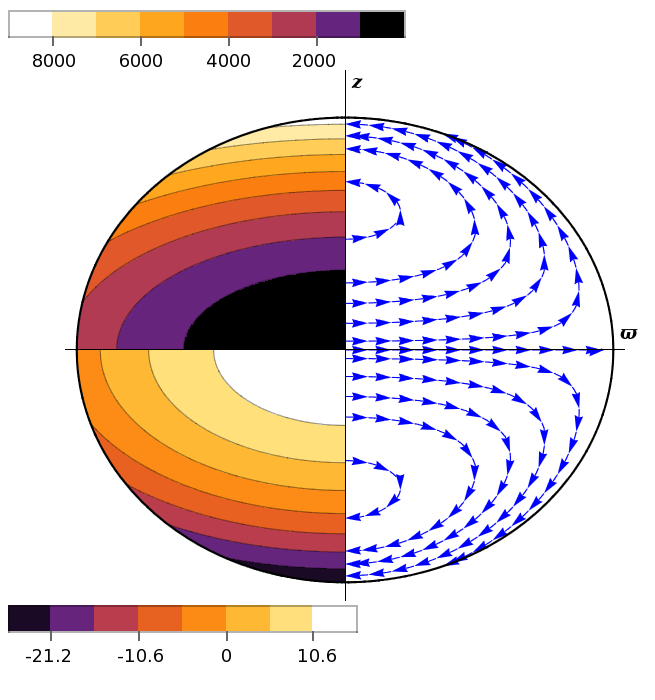}
\caption{
The $l=3$, $m=1$ retrograde inertial mode in a spheroid of unit mean radius and eccentricity $e=0.5$.
Quadrants and colour-keys are as described beneath figure \ref{fig:l2m0spatial}.
}
\label{fig:l3m1retinert}
\end{figure}

The $m=1$ modes may be written as
\begin{align} W = & \Omega \varpi \left(2+\kappa\right) \nonumber \\
& \times \left(\left(\kappa^2 - 4\right)\left(4b^2 - 5\varpi^2\right) + 20\kappa^2 z^2\right) e^{i\varphi - i\kappa \Omega t} \,, \\
\bm{u} = \bigg(i&\left(4b^2\left(4-\kappa^2\right) - 20\kappa^2 z^2 + 5\left(2+\kappa\right)\left(3\kappa-2\right)\varpi^2\right)\bm{e}_{\varpi} \nonumber \\
-&\left(4b^2\left(4 - \kappa^2\right) - 20\kappa^2 z^2 + 5\left(2+\kappa\right)\left(\kappa-6\right)\varpi^2\right) \bm{e}_{\varphi} \nonumber \\
-&40 i \kappa\left(2+\kappa\right) \varpi z \bm{e}_z
\bigg) e^{i\varphi - i \kappa \Omega t} \,,
\end{align}
and the norm of the local rate-of-strain tensor is
\begin{align}
e_{ij}^* e_{ij} = 400\big( & \left(2+\kappa\right)^2\left(7 \kappa^2 - 4 \kappa + 4\right)\varpi^2 \nonumber \\
& + 16\kappa^2\left(1+\kappa\right)^2 z^2 \big) \,.
\end{align}
There are again two surface gravity modes and two inertial modes, the frequencies of which are shown in figures \ref{fig:l3freqplot} and \ref{fig:l3kappaplot} respectively.
The surface gravity modes are shown in figures \ref{fig:l3m1prosurfgrav} and \ref{fig:l3m1retsurfgrav} and the inertial modes in figures \ref{fig:l3m1proinert} and \ref{fig:l3m1retinert}.

Inspection of figure \ref{fig:l3freqplot} shows a relatively modest change in the frequencies of the surface gravity modes between a sphere and a spheroid of $e=0.5$ (the difference between the frequencies is in fact $\approx 15\%$).
However, figures \ref{fig:l3m1prosurfgrav} and \ref{fig:l3m1retsurfgrav} shows a more significant difference in the form of the velocity fields.
The velocity of the prograde mode is reduced along the rotation axis, whereas the velocity of the retrograde mode is reduced around the equatorial belt.
The difference in $e_{ij}^* e_{ij}$ between the two modes also obeys this pattern, though less dramatically.

The difference between the two inertial modes is even more pronounced (though this is less surprising, since we should not expect these modes to be identical in the spherical limit).
The prograde mode, shown in figure \ref{fig:l3m1proinert}, has very little vertical structure in its velocity field.
We can see from figure \ref{fig:l3kappaplot} that this mode has a very low frequency ($\approx 0.17 \Omega$), therefore we should expect the flow to be very nearly organised into Taylor-Proudman columns.
With a frequency almost ten times higher, the retrograde mode, shown in figure \ref{fig:l3m1retinert}, is under no such constraint.
The flow is somewhat similar to that of the axisymmetric inertial mode shown in figure \ref{fig:l3m0inert}.

\begin{figure}
\centering
\includegraphics[scale=0.35]{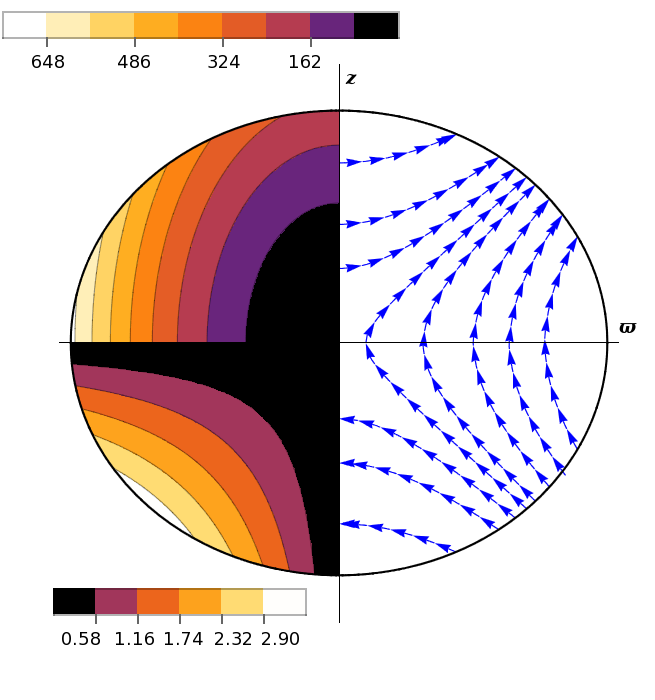}
\caption{
The $l=3$, $m=2$ prograde surface gravity mode in a spheroid of unit mean radius and eccentricity $e=0.5$.
Quadrants and colour-keys are as described beneath figure \ref{fig:l2m0spatial}.
}
\label{fig:l3m2prosurfgrav}
\end{figure}

\begin{figure}
\centering
\includegraphics[scale=0.35]{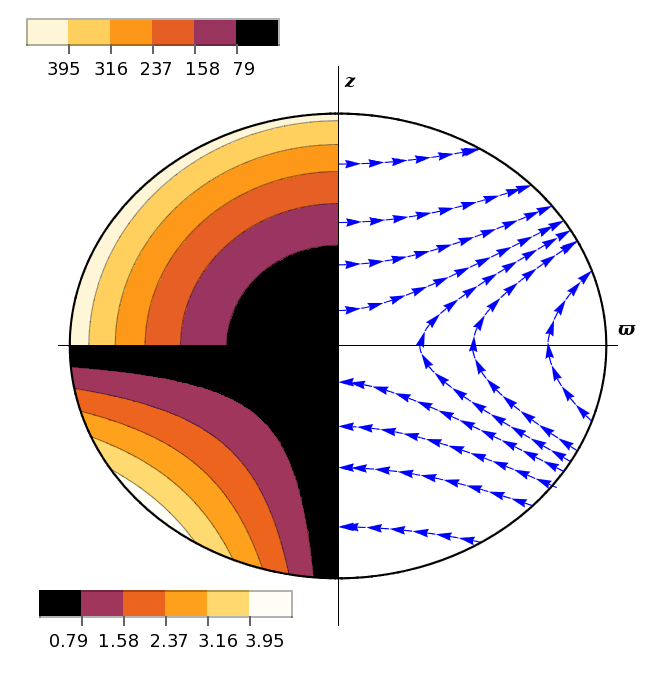}
\caption{
The $l=3$, $m=2$ retrograde surface gravity mode in a spheroid of unit mean radius and eccentricity $e=0.5$.
Quadrants and colour-keys are as described beneath figure \ref{fig:l2m0spatial}.
}
\label{fig:l3m2retsurfgrav}
\end{figure}

The $m=2$ modes may be written as
\begin{align} W & = \Omega \kappa \left(2+\kappa\right) \varpi^2 z e^{2i\varphi - i \kappa \Omega t} \,, \\
\bm{u} & = \left(
2 \kappa \varpi z (\bm{e}_{\varphi} - i \bm{e}_{\varpi})
- i(2 + \kappa) \varpi^2 \bm{e}_z
\right)e^{2i\varphi - i\kappa \Omega t} \,,
\end{align}
and the norm of the local rate-of-strain tensor is
\begin{align} e_{ij}^* e_{ij} = 16\left(\left(1+\kappa\right)^2 \varpi^2 + \kappa^2 z^2\right) \,. \end{align}
At this order there are two surface gravity modes, one prograde and one retrograde, and a single retrograde inertial mode.
By an eccentricity of $e=0.5$ the frequencies of the surface gravity modes differ by $\approx 37\%$, but figures \ref{fig:l3m2prosurfgrav} and \ref{fig:l3m2retsurfgrav} shows only modest qualitative variation between their velocity fields.
The norm of the rate-of-strain tensor shows more clearly the difference between the modes, with the strain of the prograde mode being more heavily concentrated towards the equatorial belt.
The flow associated with the inertial mode is qualitatively similar to the axisymmetric inertial mode, and the retrograde inertial mode of order $m=1$.

\begin{figure}
\centering
\includegraphics[scale=0.35]{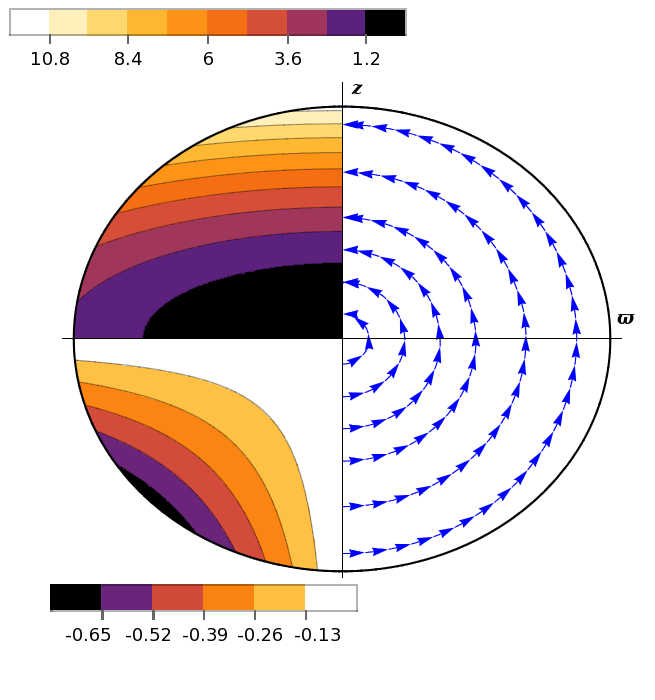}
\caption{
The $l=3$, $m=2$ inertial mode in a spheroid of unit mean radius and eccentricity $e=0.5$.
Quadrants and colour-keys are as described beneath figure \ref{fig:l2m0spatial}.
}
\label{fig:l3m2inert}
\end{figure}

Finally, note that the sectoral ($l=m=3$) modes are described in section \ref{sec:sectoral_modes}.
Despite the significant shift of frequencies, as shown in figure \ref{fig:l3freqplot}, the spatial form of the velocities and rate-of-strain tensors of these two modes will be identical.

\section{Modes of degree $l=4$}
\label{sec:degree4modes}

\begin{figure}
\centering
\input{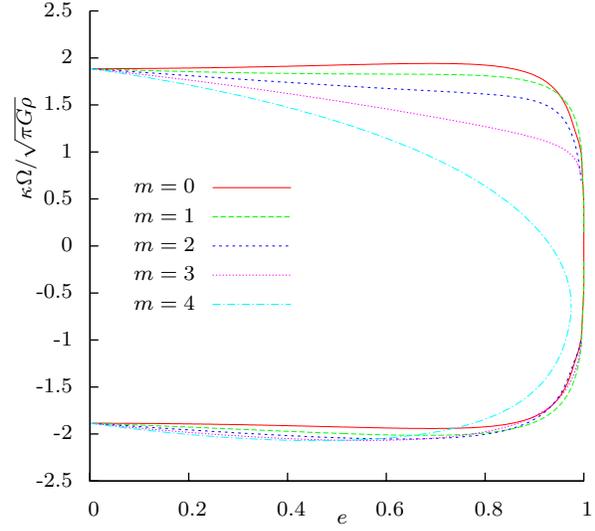}
\caption{
Mode frequencies in the frame rotating with the spheroid, in units of the dynamical frequency of the spheroid ($\kappa \Omega / \sqrt{\pi G \rho}$) versus the eccentricity of the spheroid ($e$) for the surface gravity modes of degree $l=4$.
}
\label{fig:l4freqplot}
\end{figure}

The overall view of the modes at degree 4 is largely similar to those of degree 3.
In the limit $e\rightarrow 0$ the allowed frequencies either match those of surface gravity modes (equation \eqref{sphereSurfGrav}) or inertial waves, the latter being shown in \citet{Ogilvie2013} to satisfy
\begin{align}
\label{spherel4}
42 \kappa^3 + 63 m \kappa^2 - 36(2 - m^2) \kappa - 4 m (11- 2 m^2) = 0
\end{align}
for degree 4 and $-3 \le m \le 3$.
These frequencies are plotted in figures \ref{fig:l4freqplot} and \ref{fig:l4kappaplot} respectively.

\begin{figure}
\centering
\input{figures/kappaplots/l4kappaplot/l4kappaplot}
\caption{
Mode frequencies in the frame rotating with the spheroid, in units of the angular velocity of the spheroid ($\kappa$) versus the eccentricity of the spheroid ($e$) for the inertial modes of degree $l=4$.
}
\label{fig:l4kappaplot}
\end{figure}

\begin{figure}
\centering
\includegraphics[scale=0.35]{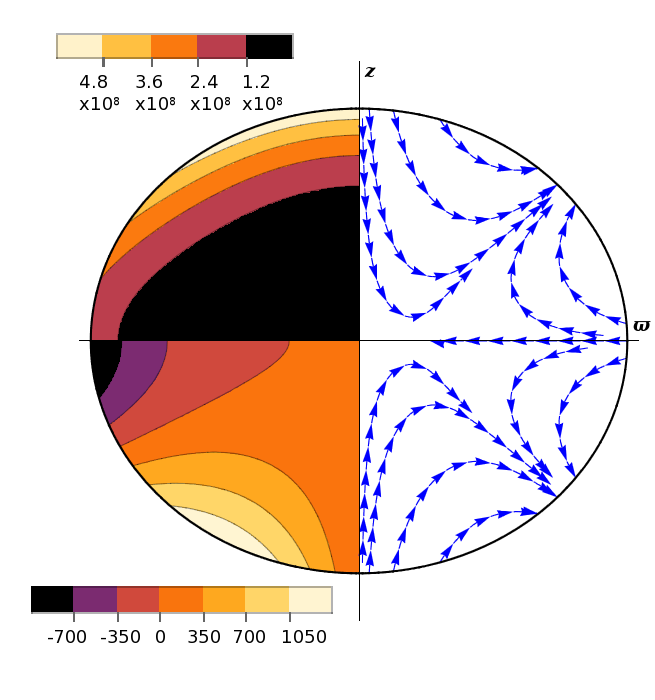}
\caption{
The $l=4$, $m=0$ surface gravity mode in a spheroid of unit mean radius and eccentricity $e=0.5$.
Quadrants and colour-keys are as described beneath figure \ref{fig:l2m0spatial}.
}
\label{fig:l4m0surfgrav}
\end{figure}
\begin{figure}
\centering
\includegraphics[scale=0.35]{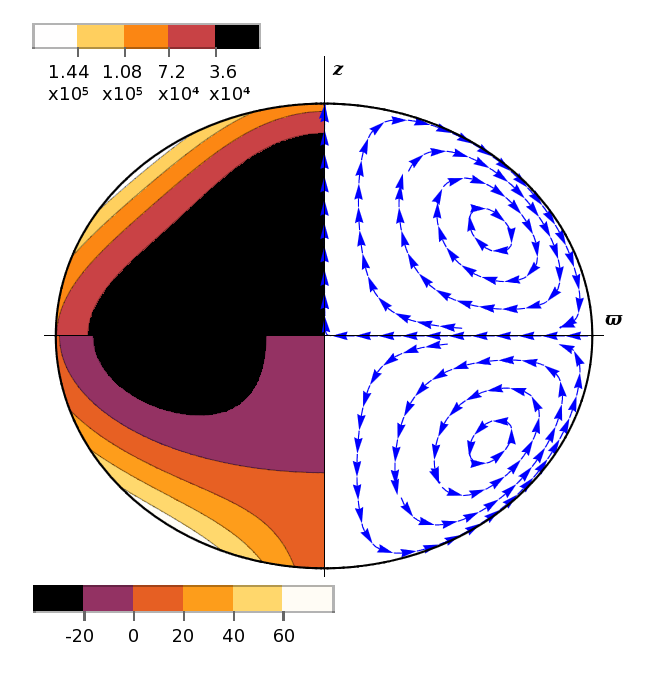}
\caption{
The $l=4$, $m=0$ inertial mode in a spheroid of unit mean radius and eccentricity $e=0.5$.
Quadrants and colour-keys are as described beneath figure \ref{fig:l2m0spatial}.
}
\label{fig:l4m0inert}
\end{figure}

The axisymmetric mode, $m=0$, may be written as
\begin{align} \label{l4m0W}
W &= \Omega \bigg( \left(\kappa^2 - 4\right) \bigg(\frac{3}{20} \left(\kappa^2 - 4\right)\left(8 b^2 - 40 b^2 \varpi^2 + 35 \varpi^4\right) \nonumber \\
& + 6 \kappa^2 \left(2b^2 - 7 \varpi^2\right)z^2 \bigg) + 14 \kappa^4 z^4 \bigg) e^{-i\kappa \Omega t} \,,
\end{align}
\begin{align} \label{l4m0u}
\bm{u} = & \left(\kappa^2-4\right)\Big( 12 b^2 \left(i\kappa\left(\varpi\bm{e}_{\varpi} - 2z\bm{e}_z\right) + 2\varpi\bm{e}_{\varphi}\right) \nonumber \\
& - 21\varpi^2\left(i\kappa\left(\varpi\bm{e}_{\varpi}-4z\bm{e}_z\right) + 2\varpi\bm{e}_{\varphi}\right)\Big) e^{-i\kappa \Omega t} \nonumber \\
+ & 28 z^2 \kappa^2\left(i\kappa\left(3\varpi\bm{e}_{\varpi} - 2z\bm{e}_z\right) + 6\varpi\bm{e}_{\varphi}\right) e^{-i\kappa\Omega t} \,,
\end{align}
and the rate-of-strain tensor has the azimuthally-averaged squared norm
\begin{align}
e_{ij}^* e_{ij} = 9\bigg( & \left(\kappa^2-4\right)^2\left(48b^4\kappa^2 - 336b^2\kappa^2\varpi^2 + 49\left(4+13\kappa^2\right)\varpi^4\right) \nonumber \\
& + 112\kappa^2\left(6b^2\kappa^2\left(\kappa^2-4\right) + 7\left(\kappa^4+16\right)\varpi^2\right)z^2 \nonumber \\
& + 2352 \kappa^6 z^4 \bigg) \,.
\end{align}
Solving \eqref{freqEqn} numerically we find two surface gravity modes and two inertial modes, whose frequencies are shown in figures \ref{fig:l4freqplot} and \ref{fig:l4kappaplot} respectively.
Since the positive and negative frequency modes are simply conjugates of each other for $m=0$, physically this is a single inertial and a single surface gravity mode.
The velocity and rate-of-strain fields of these are plotted in figures \ref{fig:l4m0surfgrav} and \ref{fig:l4m0inert}.

\begin{figure}
\centering
\includegraphics[scale=0.35]{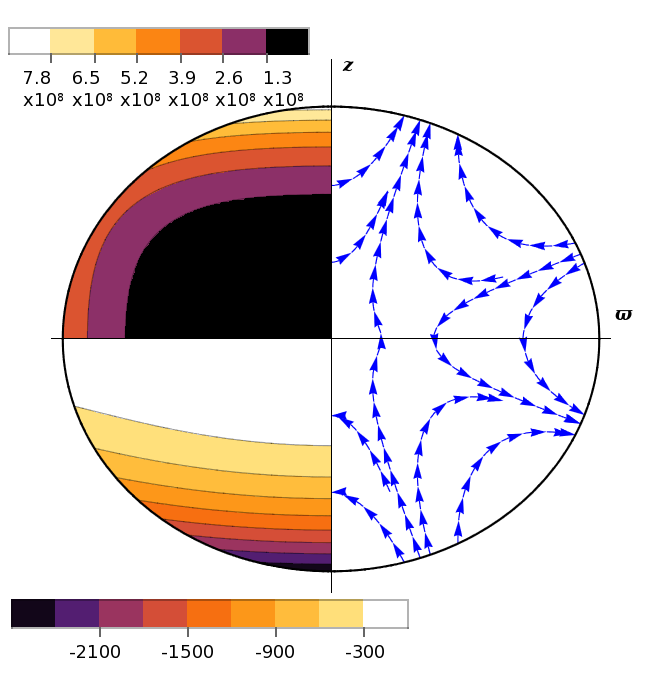}
\caption{
The $l=4$, $m=1$ prograde surface gravity mode in a spheroid of unit mean radius and eccentricity $e=0.5$.
Quadrants and colour-keys are as described beneath figure \ref{fig:l2m0spatial}.
}
\label{fig:l4m1prosurfgrav}
\end{figure}

\begin{figure}
\centering
\includegraphics[scale=0.35]{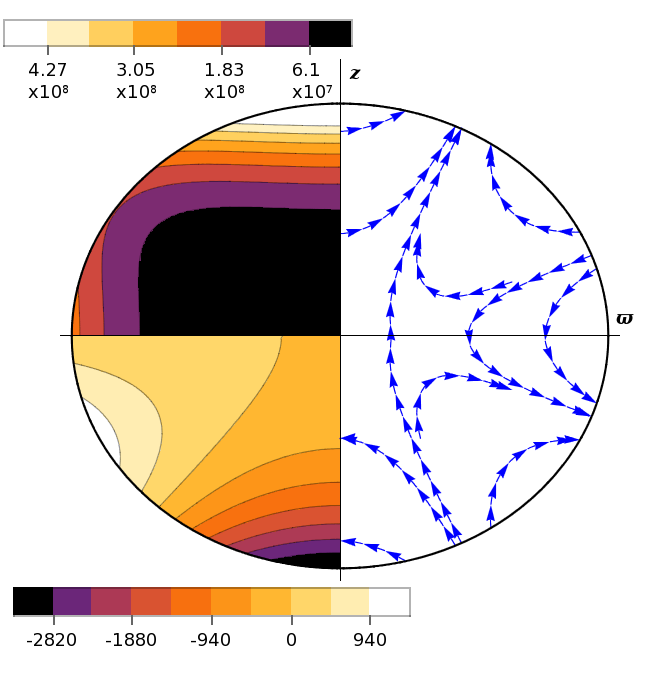}
\caption{
The $l=4$, $m=1$ retrograde surface gravity mode in a spheroid of unit mean radius and eccentricity $e=0.5$.
Quadrants and colour-keys are as described beneath figure \ref{fig:l2m0spatial}.
}
\label{fig:l4m1retsurfgrav}
\end{figure}

\begin{figure}
\centering
\includegraphics[scale=0.35]{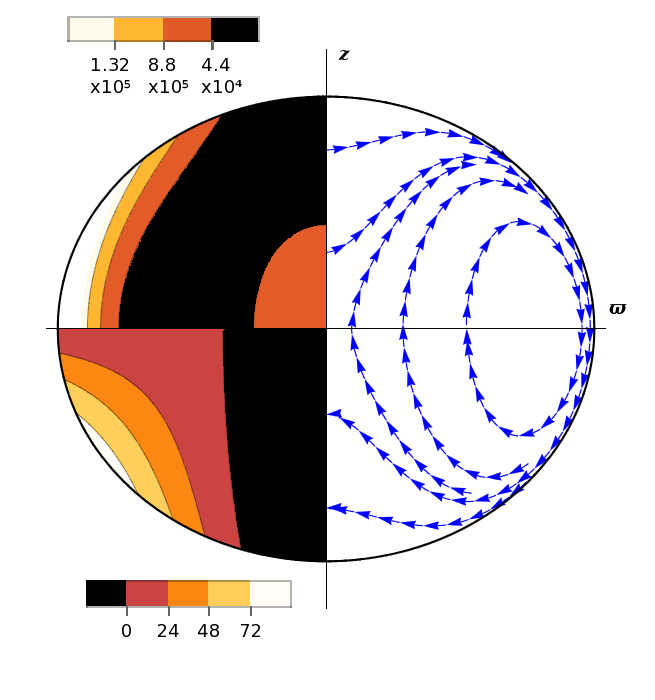}
\caption{
The $l=4$, $m=1$ prograde inertial mode in a spheroid of unit mean radius and eccentricity $e=0.5$.
Quadrants and colour-keys are as described beneath figure \ref{fig:l2m0spatial}.
}
\label{fig:l4m1proinert}
\end{figure}

\begin{figure}
\centering
\includegraphics[scale=0.35]{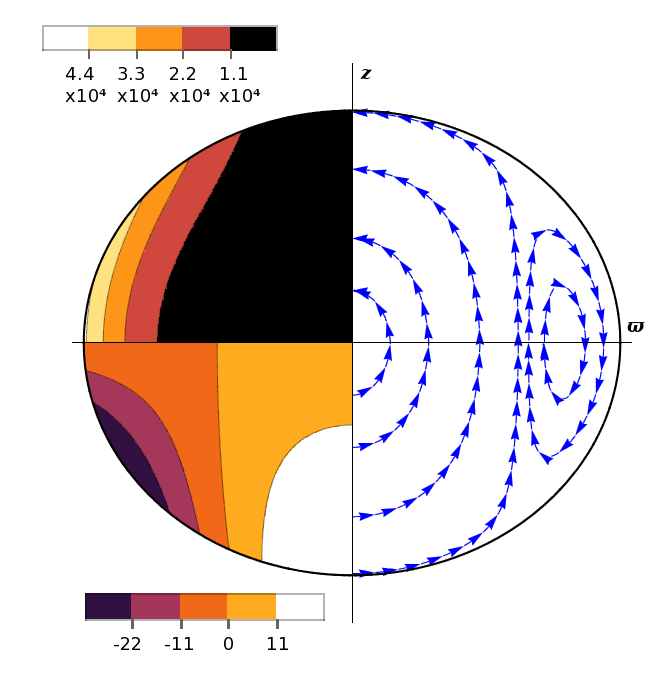}
\caption{
The $l=4$, $m=1$ slow retrograde inertial mode in a spheroid of unit mean radius and eccentricity $e=0.5$.
Quadrants and colour-keys are as described beneath figure \ref{fig:l2m0spatial}.
}
\label{fig:l4m1slowretinert}
\end{figure}

\begin{figure}
\centering
\includegraphics[scale=0.35]{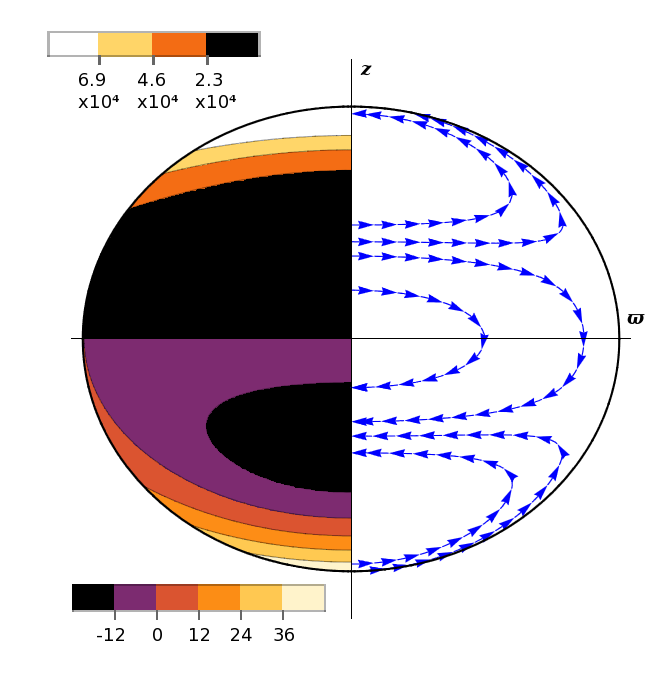}
\caption{
The $l=4$, $m=1$ fast retrograde inertial mode in a spheroid of unit mean radius and eccentricity $e=0.5$.
Quadrants and colour-keys are as described beneath figure \ref{fig:l2m0spatial}.
}
\label{fig:l4m1fastretinert}
\end{figure}

The $m=1$ mode may be written as
\begin{align} \label{l4m1W}
W = \Omega \kappa \left(\kappa + 2\right) \varpi z \big( & 3 \left(\kappa^2 - 4\right)\left(7\varpi^2 - 4b^2\right) \nonumber \\
& - 28 \kappa^2 z^2 \big) e^{i\varphi - i\kappa \Omega t} \,,
\end{align}
\begin{align} \label{l4m1u}
\bm{u} = & \bigg( \kappa z \left( 3\left(\kappa^2-4\right)\left(4 b^2 - 7 \varpi^2\right) + 28 \kappa^2 z^2 \right)\left(i \bm{e}_{\varpi} - \bm{e}_{\varphi}\right) \nonumber \\
& - 21 \kappa \left(\kappa + 2\right) z \varpi^2 \left(4 \bm{e}_{\varphi} + 2 i \kappa \bm{e}_{\varpi} \right) \nonumber \\
& + 3 i \left(\kappa + 2\right) \varpi \Big( \left(\kappa^2 - 4\right) \left(4 b^2 - 7 \varpi^2\right) \nonumber \\
& + 28 \kappa^2 z^2 \Big) \bm{e}_z \bigg) e^{i\varphi - i \kappa \Omega t} \,,
\end{align}
and the azimuthally averaged square of the norm of the rate-of-strain tensor may be written as
\begin{align}
e_{ij}^* e_{ij} = 18 \bigg( & 16 b^4 \left(\kappa^3 + \kappa^2 - 4 \kappa - 4\right)^2 \nonumber \\
& - 112 b^2 \left(1+\kappa\right)^2 \left(\kappa^2 - 4\right)\left(\left(\kappa^2 - 4\right)\varpi^2 - 2\kappa^2 z^2\right) \nonumber \\
& + 49 \Big(
\left(\kappa + 2\right)^2 \left( 5 \kappa^4 - 6 \kappa^3 - 15 \kappa^2 + 12 \kappa + 20 \right) \varpi^4 \nonumber \\
 + 4 \kappa^2 \left(\kappa + 2\right) & \left( 3 \kappa^3  + 10 \kappa^2 + 8 \kappa + 16 \right) \varpi^2 z^2
 + 16 \kappa^4 \left(1+\kappa\right)^2 z^4
\Big)
\bigg) \,.
\end{align}
Figure \ref{fig:l4freqplot} again shows that there are two surface gravity modes at this degree and order.
Their velocity fields are shown in figures \ref{fig:l4m1prosurfgrav} and \ref{fig:l4m1retsurfgrav}.
The spectrum of inertial modes consists of a single prograde mode, and a `slow' and a `fast' retrograde mode (the slow mode being the one with the less-negative frequency).
The spatial form of these is shown in figures \ref{fig:l4m1proinert}, \ref{fig:l4m1slowretinert} and \ref{fig:l4m1fastretinert} respectively.

\begin{figure}
\centering
\includegraphics[scale=0.35]{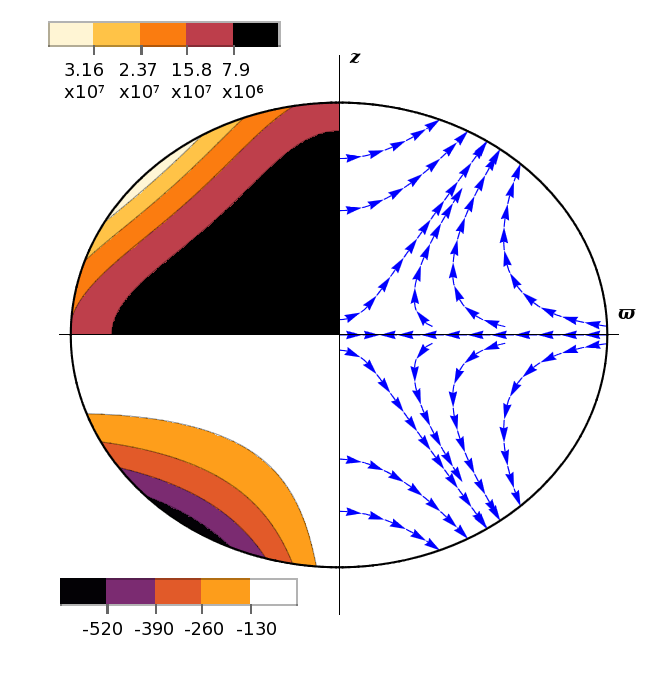}
\caption{
The $l=4$, $m=2$ prograde surface gravity mode in a spheroid of unit mean radius and eccentricity $e=0.5$.
Quadrants and colour-keys are as described beneath figure \ref{fig:l2m0spatial}.
}
\label{fig:l4m2prosurfgrav}
\end{figure}

\begin{figure}
\centering
\includegraphics[scale=0.35]{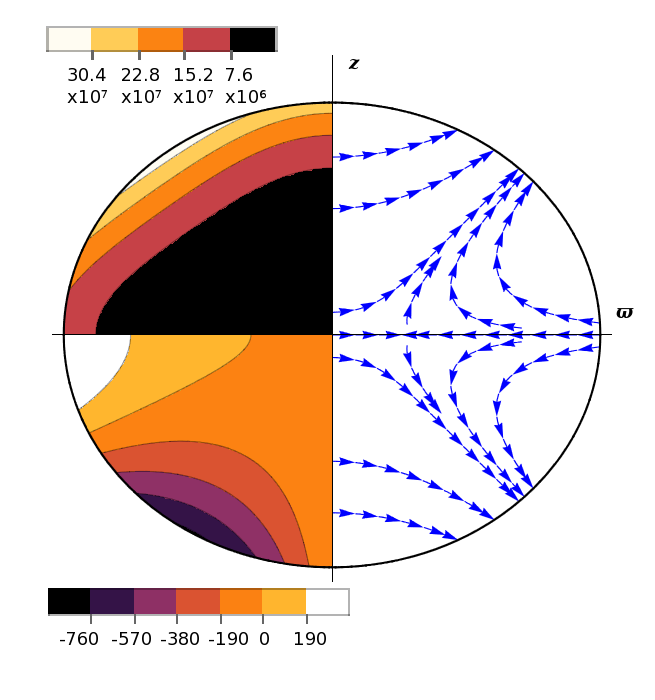}
\caption{
The $l=4$, $m=2$ retrograde surface gravity mode in a spheroid of unit mean radius and eccentricity $e=0.5$.
Quadrants and colour-keys are as described beneath figure \ref{fig:l2m0spatial}.
}
\label{fig:l4m2retsurfgrav}
\end{figure}

\begin{figure}
\centering
\includegraphics[scale=0.35]{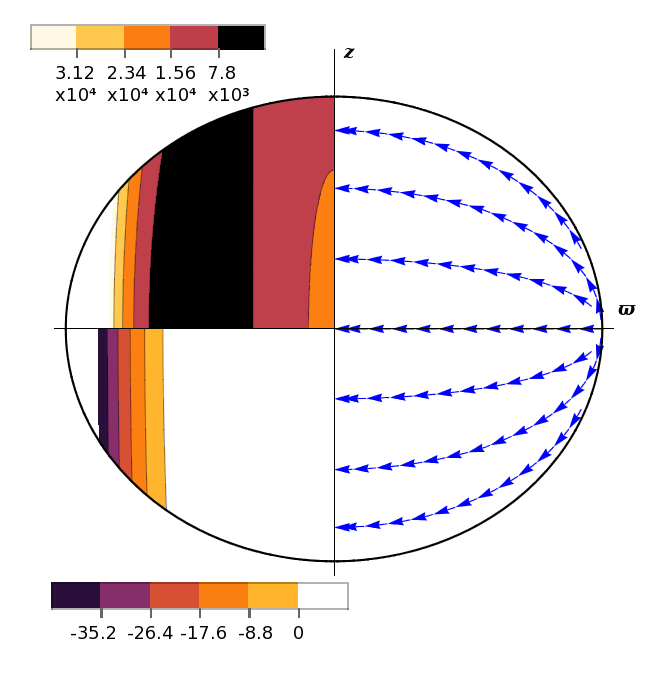}
\caption{
The $l=4$, $m=2$ prograde inertial mode in a spheroid of unit mean radius and eccentricity $e=0.5$.
Quadrants and colour-keys are as described beneath figure \ref{fig:l2m0spatial}.
}
\label{fig:l4m2proinert}
\end{figure}

\begin{figure}
\centering
\includegraphics[scale=0.35]{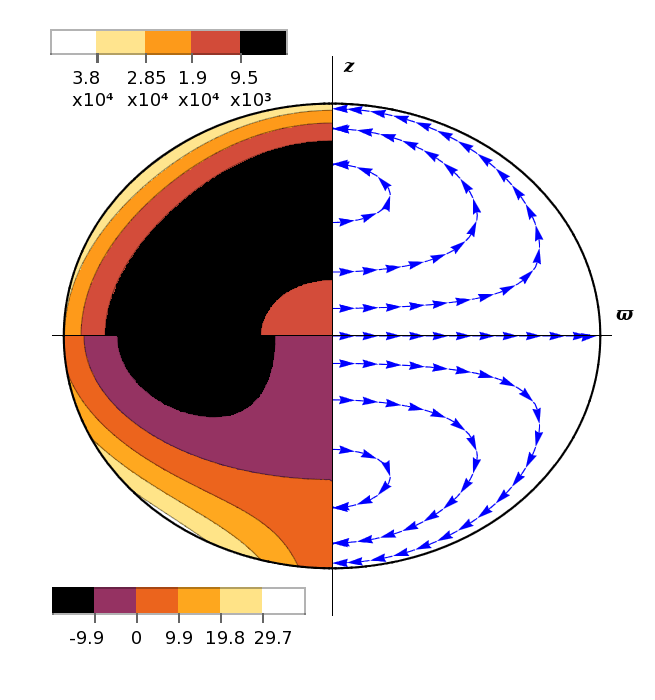}
\caption{
The $l=4$, $m=2$ retrograde inertial mode in a spheroid of unit mean radius and eccentricity $e=0.5$.
Quadrants and colour-keys are as described beneath figure \ref{fig:l2m0spatial}.
}
\label{fig:l4m2retinert}
\end{figure}

The $m=2$ mode may be written as
\begin{align} \label{l4m2W}
W = \Omega \left(\kappa + 2\right) \varpi^2 \Big( &
\left(\kappa^2 - 4\right) \left(7 \varpi^2 - 6 b^2\right) \nonumber \\
& - 42 \kappa^2 z^2
\Big) e^{2i\varphi - i \kappa \Omega t} \,,
\end{align}
\begin{align} \label{l4m2u}
\bm{u} = 2 \varpi \bigg( &
\left(6 b^2 \left(\kappa^2 - 4\right) + 42 \kappa^2 z^2 \right) \left(i \bm{e}_{\varpi} - \bm{e}_{\varphi}\right) \nonumber \\
& + 7 \left(\kappa + 2\right) \varpi^2 \left( \left(\kappa - 4\right) \bm{e}_{\varphi} - 2 \left(\kappa - 1\right) i \bm{e}_{\varpi} \right) \nonumber \\
& + 42 i \kappa \varpi z \left(\kappa + 2\right) \bm{e}_z
\bigg) e^{2i\varphi - i \kappa \Omega t} \,,
\end{align}
and the azimuthally averaged squared norm of its rate-of-strain tensor is
\begin{align}
e_{ij}^* e_{ij} = 36 & \bigg( 
8 b^4 \left(\kappa^2 - 4\right)^2 - 56 b^2 \left(\kappa^2 - 4\right)\left( \left(\kappa^2 - 4\right)\varpi^2 - \kappa^2 z^2 \right) \nonumber \\
&+ 49 \Big(
\left(\kappa+2\right)^2 \left(5 \kappa^2 - 8 \kappa + 8\right) \varpi^4 \nonumber \\
&+ 8 \kappa^2 \left( 3 \kappa^2 + 8 \kappa + 8 \right) \varpi^2 z^2 + 8 \kappa^4 z^4
\Big)
\bigg) \,.
\end{align}
The two surface gravity modes are illustrated in figures \ref{fig:l4m2prosurfgrav} and \ref{fig:l4m2retsurfgrav}, and their frequencies may be found in figure \ref{fig:l4freqplot}.
There is a single prograde and a single retrograde inertial mode, illustrated in figures \ref{fig:l4m2proinert} and \ref{fig:l4m2retinert} respectively, and whose frequencies are given in figure \ref{fig:l4kappaplot}.
This is qualitatively similar to the $l=3$, $m=1$ inertial modes;
the prograde mode is of low frequency, and hence its velocity field shows relatively little vertical variation, whilst the retrograde mode is of much higher frequency and correspondingly possesses more complicated vertical structure.

\begin{figure}
\centering
\includegraphics[scale=0.35]{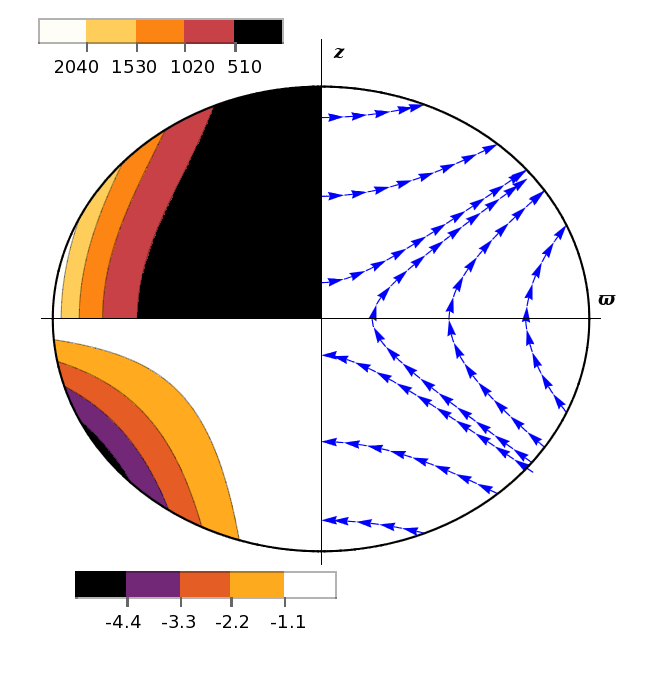}
\caption{
The $l=4$, $m=3$ prograde surface gravity mode in a spheroid of unit mean radius and eccentricity $e=0.5$.
Quadrants and colour-keys are as described beneath figure \ref{fig:l2m0spatial}.
}
\label{fig:l4m3prosurfgrav}
\end{figure}

\begin{figure}
\centering
\includegraphics[scale=0.35]{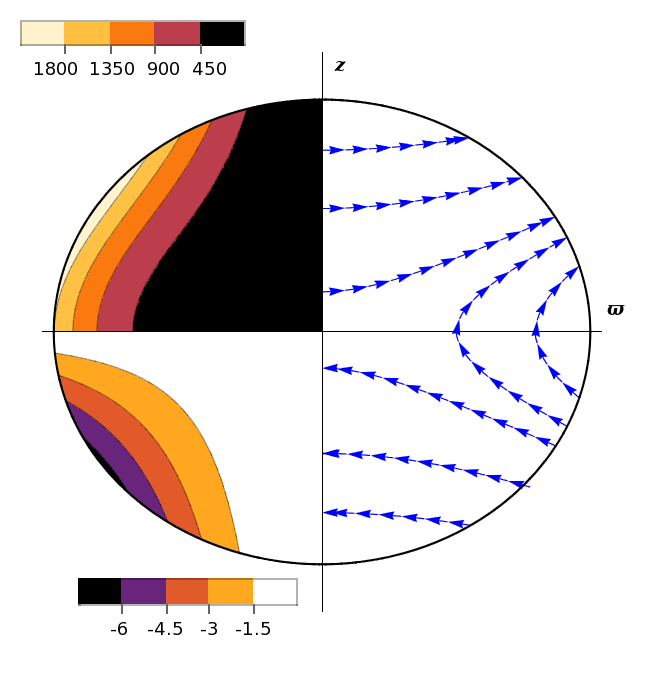}
\caption{
The $l=4$, $m=3$ retrograde surface gravity mode in a spheroid of unit mean radius and eccentricity $e=0.5$.
Quadrants and colour-keys are as described beneath figure \ref{fig:l2m0spatial}.
}
\label{fig:l4m3retsurfgrav}
\end{figure}

\begin{figure}
\centering
\includegraphics[scale=0.35]{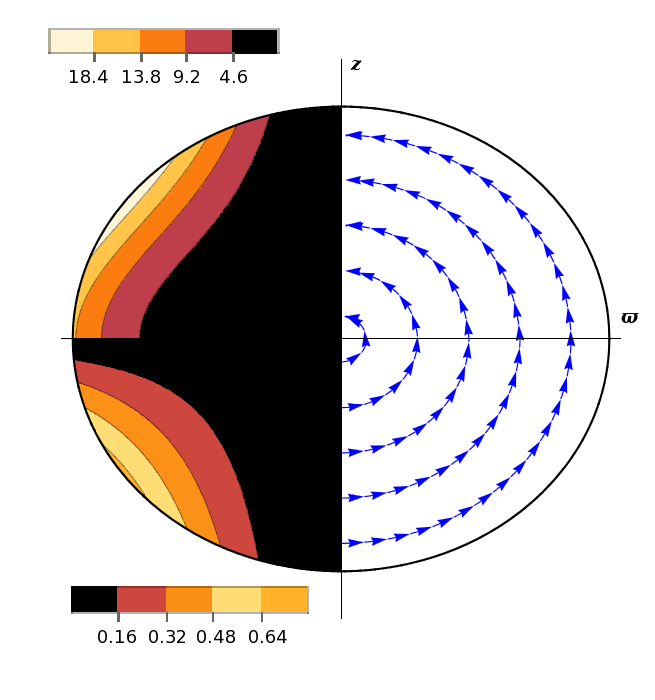}
\caption{
The $l=4$, $m=3$ inertial mode in a spheroid of unit mean radius and eccentricity $e=0.5$.
Quadrants and colour-keys are as described beneath figure \ref{fig:l2m0spatial}.
}
\label{fig:l4m3inert}
\end{figure}

The $m=3$ modes are considerably simpler and may be written as
\begin{align} \label{l4m3W}
W = -\Omega \kappa \left(2 + \kappa\right) \varpi^3 z e^{3i\varphi - i \kappa \Omega t} \,,
\end{align}
\begin{align} \label{l4m3u}
\bm{u} = \varpi^2 \left( 3 \kappa z \left( i \bm{e}_{\varpi} - \bm{e}_{\varphi} \right) + \left(2+\kappa\right) i \varpi \bm{e}_z \right) e^{3i\varphi - i \kappa \Omega t} \,,
\end{align}
and the squared norm of their rate-of-strain tensor as
\begin{align}
e_{ij}^* e_{ij} = 18 \left( \left(1+\kappa^2\right) \varpi^4 + 4 \kappa^2 \varpi^2 z^2 \right) \,.
\end{align}
There are two surface gravity modes whose frequencies are given in figure \ref{fig:l4freqplot} and whose spatial form is shown in figures \ref{fig:l4m3prosurfgrav} and \ref{fig:l4m3retsurfgrav}.
The single inertial mode is retrograde (as in the $l=3$, $m=2$ case) and is shown in figure \ref{fig:l4m3inert}.

The $m=4$ modes fit into the special class of sectoral modes, described in section \ref{sec:sectoral_modes}.
The frequencies of the two modes of this order are shown in figure \ref{fig:l4freqplot}.

\begin{figure}
\centering
\input{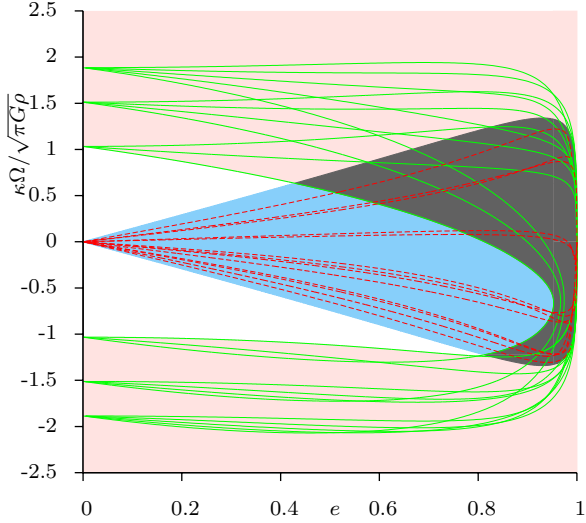}
\caption{
A summary figure of the frequencies of the modes of degree $l \le 4$.
The region having $-2 < \kappa < 2$ is shaded blue and represents the extent of the frequency range in which equation \eqref{W_Poincare} is hyperbolic and in which we may have inertial waves.
The region shaded light pink represents the region occupied by surface gravity waves.
Since the lower boundary in $\left| \kappa \right|$ of this region is set by $l=2$ modes, we do not expect it to extend to lower $\left| \kappa \right|$ with the inclusion of higher degree modes.
The dark grey beyond $e \approx 0.43$ denotes where these two regions overlap.
Frequencies of individual surface gravity modes are shown by solid green lines and those of inertial modes by dashed red lines.
}
\label{fig:freqregion}
\end{figure}

We conclude our discussion of the individual modes of the Maclaurin spheroid by illustrating in figure \ref{fig:freqregion} the frequencies of every mode we have considered in this paper (with the exception of the differential rotation and spin-over modes).
This figure contains the same information as figures \ref{fig:l2freqplot}, \ref{fig:l3freqplot}, \ref{fig:l3kappaplot}, \ref{fig:l4freqplot} and \ref{fig:l4kappaplot}, but here we have plotted the inertial mode frequencies in units of the dynamical frequency.
We caution the reader against directly comparing these with the orbital frequency of a companion, due to the Doppler shift associated with transforming from the inertial frame to the frame rotating with the spheroid, and due to the Wigner $d$-matrix transformation the must be performed if the orbit is misaligned with the rotation axis of the spheroid.
The set of mean motions at which a planet on a misaligned orbit may resonantly force inertial modes in a sphere can be found in \citet{Ogilvie2013}, and we shall consider the case of such resonant forcing of a Maclaurin spheroid in a subsequent paper.

\section{The Energy Equation and the Decay Rate of Free Modes}
\label{sec:eigenmode_problem}

We have described above the inviscid modes of a self-gravitating Maclaurin spheroid with a free surface, up to degree 4.
Suppose that we now consider a Maclaurin spheroid composed of fluid having some small Newtonian viscosity, so that the fluid within the spheroid obeys
\begin{equation} \label{viscousNS}
\partial_t \bm{u} + 2 \bm{\Omega} \times \bm{u} = - \bm{\nabla} W + \nu \nabla^2 \bm{u} \,,
\end{equation}
with the additional boundary condition that the viscous stress vanish at the surface of the spheroid.
From \eqref{viscousNS} we may form the energy equation
\begin{align} \label{energy_integral}
&\frac{\mathrm{d}}{\mathrm{d} t} \int_V \frac{1}{2} u^2 \mathrm{d}V + \int_V \bm{\nabla} \cdot \left( \bm{u} W \right) \mathrm{d}V \nonumber \\
= & \frac{\mathrm{d}}{\mathrm{d} t} \bigg( \int_V \frac{1}{2} u^2 dV + \frac{1}{2} \int_{\partial V} g_s \left(\bm{\xi}\cdot\bm{n}\right)^2\mathrm{d}S \nonumber \\
&- \frac{1}{8\pi G \rho} \int_{\mathbb{R}^3} \left|\bm{\nabla}\Phi^{\prime}\right|^2 \mathrm{d}V \bigg) \nonumber \\
= & \nu \int_V u_j \partial_i \partial_i u_j dV \nonumber \\
= & \nu \int_{\partial V} n_i u_j \partial_i u_j \mathrm{d}S - \nu \int_V \left(\partial_j u_i\right)\left(\partial_j u_i\right) \mathrm{d}V \nonumber \\
= & 2\nu \int_{\partial V} n_i u_j e_{ij} \mathrm{d}S - 2\nu \int_{V} e_{ij} e_{ij} \mathrm{d}V \,,
\end{align}
where in the final line we have used the symmetrised rate-of-strain tensor, $e_{ij} = \frac{1}{2} \left(\partial_j u_i + \partial_i u_j\right)$.

In the final line of the above expression the first term represents work done on the fluid by stress from the boundaries, which must vanish for the viscous modes due to our imposing of stress-free boundary conditions.
The second term represents viscous dissipation within the fluid.
We expect that the form of the rate-of-strain tensor in the integrand will not change over much of the spheroid between the inviscid and viscous modes, and hence that this integral will be well-approximated by using the rate-of-strain tensors of the inviscid modes calculated in sections \ref{sec:sectoral_modes} to \ref{sec:degree4modes}.
We therefore suspect that the leading order perturbation to the eigenvalue will be
\begin{align}\label{dispexpr}
& \Omega \, \delta \kappa = -i \rho \nu \int_V e_{ij}^* e_{ij} dV \Bigg/
\bigg( \frac{1}{2}\int_V \rho \left|\bm{u}\right|^2 dV \nonumber \\
& + \frac{1}{2}\int_{\partial V} \rho g_s \left|\bm{\xi}.\bm{n}\right|^2 dS - \frac{1}{8\pi G}\int_{\mathbb{R}^3} \left|\bm{\nabla} \Phi^{\prime}\right|^2 dV \bigg)\,.
\end{align}
Detailed justification of this result is given in appendix \ref{sec:decay_rate_derivation}.

We have expressions for $\bm{u}$ and $e_{ij}^* e_{ij}$ in cylindrical polar coordinates for each mode, as a function of $\kappa$, which may be found numerically from \eqref{freqEqn}.
The potential energy (the second two terms in the denominator of \eqref{dispexpr}) may be found directly for a general mode as
\begin{align}
& \frac{1}{2} \int_V \bm{\nabla} \cdot \left( \bm{\xi}^* W \right) \mathrm{d}V
= \frac{1}{2} \int_{\partial V} \left(\bm{n} \cdot \bm{\xi}\right)^* W \mathrm{d}S \nonumber \\
& = \int_{\partial V} \frac{ c \left(W - \Phi^{\prime}\right)^* W}{8 \pi G \rho \zeta_0 (1 - \zeta_0 \cot^{-1} \zeta_0 )} \mathrm{d}\varphi \, \mathrm{d}\mu \nonumber \\
& = \frac{c}{4 G \rho} \frac{\left(D_{l, m, \kappa} - A_{l, m, \kappa}\right)^* D_{l, m, \kappa}}{\zeta_0 (1- \zeta_0 \cot^{-1} \zeta_0)} \int_{-1}^{+1} P_l^m(\mu) P_l^m(\mu) \mathrm{d}\mu \nonumber \\
& = \frac{c}{2 G \rho} \frac{\left| D_{l, m, \kappa}\right|^2 B_l^m(\zeta_0)}{1 + \zeta_0 B_l^m(\zeta_0) (1 - \zeta_0 \cot^{-1} \zeta_0)} \frac{(l+m)!}{(2l+1)(l-m)!} \,.
\end{align}
In going from the first to the second line we have used the expression \eqref{p_prime_over_gradient} for $\bm{\xi} \cdot \bm{n}$ on $\partial V$, and the surface area element in oblate spheroidal coordinates, $\mathrm{d}S = c^2 \sqrt{(1+\zeta_0^2)(\zeta_0^2 + \mu^2)}$.
The solutions \eqref{gravsoln} and \eqref{Wsoln} for $\Phi^{\prime}$ and $W$ respectively have been used in going from the second to the third line.
In going from the third to the fourth line we have used the orthogonality relation for associated Legendre polynomials.
We have also used \eqref{Phibc_A_D} to write the amplitude entirely in terms of $D_{l, m, \kappa}$.
Since the choice of $D_{l, m, \kappa}$ may be read off from \eqref{W_cylind} and our explicit expressions for $W$ in sections \ref{sec:lowdegreemodes} to \ref{sec:degree4modes}, it is the more useful amplitude to work with.

\begin{figure}
\centering
\input{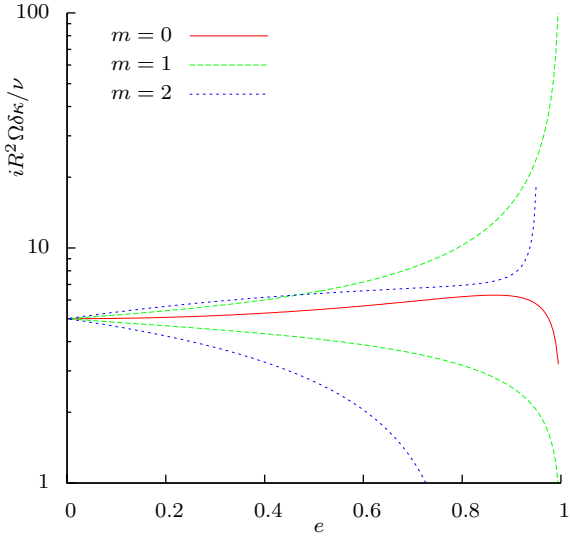}
\caption{
Non-dimensionalised decay rates ($i R^2 \Omega \delta \kappa / \nu$) plotted against eccentricity of the spheroid ($e$) for the $l=2$ modes.
In this plot the prograde and retrograde modes may be distinguished by the fact that the retrograde mode has a higher dissipation rate than the prograde mode of the same degree and order.
}
\label{fig:l2decayplot}
\end{figure}

\begin{figure}
\centering
\input{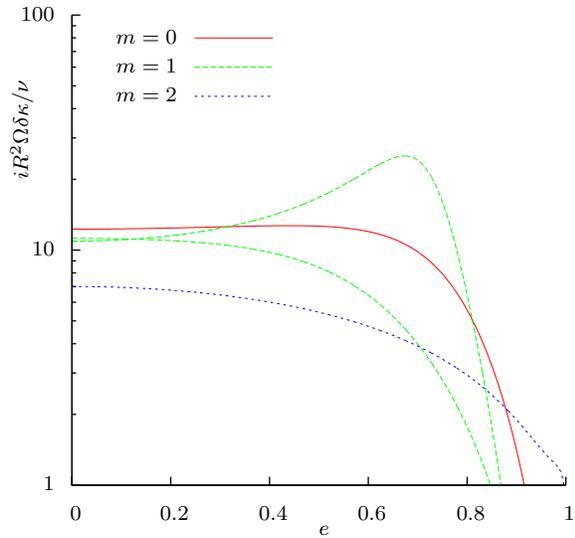}
\caption{
Non-dimensionalised decay rates ($i R^2 \Omega \delta \kappa / \nu$) plotted against eccentricity of the spheroid ($e$) for the $l=3$ inertial modes.
The $m=1$ modes may be distinguished by the fact that the retrograde mode has a lower decay rate than the prograde mode in the $e\rightarrow 0$ limit.
The decay rate of the retrograde mode become larger than that of the prograde mode at $e \approx 0.12$.
}
\label{fig:l3inertialdecayplot}
\end{figure}

\begin{figure}
\centering
\input{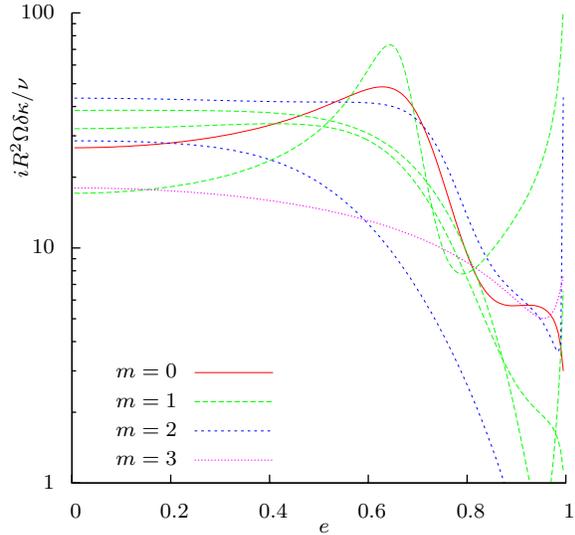}
\caption{
Non-dimensionalised decay rates ($i R^2 \Omega \delta \kappa / \nu$) plotted against eccentricity of the spheroid ($e$) for the $l=4$ inertial modes.
The $m=1$ modes may be distinguished by the fact that, in the $e \rightarrow 0$ limit, the slow retrograde mode has a higher decay rate than the prograde mode, which in turn has a higher decay rate than the fast retrograde mode.
The $m=2$ modes may be distinguished by the fact that the retrograde mode has higher decay rate than the prograde mode.
}
\label{fig:l4inertialdecayplot}
\end{figure}

Equipped with this expression for the potential energy, and using a computer algebra system to evaluate the dissipation and kinetic energy integrals, we have numerically evaluated the decay rate for all of the modes described in sections \eqref{sec:lowdegreemodes} to \ref{sec:degree4modes}.
These are plotted as figures \ref{fig:l2decayplot} to \ref{fig:l4inertialdecayplot}, in order of increasing degree, with surface gravity and inertial modes plotted separately.

In these figures some clear differences can be seen as the rotation rate of the spheroid is increased.
The decay rates of the two $l=3$, $m=1$ inertial modes differ by only $8\%$ in the spherical limit, but the retrograde mode has a decay rate $40\%$ higher than that of the prograde mode by $e=0.6$.
In the spherical limit the $l=4$, $m=1$ fast retrograde mode has only half the decay rate of the slow retrograde mode.
However, by $e=0.6$ the fast retrograde mode has the highest decay rate of any mode of this degree and order.

Note that inertial modes of high decay rate are not necessarily of more interest for tidal applications.
While the dissipation will reach a higher peak at resonance, this peak will also be sharper.
Lower decays rates will result in enhanced dissipation across a wider range of orbital periods.
We shall explore this in more detail in paper II.

\begin{figure}
\centering
\input{figures/decay_rate_plots/l3surfgrav/l3surfgrav}
\caption{
Non-dimensionalised decay rates ($i R^2 \Omega \delta \kappa / \nu$) plotted against eccentricity of the spheroid ($e$) for the $l=3$ surface gravity modes.
In this plot the prograde and retrograde modes may be distinguished by the fact that the retrograde mode has a higher dissipation rate than the prograde mode of the same degree and order.
}
\label{fig:l3surfgravdecayplot}
\end{figure}

\begin{figure}
\centering
\input{figures/decay_rate_plots/l4surfgrav/l4surfgrav}
\caption{
Non-dimensionalised decay rates ($i R^2 \Omega \delta \kappa / \nu$) plotted against eccentricity of the spheroid ($e$) for the $l=4$ surface gravity modes.
In this plot the prograde and retrograde modes may be distinguished by the fact that the retrograde mode has a higher dissipation rate than the prograde mode of the same degree and order.
}
\label{fig:l4surfgravdecayplot}
\end{figure}

\section{Conclusion}
\label{sec:conclusion}

We have reviewed in section \ref{sec:general_Bryan_mode} the original derivation of \citet{Bryan1889} of the normal modes of a spheroid of inviscid, homogeneous, incompressible fluid in solid body rotation, following the slightly clearer path taken by \citet{LI1999}.
We have shown that this notation is valid for both the inertial and the surface gravity modes, and considered the special cases of modes of frequencies $0$ (in appendix \ref{sec:zeromodes}), $2 \Omega$ and $-2\Omega$ (in appendix \ref{sec:kappa4modes}).
To allow easier computation of the flow within the spheroid we have derived the previously-unpublished expression \eqref{W_cylind} for the hydrodynamic potential.
We made use of this expression in sections \ref{sec:lowdegreemodes} to \ref{sec:degree4modes}, where we describe the properties of the low degree modes in detail.
In paper II we shall describe the response of such a spheroid to tidal forcing from an orbiting companion.
The magnitude of the forcing decreases with increasing degree of the forcing harmonic, and hence we focus on the lowest two degrees ($l=3$ and $l=4$) at which inertial waves occur.

In section \ref{sec:eigenmode_problem} we considered the energy equation of the normal modes, and proposed equation \eqref{dispexpr} as the decay rate of such modes for small viscosities.
A detailed justification of this equation may be found in appendix \ref{sec:decay_rate_derivation}.
Such decay rates will be necessary to make predictions of the evolution rates of various orbital elements, which we shall describe in paper II.
For each degree and order, a computer algebra system and the expression \eqref{W_cylind} for the hydrodynamic potential may be used to generate an algebraic expression of the decay rate.
The only numerical step required is to repeatedly solve equation \eqref{freqEqn} to find the allowed mode frequencies for each eccentricity.
For each mode of degree $l \le 4$ we plotted the decay rate against the eccentricity of the spheroid, finding that the decay rate of several inertial modes changed substantially over between $e=0$ and $e=0.6$.
Yet other inertial modes (in particular the $l=3$, $m=0$ mode and the $l=4$, $m=2$ retrograde mode) showed virtually no change over this range.

The model we have considered herein allows inertial and surface gravity modes to be treated in a self-consistent fashion, with no need to make a distinction between the dynamical and equilibrium tides.
Working beyond the slowly rotating spherical limit caused the inertial modes to move the stellar surface, allowing us to consider their direct gravitational excitation.
The excitation of both the inertial and surface gravity modes, and the dissipation that results, will be considered in paper II and is a necessary part of understanding the effect of rapid rotation.
With the conclusions of this paper, however, we can already make some statements regarding the effects of the equatorial bulge upon the response.
The frequencies of the inertial waves show little departure from linear dependence on the rotation rate of the spheroid, at least until very high eccentricities are reached (around $e \approx 0.7$).
Therefore the orbital periods at which resonances are encountered will not be changed significantly relative to the spherical approximation.
However, the fact that the decay rates of some of the inertial modes change by an order one factor between $e=0$ and $e=0.6$ implies that we must take into account the equatorial bulge if we wish to accurately estimate the rate of orbital evolution.

The surface gravity modes show more significant frequency changes, but still lie close to the dynamical frequency of the spheroid.
We the possible exception of the prograde sectoral modes at extreme eccentricities ($e \approx 0.8$), these are unlikely to be resonantly excited.
Away from resonances with inertial modes we expect the most strongly excited mode (and hence the main contributor to tidal dissipation) to be the $l=2$, $m=2$ equilibrium tide.
The decrease in frequency of the prograde mode seen in figure \ref{fig:l2freqplot} may increase the non-resonant excitation of this mode.
Examining figure \ref{fig:l2decayplot} shows that the decay rate of the prograde $l=2$, $m=2$ mode does not vary significantly over realistic eccentricities.
However, the decrease in the decay rate of the corresponding retrograde mode (a precursor to this `decay rate' becoming negative at the onset of secular instability) will decrease the dissipation associated with this mode.
Which of these effects (the decrease in frequency separation between a realistic forcing frequency and the natural frequency of the prograde mode, or the decrease in dissipation due to off-resonant excitation of the retrograde mode) is more significant may be determined once we have considered the problem of the forced spheroid in paper II.

In paper II we shall again review the work of \citeauthor{Bryan1889}, making clear how the forced response of the spheroid may be decomposed into free modes.
We shall also develop conversions between the oblate spheroidal and the spherical coordinate systems, which are necessary if we are to calculate the forcing due to, and subsequent evolution of, an orbit with a non-zero inclination.

\section*{Acknowledgements}

We are grateful to the anonymous reviewer for helpful comments.
This research was supported by the STFC.
Harry Braviner is partially supported by Trinity College.

\FloatBarrier

\appendix

\section{The Oblate Spheroidal Coordinates}
\label{sec:maclaurin_spheroid_obl_coords}

In order to solve Laplace's equation \eqref{Poisson_eqn} in such a way that we can easily apply the boundary conditions \eqref{pbc} and \eqref{Phibc} at the surface of the spheroid, we must employ a coordinate system in which the both the Poisson equation and the surface of the spheroid take a simple form.
We define the \emph{oblate spheroidal coordinates}, $(\zeta, \mu, \varphi)$, by
\begin{align}\label{obl_spher_def} \varpi = c \sqrt{(1 + \zeta^2)(1 - \mu^2)}, \;\;\;\; z = c\zeta\mu \,, \end{align}
where $(\varpi, z, \varphi)$ are the usual cylindrical polar coordinates and $c$ is the focal radius defined previously.
Using equation \eqref{spheroid_cylindricals} we find that the surface of the spheroid is $\zeta = \zeta_0$, $-1 \le \mu \le 1$.

\begin{figure}
\input{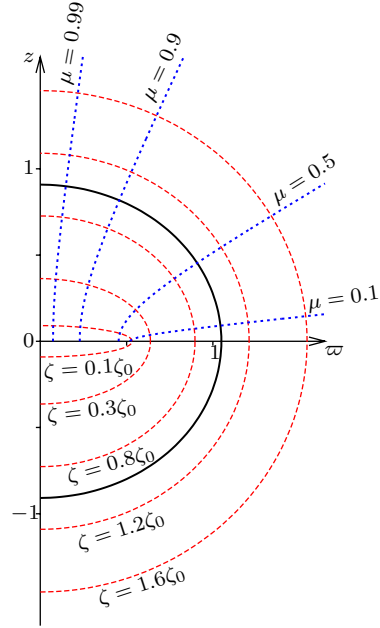}
\caption{Surfaces of constant $\zeta$ and $\mu$, illustrating how the oblate spheroidal coordinates cover all of space for $\zeta \ge 0$ and $-1\le \mu \le 1$.
The solid line denotes the surface of the spheroid, and the coordinates plotted are for a spheroid of unit mean radius and an eccentricity of $0.5$.
The $z<0$ region is covered by $\mu<0$.
}
\label{fig:oblate_spher}
\end{figure}

The surfaces of constant $\zeta$ form a family of confocal oblate spheroidal shells,
\begin{equation}
\left(\frac{1 + \zeta_0^2}{1 + \zeta^2}\right) \frac{\varpi^2}{R_e^2} + \left(\frac{\zeta_0^2}{\zeta^2}\right) \frac{z^2}{R_p^2} = 1 \,,
\end{equation}
which cover the interior of the spheroid for $0 \le \zeta \le \zeta_0$ and the exterior for $\zeta_0 \le \zeta \le + \infty$.
Note that these are \emph{not} equipotentials of either the gravitational potential or the combined gravito-centrifugal potential.
For $-1\le\mu\le1$ the surfaces of constant $\mu$ form a family of confocal hyperboloids,
\begin{equation}
\left(\frac{1+\zeta_0^2}{1-\mu^2}\right) \frac{\varpi^2}{R_e^2} - \left(\frac{\zeta_0^2}{\mu^2}\right) \frac{z^2}{R_p^2} = 1 \,,
\end{equation}
covering the entirety of space.
These surfaces are illustrated in figure \ref{fig:oblate_spher} for a spheroid of eccentricity $0.5$.

From the definition \eqref{obl_spher_def} it may easily be shown that the line element in these coordinates is
\begin{equation}
ds^2 = c^2 \frac{\zeta^2 + \mu^2}{1 + \zeta^2} d\zeta^2 + c^2 \frac{\zeta^2 + \mu^2}{1 - \mu^2} d\mu^2 + c^2\left(1+\zeta^2\right)\left(1-\mu^2\right) d\varphi^2 \,.
\end{equation}
From this we may read off the metric components, $g_{ab}$, and calculate the metric determinant, $g$, and the metric inverse, $g^{ab}$. The covariant divergence formula, $ \nabla^2 \Phi = \frac{1}{\sqrt{g}} \partial_a \left(\sqrt{g} g^{ab} \partial_b \Phi \right)$, then gives
\begin{align}
\label{laplacianObl} \nabla^2 \Phi^{\prime} =& \frac{1}{c^2\left(\zeta^2 + \mu^2\right)} \frac{\partial}{\partial \zeta} \left( \left(1 + \zeta^2\right) \frac{\partial \Phi^{\prime}}{\partial \zeta}\right) \nonumber \\
&+ \frac{1}{c^2\left(\zeta^2 + \mu^2\right)} \frac{\partial}{\partial \mu} \left( \left(1 - \mu^2\right) \frac{\partial \Phi^{\prime}}{\partial \mu}\right) \nonumber \\
&+ \frac{1}{c^2 \left(1+\zeta^2\right)\left(1-\mu^2\right)} \frac{\partial^2 \Phi^{\prime}}{\partial \varphi^2} \,.
\end{align}
We make use of this in section \ref{sec:general_Bryan_mode}.

Finally, figure \ref{fig:oblate_spher} suggests that at large radii the $\zeta$ and $\mu$ coordinates may approximate the $r$, $\theta$ coordinates of spherical polars. This is indeed the case, with
\begin{align} \label{zeta_large_r}
\zeta = \frac{r}{c}\left(1 - \frac{1}{2}\sin^2\theta\frac{c^2}{r^2} + \mathcal{O}\left(\frac{c^4}{r^4}\right)\right)
\end{align}
and
\begin{align} \label{mu_large_r}
\mu = \cos \theta\left(1 + \frac{1}{2} \sin^2 \theta \frac{c^2}{r^2} + \mathcal{O}\left(\frac{c^4}{r^4}\right)\right) \,.
\end{align}
We will expand upon this relation in a subsequent paper, where we shall make use of a conversion between spherical and oblate spheroidal harmonics to calculate the tidal interaction between a fluid Maclaurin spheroid and an orbiting companion.

\section{The Bi-Spheroidal Coordinates}
\label{sec:maclaurin_spheroid_bi_spheroidal_coords}

In section \ref{sec:general_Bryan_mode} we saw that the inviscid Euler equations could be reduced to the Poincar\'{e} equation \eqref{Poisson_eqn} for the hydrodynamic potential.
The differential operator in this equation depends upon the frequency, $\kappa$, of the mode we are seeking.
However, we could bring it into the form of a (frequency-independent) Laplacian operator in a (frequency-dependent) coordinate system by defining $(x,y,z^{\prime})$ such that $z^2 = (1 - 4/\kappa^2) (z^{\prime})^2$.
Defining oblate spheroidal coordinates as in the previous appendix, but now by their relation to $(x,y,z^{\prime})$, would then bring \eqref{Poisson_eqn} into a separable form.

Whilst this will work algebraically, it is worth pausing to consider what is happening geometrically.
In the case that $\kappa^2>4$ the Poincar\'{e} equation is elliptic and the $z^{\prime}$ coordinate is real.
In this new coordinate system the fluid occupies a spheroidal volume that is either prolate or oblate when $\kappa^2 < 4\left(1+\zeta_0^2\right)$ or $\kappa^2 > 4\left(1+\zeta_0^2\right)$ respectively.
This is what Bryan calls the \emph{auxiliary spheroid}; in these two $\kappa$ intervals the focal radius is respectively $ib$ or $b$ where
\begin{equation}
\label{bdef}
b = \frac{R}{\left(\zeta_0\left(1+\zeta_0^2\right)\right)^{1/3}} \sqrt{\frac{4\left(1+\zeta_0^2\right) - \kappa^2}{4-\kappa^2}} \,.
\end{equation}

Based on the above argument, we define the \emph{bi-spheroidal coordinates}, $\left(\xi, \tilde{\mu}, \varphi \right)$, by
\begin{align}\label{bi_spher_def}
\varpi = b \sqrt{(1-\xi^2)(1-\tilde{\mu}^2)}, \;\;\;\;  z = b \frac{\sqrt{4-\kappa^2}}{\kappa} \xi \tilde{\mu} \,.
\end{align}
We will see shortly that the name `bi-spheroidal' is only appropriate for $\kappa^2 < 4$. However, in the interests of consistency we shall use it to refer to this coordinate system regardless of the value of $\kappa$.

\begin{figure*}
\input{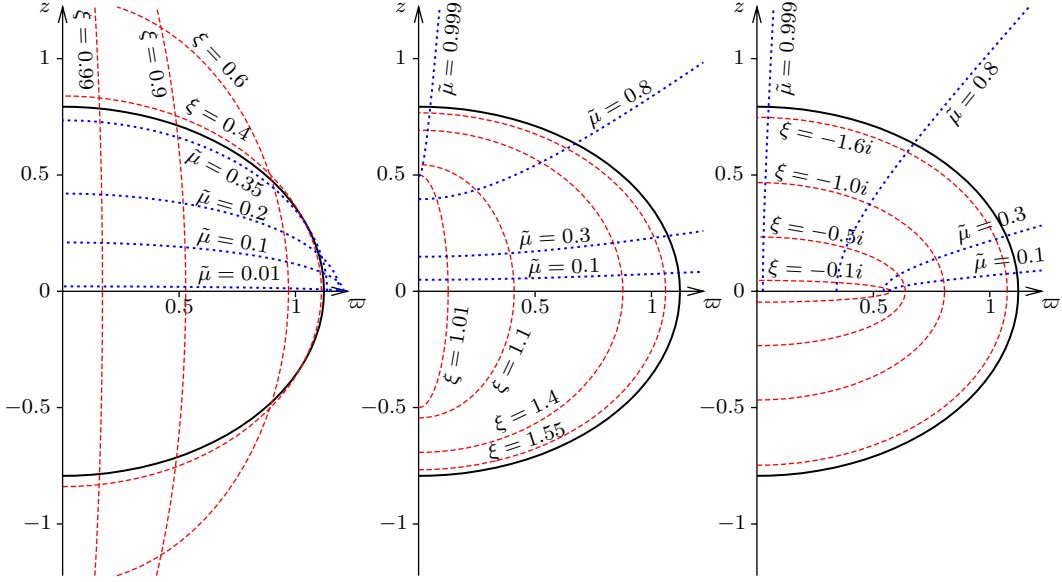}
\caption{Surfaces of constant $\xi$ and $\tilde{\mu}$, illustrating how the bi-spheroidal coordinates defined in \eqref{bi_spher_def} cover the interior of the spheroid for different values of $\kappa$.
The solid line denotes the surface of the spheroid, and the coordinates plotted are for a spheroid of unit mean radius and an eccentricity of $2/3$, for $\kappa = 1$, $2.4$ and $3.5$ (from left to right).
In all cases the $z<0$ region is covered by $\mu<0$.
}
\label{fig:bi_spher}
\end{figure*}

Note that \eqref{bi_spher_def} is invariant under exchanging $\xi$ and $\tilde{\mu}$, therefore these coordinates will be distinguished from one another only by the ranges they take.
In describing these ranges it will be helpful to define
\begin{align} \label{xi0def}
\xi_0 = \frac{\zeta_0 \kappa }{\sqrt{4\left(1+\zeta_0^2\right) - \kappa^2}} \,.
\end{align}
We have largely retained the notation of \cite{LI1999}, who were only interested in modes for which $\kappa^2 < 4$.
We shall now show how these coordinates may be applied for all real $\kappa$, with the coordinate ranges changing as $\kappa^2$ passes $4$ and $4\left(1+\zeta_0^2\right)$.

Consider first the location of the surface of the spheroid in these coordinates.
Using \eqref{bi_spher_def} we may write the surface as
\begin{equation}\label{ximusurf}
0 = \frac{\varpi^2}{R_e^2} + \frac{z^2}{R_p^2} - 1
= \frac{\left(\xi^2 - \xi_0^2\right)\left(\tilde{\mu}^2 - \xi_0^2\right)}{\zeta_0^2 \kappa^2 \left(1+\zeta_0^2\right)\left(4 - \kappa^2\right)\left(4\left(1+\zeta_0^2\right)-\kappa^2\right)^2} \,,
\end{equation}
from which we can see that a point on the surface must have at least one of $\xi^2$ or $\tilde{\mu}^2$ equal to $\xi_0^2$.
If we set $\xi = \xi_0$, then $z = R_e \tilde{\mu}$ and thus it appears that we may write the surface as $\left\{\xi = \xi_0,\; -1\le \tilde{\mu} \le +1\right\}$.
This works for $\kappa^2 > 4$, but does not connect to coordinates that smoothly cover the interior of the spheroid for $\kappa^2 < 4$.
To see this, note that $0 < \xi_0^2 < 1$ for $\kappa^2 < 4$.
Consider a curve beginning on the equator end ending at the north pole, lying entirely inside the spheroid.
Along such a curve $\tilde{\mu}$ increases from $0$ to $+1$.
If the coordinates cover the interior of the spheroid continuously then $\tilde{\mu}$ will pass through $\xi_0$ at some point along this curve, before it has reached the pole, despite us having shown in equation \eqref{ximusurf} that this implies that such a point lies on the surface of the spheroid.
Therefore such a continuous covering of the interior with $\xi = \xi_0$ on the entirety of the surface is not possible for $\kappa^2 < 4$.

It turns out that it is still possible to cover the interior continuously in the $\kappa^2 < 4$ case.
The coordinates take the range $\left|\xi_0\right| \le \xi \le +1$, $-\left|\xi_0\right| \le \tilde{\mu} \le +\left|\xi_0\right|$.
The surface is now split into three regions: an equatorial band in which $\xi = \left|\xi_0\right|$ and $-\left|\xi_0\right| \le \tilde{\mu} \le +\left|\xi_0\right|$, and north and south polar `caps' in which $\tilde{\mu} = \pm \left|\xi_0\right|$ and $\left|\xi_0\right| \le \xi \le +1$
In this case both the surfaces of constant $\xi$ and those of constant $\tilde{\mu}$ are spheroids, hence naming the coordinate system `bi-spheroidal'.
These surfaces are illustrated in the left-hand plot of figure \ref{fig:bi_spher}.

In the case that $ 4 < \kappa^2 < 4\left(1+\zeta_0^2\right)$ we can cover the spheroid with the coordinate ranges $1 \le \xi \le \xi_0$, $-1 \le \tilde{\mu} \le +1$.
The surface is now $\xi = \xi_0$, $-1 \le \tilde{\mu} \le +1$.
Note that in this case $b$ is purely imaginary, but that $\varpi$, $z$, $\xi$ and $\tilde{\mu}$ are real.
The surfaces of constant $\xi$ are now spheroids, whereas those of constant $\tilde{\mu}$ are hyperboloids.
We will still refer to this coordinate system as the `bi-spheroidal coordinates', even though this name is now longer really appropriate.
An example of the constant $\xi$, $\tilde{\mu}$ surfaces for $\kappa$ within this range can be seen in the central plot of figure \ref{fig:bi_spher}.

For $\kappa^2 > 4\left(1 + \zeta_0^2\right)$, $b$ is again real, but now $\xi_0$ is imaginary.
Formally, the coordinates \eqref{bi_spher_def} still work if we take $\xi$ to range over purely imaginary values.
In this case the coordinate patch $ -\left|\xi_0\right| \le -i \xi \le 0$, $-1 \le \tilde{\mu} \le +1$ covers the spheroid and $i\xi = \left|\xi_0\right|$, $-1\le \tilde{\mu} \le+1$ is the surface of the spheroid.
The surfaces of constant $\xi$ are spheroids, whereas those of constant $\tilde{\mu}$ are hyperboloids.
An example of these surfaces for this $\kappa$ range can be seen in the right-hand plot of figure \ref{fig:bi_spher}.
One new issue arises in this range of $\kappa$ - the $\xi$, $\tilde{\mu}$ coordinates possess a degeneracy on the $\xi = 0$ disc.
Specifically, the point $\left(\xi = 0, \tilde{\mu}\right)$ is the same as the point $\left(\xi = 0, -\tilde{\mu}\right)$.
This will prove useful to rule out certain solutions for the hydrodynamic potential.

Unlike the oblate spheroidal coordinates of the previous section, these coordinates are not orthogonal.
They have line element
\begin{align}
ds^2 = b^2 \Bigg(
& \frac{4\xi^2\left(1 - \tilde{\mu}^2\right) + \kappa^2\left(\tilde{\mu}^2 - \xi^2\right)}{\kappa^2\left(1-\tilde{\mu}^2\right)} d\tilde{\mu}^2 \nonumber \\
&+ \frac{4\tilde{\mu}^2\left(1 - \xi^2\right) + \kappa^2\left(\xi^2 - \tilde{\mu}^2\right)}{\kappa^2\left(1-\xi^2\right)} d \xi^2 \nonumber \\
&+ 2 \frac{4}{\kappa^2} \xi \tilde{\mu} d\xi d\tilde{\mu} + \left(1-\xi^2\right)\left(1-\tilde{\mu}^2\right) d\varphi^2 \Bigg) \,.
\end{align}

To convert the Poincar\'{e} equation into these coordinates we employ a rescaled metric $\tilde{g}_{a b}$, with respect to which $W$ obeys Laplace's equation and the $(\xi, \tilde{\mu}, \varphi)$ coordinates are orthogonal, 
\begin{align}
&d\tilde{s}^2 = \tilde{g}_{a b} dx^a dx^b = dx^2 + dy^2 + d z^{\prime 2} \nonumber \\
&= b^2 \frac{(\xi^2 - \tilde{\mu}^2)}{1-\xi^2} d \xi^2 + b^2 \frac{(\tilde{\mu}^2 - \xi^2)}{1-\tilde{\mu}^2} d \tilde{\mu}^2 + b^2 (1-\xi^2)(1-\tilde{\mu}^2) d \varphi^2 \,.
\end{align}
We may now apply the covariant divergence formula to write
\begin{align} \label{PoinBiSpher}
&\left( \partial_x^2 + \partial_y^2 + (1-4/\kappa^2) \partial_z^2\right) W =
\frac{1}{\sqrt{\tilde{g}}} \partial_a \left( \sqrt{\tilde{g}} \tilde{g}^{a b} \partial_b W \right) \nonumber \\
=& \frac{1}{b^2(\xi^2 - \tilde{\mu}^2)} \frac{\partial}{\partial \xi} \left( \left(1 - \xi^2\right) \frac{\partial W}{\partial \xi} \right) \nonumber \\
& + \frac{1}{b^2(\tilde{\mu}^2 - \xi^2)} \frac{\partial}{\partial \tilde{\mu}} \left( \left(1 - \tilde{\mu}^2\right) \frac{\partial W}{\partial \tilde{\mu}} \right) \nonumber \\
& + \frac{1}{b^2(1-\xi^2)(1-\tilde{\mu}^2)} \frac{\partial^2 W}{\partial \varphi^2} \,.
\end{align}

In section \ref{sec:general_Bryan_mode} we will need to convert the differential operator $\partial^2 / \partial \xi \partial \tilde{\mu}$ into cylindrical polar coordinates.
To do so we will need to employ the partial derivatives
\begin{align} \label{cylind_bi_spher_derivs}
\left.\frac{\partial \varpi}{\partial \xi}\right|_{\tilde{\mu}} &= - b \frac{\sqrt{1-\tilde{\mu}^2}}{\sqrt{1-\xi^2}} \xi, &
\left.\frac{\partial \varpi}{\partial \tilde{\mu}}\right|_{\xi} &= - b \frac{\sqrt{1-\xi^2}}{\sqrt{1-\tilde{\mu}^2}} \tilde{\mu} \, \nonumber \\
\left.\frac{\partial z}{\partial \xi}\right|_{\tilde{\mu}} &= b \frac{\sqrt{4-\kappa^2}}{\kappa} \tilde{\mu}, &
\left.\frac{\partial z}{\partial \tilde{\mu}}\right|_{\xi} &= b \frac{\sqrt{4-\kappa^2}}{\kappa} \xi \,.
\end{align}
Later in section \ref{sec:general_Bryan_mode} will also convert the differential operators $\partial / \partial \varpi$ and $\partial / \partial z$ into bi-spheroidal coordinates.
To do so we require the partial derivatives
\begin{align} \label{bi_spher_cylind_derivs}
\left.\frac{\partial \xi}{\partial \varpi}\right|_z & = \frac{\xi \sqrt{\left(1-\xi^2\right)\left(1 - \tilde{\mu}^2\right)}}{b \left(\tilde{\mu}^2 - \xi^2 \right)} \,, \nonumber \\
\left.\frac{\partial \tilde{\mu}}{\partial \varpi}\right|_z & = - \frac{\tilde{\mu} \sqrt{\left(1-\xi^2\right)\left(1 - \tilde{\mu}^2\right)}}{b \left(\tilde{\mu}^2 - \xi^2 \right)} \,, \nonumber \\
\left.\frac{\partial \xi}{\partial z}\right|_{\varpi} & = \frac{\kappa}{b\sqrt{4-\kappa^2}} \frac{\tilde{\mu}\left(1-\xi^2\right)}{\tilde{\mu}^2 - \xi^2} \,, \nonumber \\
\left.\frac{\partial \tilde{\mu}}{\partial z}\right|_{\varpi} & =-\frac{\kappa}{b\sqrt{4-\kappa^2}} \frac{\xi\left(1-\tilde{\mu}^2\right)}{\tilde{\mu}^2 - \xi^2} \,.
\end{align}

\FloatBarrier

\section{Modes with $\kappa^2 = 0$}
\label{sec:zeromodes}

Bryan's treatment, which we described in section \ref{sec:general_Bryan_mode}, is not valid for zero-frequency modes.
On physical grounds we should expect such modes to exist.
For example, an axisymmetric differential rotation profile combined with an appropriate surface deformation, should be a steady flow in the rotating frame (including the simplest case, that of spinning-up the spheroid combined with an appropriate increase in its eccentricity).
We should also expect to find the vertical translation mode of the spheroid.
We might also expect to find a zero frequency mode at the onset of the secular instability.
However, we shall see that only constant deformations (as opposed to constant velocities) are permitted in the latter case.

When $\kappa=0$ the inviscid Euler equations \eqref{inviscidNS} reduce to
\begin{align}
- 2 \Omega u_{\varphi} & = - \partial_{\varpi} W \,, \\
\label{kappa0u_varpi}
2 \Omega u_{\varpi} & = - \varpi^{-1} \partial_{\varphi} W \,, \\
0 & = \partial_z W \,.
\end{align}
The bi-spheroidal coordinate system described in appendix \ref{sec:maclaurin_spheroid_bi_spheroidal_coords} no longer covers the spheroid, since $z$ is now undefined.
However, for a zero frequency mode neither $\bm{u}$ nor $W$ depend on $z$ (incompressibility must be used to show that $\partial_z u_z = 0$) and inspection of appendix \ref{sec:maclaurin_spheroid_bi_spheroidal_coords} suggests that it is natural to write $\varpi = c \sqrt{1+\zeta_0^2} \sqrt{1- \xi^2}$, where $0 \le \xi \le 1$, and to take $W$ to be
\begin{align}
W = D_{l,m,0} P_l^m(\xi) e^{im\varphi} \,.
\end{align}
This determines the horizontal velocity components
\begin{align} \label{kappa0uphi}
u_{\varphi} &= -\frac{\sqrt{1-\xi^2} D_{l,m,0}}{2\Omega c \: \xi \sqrt{1+\zeta_0^2}} \frac{d P_l^m(\xi)}{d \xi} e^{im\varphi} \,, \\
u_{\varpi} &= - \frac{i m D_{l,m,0}}{2 \Omega c \sqrt{1-\xi^2} \sqrt{1 + \zeta_0^2}} P_l^m(\xi) e^{im\varphi} \,.
\end{align}

We must still satisfy the boundary conditions \eqref{pbc} and \eqref{Phibc}, and may again combine these to write
\begin{align} \label{gravbcsimp}
\left[\bm{n} \cdot \bm{\nabla} \Phi^{\prime}\right]^{\partial V^+}_{\partial V^-} &= - 4 \pi G \rho^2 \left(W - \Phi^{\prime} \right) / \bm{n}\cdot \bm{\nabla} p \,.
\end{align}
In addition to static displacements we wish to consider constant velocities, so we generalise the gravitational solution \eqref{gravsoln} by replacing $A_{l,m,0}$ with $A_{l,m,0} + \widehat{A}_{l,m} \Omega t$ (we do not similarly generalise $W$, since this would produce time dependent velocities).
Equation \eqref{gravbcsimp} reduces to
\begin{align} \label{kappa0gravbc1}
\left(B_l^m(\zeta_0) \zeta_0 \left(1- \zeta_0 \cot^{-1}\zeta_0\right) + 1\right) A_{l,m,0} & = D_{l,m,0} \,, \\
\label{kappa0gravbc2}
\left(B_l^m(\zeta_0) \zeta_0 \left(1- \zeta_0 \cot^{-1}\zeta_0\right) + 1\right) \widehat{A}_{l,m} & = 0 \,.
\end{align}
Note that the first of these is identical to \eqref{Phibc_A_D}, the result for harmonic time dependence.

The pressure boundary condition is more complicated, since we must first compute $\bm{\xi}\cdot \bm{n}$ at the surface.
Using \eqref{kappa0u_varpi} and the fact that $\xi = \left| \mu \right|$ on the surface we write
\begin{align}
\bm{\xi} \cdot \bm{n} = &  \frac{1}{\sqrt{\zeta_0^2 + \mu^2}} \bigg(
\frac{-i m \zeta_0 D_{l,m,0} P_l^m(\left|\mu\right|) e^{im\varphi}t}{2 \Omega c \sqrt{1+\zeta_0^2}} \nonumber \\
&+ \zeta_0 \sqrt{1-\mu^2} \xi_{\varpi}\left(t=0\right)
+ \mu\sqrt{1+\zeta_0^2} \left( u_z t + \xi_z \left(t=0\right) \right)
\bigg) \,.
\end{align}
To match the spatial dependence of the time-dependent terms in $p^{\prime}$ we require that either $l+m$ is even (so that $P_l^m(\left|\mu\right|) = P_l^m(\mu)$) or that $m=0$, and that
\begin{align} \label{kappa0uz}
\left.u_z\right|_{\partial V} = \frac{\zeta_0 Z_{l,m} P_l^m(\mu) e^{im\varphi}}{2 \Omega c (1+\zeta_0^2) \mu} \,.
\end{align}
Neglecting any constant displacement parallel to the surface, we may write
\begin{align}
\left. \xi_{\varpi} (t=0) \right|_{\partial V} &= \frac{\widehat{D}_{l,m} P_l^m(\mu) e^{i m \varphi}}{2 \Omega^2 c \sqrt{1+\zeta_0^2}\sqrt{1-\mu^2}} \,, \\
\left. \xi_z (t=0) \right|_{\partial V} &= \frac{\zeta_0 \widehat{Z}_{l,m} P_l^m(\mu) e^{im \varphi}}{2 \Omega^2 c (1+\zeta_0^2) \mu} \,.
\end{align}
The pressure boundary condition then reduces to
\begin{align}
\label{kappa0pbc1}
D_{l,m,0} - A_{l,m,0} &= \frac{\zeta_0\left(1 - \zeta_0 \cot^{-1}\zeta_0\right)}{\left((1+3\zeta_0^2)\cot^{-1}\zeta_0 - 3\zeta_0\right)} \left(\widehat{D}_{l,m} + \widehat{Z}_{l,m}\right) \,, \\
\label{kappa0pbc2}
- \widehat{A}_{l,m} &= \frac{\zeta_0\left(1 - \zeta_0\cot^{-1}\zeta_0\right)}{\left((1+3\zeta_0^2)\cot^{-1}\zeta_0 - 3\zeta_0\right)} \left( Z_{l,m} - im D_{l,m,0}\right) \,.
\end{align}


Equipped with \eqref{kappa0gravbc1}, \eqref{kappa0gravbc2}, \eqref{kappa0pbc1} and \eqref{kappa0pbc2} we may ask what zero modes are permitted.
Inspection of \eqref{kappa0gravbc1} and \eqref{kappa0gravbc2} suggests that $B_l^m\left(\zeta_0\right) \zeta_0 \left(1 - \zeta_0 \cot^{-1} \zeta_0\right) + 1 = 0$ should be considered as a special case.
In fact it turns out that these values of $\zeta_0$ are exactly the eccentricities at which an otherwise non-zero mode frequency passes through zero, examples of which can been seen in figures \ref{fig:l2freqplot}, \ref{fig:l3freqplot} and \ref{fig:l4freqplot}.
The lowest eccentricity satisfying this condition (which occurs for $l=2$, $m=0$) is the eccentricity at which the onset of secular instability occurs.
Indeed, these eccentricities allow $\widehat{A}_{l,m}$ to be non-zero, corresponding to an increase in gravitational perturbation (and hence surface displacement) that is linear in time, and related to the vertical velocity through \eqref{kappa0pbc2}.
However, such eccentricities also impose $D_{l,m,0}=0$, and hence $u_{\varpi} = u_{\varphi} = 0$.
This might seem surprising, since there is no restriction on $\widehat{D}_{l,m} + \widehat{Z}_{l,m}$, only a relation to the gravitational potential through \eqref{kappa0pbc1}.
Constant horizontal displacements are permitted, but constant horizontal velocities are not.

This is due to such modes having $W=0$, and hence the only force on a control volume of fluid being $-2 \bm{\Omega} \times \bm{u}$.
A static displacement does not feel this force.
Such eccentricities do not signal the onset of the dynamical instability of these modes.
We comment on this in section \ref{sec:sectoral_modes}, where we show that between the onset of the secular and dynamical instabilities the stabilisation is indeed due to the Coriolis force.
We also note here that $B_l^m\left(\zeta_0\right) \zeta_0 \left(1 - \zeta_0 \cot^{-1} \zeta_0\right) + 1$ always vanishes for $l=1$, $m=0$.
This does not signal an instability, it is simply the vertical translation mode of the spheroid.



What happens for a general eccentricity at which $B_l^m\left(\zeta_0\right) \zeta_0 \left(1 - \zeta_0 \cot^{-1} \zeta_0\right) + 1 \ne 0$?
We must have $\widehat{A}_{l,m} = 0$ to satisfy \eqref{kappa0gravbc2}, and \eqref{kappa0pbc2} then fixes $Z_{l,m} = i m D_{l,m}$.
Recall that the equations of motion require $\partial_z u_z = \partial_z u_{\varpi} = 0$.
The surface expression \eqref{kappa0uz} must therefore be identical on the upper and lower surfaces of the spheroid, which requires $l+m$ to be odd for non-zero $Z_{l,m}$, and we saw above that non-zero $D_{l,m,0}$ requires $l+m$ to be even or $m$ to be zero.
Therefore the only zero modes at a general eccentricity are axisymmetric ($m=0$) and have no vertical or radial flow.
Such modes have $Z_{0,0} = 0$ and $D_{l,0,0} \ne 0$.
These purely azimuthal flow fields represent a differentially rotating flow within the spheroid.
In these cases we may invert \eqref{kappa0uphi} to obtain the $D_{l,0,0}$ coefficients
\begin{align} \label{diffrotD}
D_{l,0,0} = c \Omega & \sqrt{1+\zeta_0^2} \frac{2l+1}{l(l+1)} \nonumber \\
& \times \int_{-1}^{+1} P_l(\xi) \frac{d}{d\xi} \left(\xi \sqrt{1-\xi^2} u_{\varphi}(\left|\xi\right|)\right) \mathrm{d} \xi
\end{align}
for non-zero $l$.
$D_{0,0,0}$ is the addition of a constant to the hydrodynamic potential and has no physical significance, and $D_{l,0,0}$ vanishes for all odd $l$.
From \eqref{diffrotD} we can then calculate the perturbation to the gravitational field via \eqref{kappa0gravbc1}, and hence the surface displacement using \eqref{p_prime_over_gradient}.
The relation between the gravitational perturbation and the axisymmetric zero-modes has been studied previously by \citet{Kong2012}.
We have shown that these modes can be formally labelled by a degree and order, as with all of Bryan's other modes.


\section{Modes with $\kappa = \pm 2$}
\label{sec:kappa4modes}

Bryan's method also fails to cover the case that $\kappa = \pm 2$.
In this case equation \eqref{inviscidNS} becomes
\begin{align}
\mp 2 i \Omega u_{\varpi} - 2 \Omega u_{\varphi} &= - \partial_{\varpi} W \,, \label{kpm2NS1} \\
\mp 2 i \Omega u_{\varphi} + 2 \Omega u_{\varpi} &= - \varpi^{-1} \partial_{\varphi} \label{kpm2NS2} W \,, \\
\mp 2 i \Omega u_z &= - \partial_z W \label{kpm2NS3} \,,
\end{align}
which may not be inverted to find $\bm{u}$ due to the linear dependence of the left hand sides of equations \eqref{kpm2NS1} and \eqref{kpm2NS2}.
To proceed, we note that $W$ must satisfy $\partial_{\varpi} W = \pm \frac{m}{\varpi} W$, allowing us to write
\begin{align} \label{Wsolnkappapm2}
W = f(z) \varpi^{\pm m} e^{\mp 2 i \Omega t + i m \varphi} \,.
\end{align}
From now on we shall assume that $\kappa = +2$;
the solutions for $\kappa = -2$ may be obtained by complex conjugation.
If $f(z)$ is not identically zero, then regularity on the $z$-axis requires $m \ge 0$.
Using \eqref{kpm2NS2} to eliminate $u_{\varphi}$ we may write the incompressibility condition as
\begin{align}
\frac{1}{\varpi} \partial_{\varpi} \left( \varpi u_{\varpi} \right) + \frac{m}{\varpi} u_{\varpi} + \frac{i}{2 \Omega} \left( m^2 \varpi^{m-2} f(z) - \varpi^m f^{\prime \prime} (z) \right) = 0 \,.
\end{align}
This has the general solution
\begin{align} \label{uvarpikappapm2}
u_{\varpi} = \bigg( & \frac{i}{2 \Omega} \left( \frac{1}{m+1} f^{\prime \prime}(z) \varpi^{m+1} - m f(z) \varpi^{m-1} \right) \nonumber \\
& + C \varpi^{-1 -m} \bigg) e^{-2 i \Omega t + i m \varphi} \,.
\end{align}
For now we shall assume that $f \ne 0$, requiring $m \ge 0$ and $C = 0$.

We must now impose the boundary conditions. The pressure and gravitational boundary conditions, \eqref{pbc} and \eqref{Phibc}, may be combined to obtain
\begin{align} \label{gravbcsimp2}
\left[\bm{n} \cdot \bm{\nabla} \Phi^{\prime}\right]^{\partial V^+}_{\partial V^-} &= - 4 \pi G \rho^2 \left(W - \Phi^{\prime} \right) / \bm{n}\cdot \bm{\nabla} p \,.
\end{align}
Note that it must still be possible to decompose $\Phi^{\prime}$ into the solutions \eqref{gravsoln} found in section \ref{sec:general_Bryan_mode} and so we write
\begin{align} \label{gravsolnkpm2}
\Phi^{\prime} &= \left\{ \begin{matrix} \sum_{l=m}^{\infty} A_{l,m,2} \frac{P_l^m\left(i\zeta\right)}{P_l^m\left(i\zeta_0\right)} P_l^m\left(\mu\right) e^{- 2 i \Omega t + i m \varphi} \;\mathrm{for}\;\; \zeta \le \zeta_0 \,, \\
\,\, \sum_{l=m}^{\infty} A_{l,m,2} \frac{Q_l^m\left(i\zeta\right)}{Q_l^m\left(i\zeta_0\right)} P_l^m\left(\mu\right) e^{- 2 i \Omega t + i m \varphi} \;\mathrm{for}\;\; \zeta \ge \zeta_0 . \end{matrix} \right.
\end{align}
We may substitute \eqref{gravsolnkpm2} into \eqref{gravbcsimp2} to obtain
\begin{align} \label{Wkappapm2LegendreP}
W = \sum_{l=m}^{\infty} A_{l,m,2} & \left(1 + B_l^m(\zeta_0) \zeta_0 (1 - \zeta_0 \cot^{-1} \zeta_0) \right) \nonumber \\
& \times P_l^m(\mu) e^{-2 i \Omega t + i m \varphi} \, ,
\end{align}
where $B_l^m(\zeta_0)$ is defined in equation \eqref{Bdef}.
Note that \eqref{gravbcsimp2} only holds on the surface, $\partial V$, of the spheroid, and that this may be labelled by $\mu = z / c \zeta_0$ as described in appendix \ref{sec:maclaurin_spheroid_obl_coords}.
Motivated by this we define a new function, $F$, by $F(\mu) = f(c \zeta_0 \mu)$ throughout the spheroid.
This is purely for notational convenience;
$f$ may be recovered immediately as $f(z) = F(z/c\zeta_0)$.
Recalling also from appendix \ref{sec:maclaurin_spheroid_obl_coords} that $\varpi$ may be written as $\varpi = c \sqrt{1 + \zeta_0^2} \sqrt{1 - \mu^2}$ on the surface, we may use \eqref{Wsolnkappapm2} and \eqref{Wkappapm2LegendreP} to write
\begin{align} \label{Fkappapm2LegendreP}
F(\mu) = & c^{-m} (1+\zeta_0^2)^{-m/2} (1 - \mu^2)^{-m/2} \nonumber \\
& \sum_{l=m}^{\infty} A_{l,m,2} \left(1 + B_l^m(\zeta_0) \zeta_0 (1 - \zeta_0 \cot^{-1} \zeta_0) \right) \times P_l^m(\mu) \,.
\end{align}

We now apply the pressure boundary condition \eqref{pbc}.
To do so we shall need to use \eqref{surfacenormal}, \eqref{uvarpikappapm2} and $u_z = - i f^{\prime}(z) \varpi^m e^{-2i\Omega t + i m \varphi} / 2 \Omega$.
This condition may be written as
\begin{align} \label{pbcF}
W - \Phi^{\prime} = c^m & (1+\zeta_0^2)^{m/2} (1 - \mu^2)^{m/2} (1 - \zeta_0 \cot^{-1} \zeta_0) \frac{\pi G \rho}{\Omega^2} \nonumber \\
\times \Bigg( & - \frac{(1 + \zeta_0^2) (1 - \mu^2)}{2(m+1)} F^{\prime \prime}(\mu)
+ \frac{m}{2} \zeta_0^2 F(\mu) \nonumber \\
 & + \mu (1+\zeta_0^2) F^{\prime} (\mu)
\Bigg) \,.
\end{align}
Ideally we would like an expression whose only $\mu$ dependence is in $F$ and in $\Phi^{\prime}$, since these are composed of sums over associated Legendre polynomials of the same order, which are known to be orthogonal. 
This turns out to be the case.
We may eliminate $F^{\prime}$ and $F^{\prime \prime}$ by noting that
\begin{align}
& c^m(1+\zeta_0^2)^{\frac{m}{2}+1} (1 - \mu^2)^{\frac{m}{2}} \left( -\frac{(1-\mu^2)}{2(m+1)} F^{\prime \prime}(\mu) + \mu F^{\prime}(\mu) \right) \nonumber \\
& = -\frac{1 + \zeta_0^2}{2(m+1)} \sum_{l=m}^{\infty} A_{l,m,2} \big( 1 + B_l^m(\zeta_0) \zeta_0 (1 - \zeta_0 \cot^{-1} \zeta_0) \big) \nonumber \\
& \times \Bigg( (1-\mu^2) \frac{d^2 P_l^m(\mu)}{d \mu^2} - 2 \mu \frac{d P_l^m(\mu)}{d \mu} \nonumber \\
& \;\;\;\;\;\;\; + \left( m(m+1) -\frac{m^2}{1 - \mu^2} \right) P_l^m(\mu) \Bigg) \nonumber \\
& = (1+\zeta_0^2) \sum_{l=m}^{\infty} A_{l,m,2} (1 + \zeta_0 B_l^m(\zeta_0) (1 - \zeta_0 \cot^{-1} \zeta_0)) \nonumber \\
& \;\;\;\;\; \times \frac{l(l+1) - m(m+1)}{2(m+1)} P_l^m(\mu) \,,
\end{align}
where the associated Legendre equation has been used in the final manipulation.
Substitution of this result into \eqref{pbcF} gives
\begin{align}
\sum_{l=m}^{\infty} A_{l,m,2} \Bigg[ & \zeta_0 B_l^m(\zeta_0) + \frac{\pi G \rho}{\Omega^2} (1 + \zeta_0 B_l^m(\zeta_0) (1 - \zeta_0 \cot^{-1}\zeta_0) ) \nonumber \\
& \times \left( \frac{m}{2} - \frac{l (l+1)}{2(m+1)} (1 + \zeta_0^2) \right)
\Bigg] P_l^m(\mu) = 0 \,.
\end{align}
Such $l= \pm 2$, $\kappa = \pm 2$ modes will therefore only occur at specific eccentricities of the spheroid, and we do not consider them further.

Let us now return to the case that $\kappa = +2$, $m < 0$, $u_{\varpi} = C \varpi^{-1-m}$ and $W=0$.
We must still be able to decompose the gravitational potential as \eqref{gravsolnkpm2}, and applying the boundary condition \eqref{gravbcsimp2} gives
\begin{align} \label{inertialcirclegravbc}
\sum_{l=m}^{\infty} A_{l,m,2} \left( 1 + B_l^m(\zeta_0) \zeta_0 \left(1 - \zeta_0 \cot^{-1} \zeta_0 \right) \right) P_l^m(\mu) = 0 \,.
\end{align}
Using \eqref{surfacenormal} and $u_z=0$, we apply the pressure boundary condition, \eqref{pbc}, relates the velocity to the gravitational perturbation via
\begin{align} \label{inertialcirclepressurebc}
C = \frac{i}{2} \frac{\Omega}{\pi G \rho} \frac{c^2 (1-\mu^2)^{m/2} (1+ \zeta_0^2)^{m/2}}{\zeta_0^2 (1 - \zeta_0 \cot^{-1} \zeta_0)} \sum_{l=m}^{\infty} A_{l,m,2} P_l^m(\mu) \,.
\end{align}
To have a non-zero velocity, clearly we must have at least one of the $A_{l,m,2}$ non-zero.
From \eqref{inertialcirclegravbc} this requires $1 + B_l^m(\zeta_0) \zeta_0 \left(1 - \zeta_0 \cot^{-1} \zeta_0 \right) = 0$, which is in fact the same condition that we found in appendix \ref{sec:zeromodes} for the eccentricity at which a mode frequency passes through zero.
However, we have a second restriction;
we require that $C$, as given by \eqref{inertialcirclepressurebc}, really is a constant.
This can only occur for $m=-l$, so all other $A_{l,m,2}$ must vanish.
Thus the $l= \mp 2$, $\kappa = \pm 2$ modes may also occur only at specific eccentricities of the spheroid.
We do not consider them further in this paper.

\section{Tables of $\kappa$ values for modes of degree $l\le4$}
\label{sec:kappa_tables}

We list in table \ref{tab:freqtable} the frequencies in units of the rotation rate ($\kappa$) and in units of the dynamical frequency of the spheroid ($\kappa \Omega / \sqrt{\pi G \rho}$) of each of the modes described in this paper.
These are the frequencies in the frame rotating with the spheroid.
The frequency of a mode (in units of the spheroid dynamical frequency) in the inertial frame may be found by $\kappa \Omega / \sqrt{\pi G \rho} + m \Omega$.

Only modes with $m \ge 0$ are listed, and for axisymmetric ($m=0$) modes only non-negative frequencies are listed.
Therefore, for every mode in this table with $\kappa \ne 0$ there exists a corresponding mode of order $-m$ and frequency $-\kappa$.
However, since these are the complex conjugates of the modes listed here, table \ref{tab:freqtable} does list all of the physical modes of degree $l \le 4$.
The frequencies in the $e=0$ column were computed using the results for a sphere, given in equations \eqref{sphereSurfGrav}, \eqref{spherel3} and \eqref{spherel4}.

Note that these frequencies are also plotted against the eccentricity of the spheroid in figures \ref{fig:l2freqplot}, \ref{fig:l3freqplot}, \ref{fig:l3kappaplot}, \ref{fig:l4freqplot} and \ref{fig:l4kappaplot}.

\begin{table*}
\begin{minipage}{126mm}
\hspace{-1.5cm}
\begin{tabular}{|>{$}l<{$} | >{$}l<{$} | >{$}r<{$} | >{$}r<{$}| >{$}r<{$}| >{$}r<{$}| >{$}r<{$}| >{$}r<{$}| >{$}r<{$}| >{$}r<{$}| l |}
\hline
l & m & \multicolumn{2}{c|}{$e = 0$} & \multicolumn{2}{c|}{$e = 0.1$} & \multicolumn{2}{c|}{$e = 0.2$} & \multicolumn{2}{c|}{$e = 0.5$} & Comments \\
\hhline{|~|~|-|-|-|-|-|-|-|-|~|}
  &   & \kappa & \kappa \Omega / \sqrt{\pi G \rho} & \kappa & \kappa \Omega / \sqrt{\pi G \rho} & \kappa & \kappa \Omega / \sqrt{\pi G \rho} & \kappa & \kappa \Omega / \sqrt{\pi G \rho} &   \\
\hhline{|=|=|=|=|=|=|=|=|=|=|=|}
1 & 0 & 0 & 0 & 0 & 0 & 0 & 0 & 0 & 0 & vert. trans. \\
\hline
1 & 1 & -1 & 0 & -1 & -0.07308 & -1 & -0.1465 & -1 & -0.3715 & horiz. trans. \\
\hhline{|=|=|=|=|=|=|=|=|=|=|=|}
2 & 0 & 0 & 0 & 0 & 0 & 0 & 0 & 0 & 0 & spin-up \\
2 & 0 & \infty & 1.033 & 14.16 & 1.036 & 7.139 & 1.046 & 3.005 & 1.116 & SGM \\
\hline
2 & 1 & -1 & 0 & -1 & -0.07308 & -1 & -0.1465 & -1 & -0.3715 & spin-over \\
2 & 1 & -\infty & -1.033 & -14.65 & -1.071 & -7.589 & -1.112 & -3.372 & -1.253 & RSGW \\
2 & 1 & \infty & 1.033 & 13.65 & 0.9976 & 6.589 & 0.9651 & 2.372 & 0.8812 & PSGW \\
\hline
2 & 2 & -\infty & -1.033 & -15.08 & -1.102 & -7.938 & -1.163 & -3.469 & -1.289 & RSGW \\
2 & 2 & \infty & 1.033 & 13.08 & 0.9556 & 5.938 & 0.8698 & 1.469 & 0.5457 & PSGW \\
\hhline{|=|=|=|=|=|=|=|=|=|=|=|}
3 & 0 & 0.8944 & 0 & 0.8963 & 0.06551 & 0.9022 & 0.1322 & 0.9492 & 0.3526 & IM \\
3 & 0 & \infty  & 1.512 & 20.72 & 1.514 & 10.38 & 1.521 & 4.223 & 1.569 & SGM \\
\hline
3 & 1 & -1.510 & 0 & -1.512 & -0.1105 & -1.518 & -0.2224 & -1.570 & -0.5833 & RIW \\
3 & 1 & 0.1766 & 0 & 0.1762 & 0.01288 & 0.1751 & 0.02565 & 0.1653 & 0.06141 & PIW \\
3 & 1 & -\infty & -1.512 & -21.04 & -1.538 & -10.69 & -1.566 & -4.462 & -1.658 & RSGW \\
3 & 1 & \infty & 1.512 & 20.37 & 1.489 & 10.03 & 1.470 & 3.867 & 1.437 & PSGW \\
\hline
3 & 2 & -0.6667 & 0 & -0.6697 & -0.04895 & -0.6791 & -0.09947 & -0.7514 & -0.2791 & RIW \\
3 & 2 & -\infty & -1.512 & -21.34 & -1.559 & -10.95 & -1.604 & -4.605 & -1.711 & RSGW \\
3 & 2 & \infty & 1.512 & 20.01 & 1.462 & 9.628 & 1.410 & 3.357 & 1.247 & PSGW \\
\hline
3 & 3 & -\infty & -1.512 & -21.61 & -1.579 & -11.17 & -1.636 & -4.645 & -1.726 & RSGW \\
3 & 3 & \infty & 1.512 & 19.61 & 1.433 & 9.167 & 1.343 & 2.645 & 0.9827 & PSGW \\
\hhline{|=|=|=|=|=|=|=|=|=|=|=|}
4 & 0 & 0 & 0 & 0 & 0 & 0 & 0 & 0 & 0 & diff. rot. \\
4 & 0 & 1.309 & 0 & 1.312 & 0.09590 & 1.321 & 0.1935 & 1.391 & 0.5166 & IM \\
4 & 0 & \infty & 1.886 & 25.82 & 1.887 & 12.92 & 1.892 & 5.179 & 1.924 & SGM \\
\hline
4 & 1 & -1.708 & 0 & -1.710 & -0.1250 & -1.716 & -0.2513 & -1.760 & -0.6538 & RIW \\
4 & 1 & -0.6120 & 0 & -0.6146 & -0.04491 & -0.6225 & -0.09118 & -0.6871 & -0.2552 & RIW \\
4 & 1 & 0.8200 & 0 & 0.8222 & 0.06009 & 0.8289 & 0.1214 & 0.8833 & 0.3281 & PIW \\
4 & 1 & -\infty & -1.886 & -26.07 & -1.905 & -13.15 & -1.926 & -5.364 & -1.993 & RSGW \\
4 & 1 & \infty & 1.886 & 25.57 & 1.869 & 12.66 & 1.854 & 4.928 & 1.831 & PSGW \\
\hline
4 & 2 & -1.232 & 0 & -1.235 & -0.09029 & -1.246 & -0.1825 & -1.331 & -0.4945 & RIW \\
4 & 2 & 0.232 & 0 & 0.2318 & 0.01694 & 0.2314 & 0.03390 & 0.2269 & 0.08430 & PIW \\
4 & 2 & -\infty & -1.886 & -26.29 & -1.922 & -13.36 & -1.956 & -5.490 & -2.039 & RSGW \\
4 & 2 & \infty & 1.886 & 25.30 & 1.849 & 12.37 & 1.812 & 4.594 & 1.707 & PSGW \\
\hline
4 & 3 & -0.5 & 0 & -0.5032 & -0.03678 & -0.5131 & -0.07515 & -0.5931 & -0.2203 & RIW \\
4 & 3 & -\infty & -1.886 & -26.51 & -1.937 & -13.53 & -1.982 & -5.560 & -2.065 & RSGW \\
4 & 3 & \infty & 1.886 & 25.01 & 1.828 & 12.05 & 1.765 & 4.153 & 1.543 & PSGW \\
\hline
4 & 4 & -\infty & -1.886 & -26.71 & -1.952 & -13.69 & -2.005 & -5.568 & -2.068 & RSGW \\
4 & 4 & \infty & 1.886 & 24.71 & 1.806 & 11.69 & 1.712 & 3.568 & 1.325 & PSGW \\
\hline
\end{tabular}
\end{minipage}
\caption{
Frequencies of modes of degree $l \le 4$ in the frame rotating with the spheroid, in units of the rotation rate of the spheroid ($\kappa$) and in units of the dynamical frequency of the spheroid ($\kappa \Omega / \sqrt{\pi G \rho}$), for Maclaurin spheroid of eccentricity $e=0$, $0.1$, $0.2$ and $0.5$.
Only modes having $m \ge 0$ are listed, and modes having $m=0$ are only listed if they have non-negative $\kappa$ (that is, only one member of complex conjugate pairs of modes are listed, and there is a one to one correspondence between physical modes and the modes listed here).
In the right hand column the abbreviations SGM, SGW, IM and IW respectively stand for `surface gravity mode', `surface gravity wave', `inertial mode' and `inertial wave'.
P and R respectively denote whether a wave is prograde or retrograde in the frame rotating with the spheroid.
}
\label{tab:freqtable}
\end{table*}

\section{Derivation of the decay rates}
\label{sec:decay_rate_derivation}

In section \ref{sec:eigenmode_problem} we conjectured that introducing a small Newtonian viscosity caused the modes described in this paper to decay at a rate given by \eqref{dispexpr}.
This might be surprising, since for each of the inviscid modes, the integral $\int_V u_j \partial_i \partial_i u_j dV$ was shown to vanish by \cite{Zhang2004}.
However, the solutions to the viscous problem will have a slightly different spatial form - we expect that near the surface of the spheroid there will be some boundary layer of thickness $\delta \ll R$ over which the stress adjusts to zero.
Within this region, the gradient of the inviscid modes will scale like $\partial_i u_j \sim U / R$ where $U$ is a typical velocity scale of the inviscid mode (note that this is consistent with the two terms in the final line of \eqref{energy_integral} cancelling).
In this appendix we shall show that, if modification of the spatial form of the modes is confined to a thin boundary layer, then the decay rates are indeed given by \eqref{dispexpr}.

Let us first rewrite the inviscid problem as the eigenvalue problem
\begin{align}\label{inviscidprob}
& \lambda_{0 \alpha} \mathrm{M} \bm{x}_{0 \alpha} = \mathrm{K}_0 \bm{x}_{0 \alpha} \,,
\end{align}
where the eigenvalue and vector are
\begin{align}
\lambda_{0 \alpha} = \kappa_{0 \alpha} \Omega,
\;\; \bm{x}_{0 \alpha} = \left( \begin{array}{c} \bm{\xi}_{0 \alpha} \\ \bm{u}_{0 \alpha} \end{array} \right) \,,
\end{align}
and we define the matrices
\begin{align}
\mathrm{M} = \left(\begin{array}{c c} \mathrm{C} & 0 \\ 0 & 1 \end{array}\right),
\;\; \mathrm{K_0} = \left(\begin{array}{c c} 0 & i \mathrm{C} \\ -i\mathrm{C} & -2 i \bm{\Omega}\times \end{array}\right) \,.
\end{align}
The subscript `0' denotes that a quantity refers to the inviscid problem, and the subscript $\alpha$ labels distinct modes.
The $\mathrm{C}$ operator is defined through its action on $\bm{\xi}$, namely $\mathrm{C} \bm{\xi} = \bm{\nabla} W$, and can be seen to be Hermitian,
\begin{align} \label{potentialEnergyIntegral}
\int_{V} & \bm{\xi}_{\beta}^* \cdot \bm{\nabla} W_{\alpha} \mathrm{d}V
= \int_{\partial V} \frac{p_{\alpha}^{\prime}}{\rho} \bm{\xi}_{\beta}^* \cdot \bm{n} \mathrm{d}S
+ \int_{\partial V} \Phi_{\alpha}^{\prime} \bm{\xi}_{\beta}^* \cdot \bm{n} \mathrm{d}S \nonumber \\
&= \int_{\partial V} g_s \left(\bm{\xi}_{\alpha}\cdot\bm{n}\right) \left(\bm{\xi}_{\beta}^* \cdot \bm{n}\right) \mathrm{d}S
+ \left[ \int_{\partial V} \Phi^{\prime}_{\alpha} \frac{\bm{n}\cdot\bm{\nabla} \Phi^{\prime *}_{\beta}}{4\pi G\rho} \mathrm{d}S \right]_{\partial V^-}^{\partial V^+} \nonumber \\
&= \int_{\partial V} g_s \left(\bm{\xi}_{\alpha}\cdot\bm{n}\right) \left(\bm{\xi}_{\beta}^* \cdot \bm{n}\right) \mathrm{d}S
- \frac{1}{4\pi G \rho} \int_{\mathbb{R}^3} \bm{\nabla} \Phi_{\alpha}^{\prime} \cdot \bm{\nabla} \Phi_{\beta}^{\prime *} \mathrm{d} V \,,
\end{align}
where we have used $g_s = - \bm{n} \cdot \bm{\nabla} p$ to denote the effective surface gravity of the undisturbed spheroid, including the centrifugal force.
This integral gives (up to a factor of $\frac{1}{2} \rho$) the potential energy of the mode.
The fact that $\mathrm{M}$ and $\mathrm{K}_0$ are both Hermitian provides us with an orthogonality relation between the free modes.
Using \eqref{inviscidprob} and its conjugate, we may write
\begin{align}
& \left(\lambda_{0 \alpha} - \lambda_{0 \beta}^*\right) \int_V \bm{x}_{0 \beta}^{\dagger} \mathrm{M} \bm{x}_{0 \alpha} dV
\nonumber \\
& = \int_V \left( \bm{x}_{0\beta}^{\dagger} \mathrm{K}_0 \bm{x}_{0 \alpha} - \bm{x}_{0\beta}^{\dagger} \mathrm{K}_0^{\dagger} \bm{x}_{0 \alpha} \right) \mathrm{d} V
= 0 \,.
\end{align}

Having described in section \ref{sec:general_Bryan_mode} the solutions of \eqref{inviscidprob}, we now add a perturbation to produce a modified eigenvalue problem
\begin{align}\label{viscousprob}
\left(\lambda_{0 \alpha} + \delta \lambda_{\alpha}\right) \mathrm{M} \left(\bm{x}_{0 \alpha} + \bm{\delta x}_{\alpha}\right) = \left(\mathrm{K}_0 + \delta\mathrm{K}\right) \left(\bm{x}_{0 \alpha} + \bm{\delta x}_{\alpha}\right) \,,
\end{align}
where
\begin{align}
\delta \mathrm{K} = \left(\begin{array}{c c} 0 & 0 \\ 0 & i \nu \nabla^2 \end{array}\right) \,,
\end{align}
for $\delta \lambda_{\alpha}$ and for $\bm{\delta x}_{\alpha}$, which we expect to possess an expansion of the form
\begin{align}\label{deltaxexpansion}
\bm{\delta x}_{\alpha} = \sum_{\beta} \epsilon_{\alpha \beta} \bm{x}_{0 \beta} \,.
\end{align}
This is equivalent to the viscous Navier-Stokes equations \eqref{viscousNS}.
Expanding out \eqref{viscousprob}, using \eqref{inviscidprob} and \eqref{deltaxexpansion}, and contracting with $\bm{x}_{0 \alpha}^{\dagger}$ gives
\begin{align}\label{pert1}
& \delta \lambda_{\alpha} \bm{x}_{0 \alpha}^{\dagger} \mathrm{M} \bm{x}_{0 \alpha} + \sum_{\beta} \epsilon_{\alpha \beta} \lambda_{0 \alpha} \bm{x}_{0 \alpha}^{\dagger} \mathrm{M} \bm{x}_{0 \beta} + \delta \lambda_{\alpha} \bm{x}_{0 \alpha}^{\dagger} \mathrm{M} \bm{\delta x}_{\alpha} \nonumber \\
= & \sum_{\beta} \epsilon_{\alpha \beta} \bm{x}_{0 \alpha}^{\dagger} \mathrm{K}_0 \bm{x}_{0 \beta} + \bm{x}_{0 \alpha}^{\dagger} \delta \mathrm{K} \bm{x}_{0 \alpha} + \bm{x}_{0 \alpha}^{\dagger} \delta \mathrm{K} \bm{\delta x}_{\alpha} \,.
\end{align}
Applying the complex conjugate of \eqref{inviscidprob} to the first term on the right hand side of \eqref{pert1}, using the fact that
\begin{equation}
\sum_{\beta} \epsilon_{\alpha \beta} \left(\lambda_{0 \alpha} - \lambda_{0 \beta}^*\right) \int_V \bm{x}_{0 \alpha}^{\dagger} \mathrm{M} \bm{x}_{0 \beta} dV = 0 \,,
\end{equation}
and assuming that the spheroid is dynamically stable (ie. that $\lambda_{0 \beta}^* = \lambda_{0 \beta}$), we may write
\begin{align}\label{pert2}
\delta \lambda_{\alpha} \int_V \bm{x}_{0 \alpha}^{\dagger} \mathrm{M} \left(\bm{x}_{0 \alpha} + \bm{\delta x}_{0 \alpha}\right) dV
= \int_V \bm{x}_{0 \alpha}^{\dagger} \delta \mathrm{K} \left(\bm{x}_{0 \alpha} + \bm{\delta x}_{\alpha}\right) dV \,.
\end{align}

It is \emph{not} sufficient to retain only the terms that are linear in the perturbation.
Doing so would give a perturbation to the eigenvalue of
\begin{align}
\int_V \bm{x}_{0 \alpha}^{\dagger} \delta \mathrm{K} \bm{x}_{0 \alpha} dV \Bigg/ \int_V \bm{x}_{0 \alpha}^{\dagger} \mathrm{M} \bm{x}_{0 \alpha} dV \,.
\end{align}
The numerator of this expression is $i \nu \int_V \bm{u}_{0 \alpha}^* \cdot \nabla^2 \bm{u}_{0 \alpha} dV$, which vanishes for every inviscid mode.
Clearly we must include at least some of the second order terms to find the leading order perturbation to the eigenvalue.
The energy equation \eqref{energy_integral} and the discussion in the paragraph following it act as a guide for how to proceed.
Letting $e_{0\alpha ij}$, $\delta e_{\alpha ij}$ and $e_{\alpha ij}$ respectively denote the rate-of-strain tensors of the inviscid mode, the perturbation, and the sum of these two flows, we rewrite \eqref{pert2} as
\begin{align}\label{pert3}
\delta \lambda_{\alpha} \int_V & \bm{x}_{0 \alpha}^{\dagger} \mathrm{M} \left(\bm{x}_{0 \alpha} + \bm{\delta x}_{0 \alpha}\right) dV
= 2i\nu \bigg( - \int_V e_{0\alpha ij}^* e_{0\alpha ij} dV \nonumber \\
& - \int_V e_{0\alpha ij}^* \delta e_{\alpha ij} dV + \int_{\partial V} \left(u_{0 \alpha}\right)_i^* n_j e_{\alpha ij} dS \bigg) \,.
\end{align}
The surface integral on the right hand side of \eqref{pert3} vanishes by the stress-free boundary condition.
If, as conjectured earlier, the velocity perturbation $\bm{\delta u}$ is appreciably non-zero only within a boundary layer of thickness $\delta \ll R$, then the second volume integral in \eqref{pert3} is smaller than the first by a factor of $\delta/R$.
To summarise: the perturbation to $\bm{u}_{\alpha}$ is \emph{not} small within the boundary layer, and its surface stress integral cancels that of the inviscid flow field. 
However, away from this boundary layer the perturbation to $\bm{u}_{\alpha}$ (and its first derivatives) \emph{is} small, and hence the volume integral of $e_{0\alpha ij} \delta e_{\alpha ij}$ is $\mathcal{O}\left(\delta/R\right)$ relative to the contribution from the inviscid field.

The integral on the left hand side of \eqref{pert3} may be rewritten using the fact that $\mathrm{C}$ is Hermitian as
\begin{align}
\int_V \bm{x}_{0 \alpha}^{\dagger} \mathrm{M} \bm{x}_{0 \alpha} dV
+ \int_V \bm{u}_{0 \alpha}^* \cdot \bm{\delta u}_{\alpha} dV + \left(\int_V \bm{\delta \xi}_{\alpha}^* \cdot C \bm{\xi}_{0 \alpha} dV \right)^* \,.
\end{align}
The first term is the total energy of the inviscid mode, which does not vanish, and the second two integrals are $\mathcal{O}\left(\delta/R\right)$ relative to the first.

Therefore we may write the leading order perturbation to the eigenvalue as
\begin{align}
& \delta \kappa_{\alpha} \Omega  = -i \rho \nu \int_V e_{0\alpha ij}^* e_{0\alpha ij} dV \Bigg/
\bigg( \frac{1}{2}\int_V \rho \left|\bm{u}_{0 \alpha}\right|^2 dV \nonumber \\
& + \frac{1}{2}\int_{\partial V} \rho g_s \left|\bm{\xi}_{0 \alpha}.\bm{n}\right|^2 dS - \frac{1}{8\pi G}\int_{\mathbb{R}^3} \left|\bm{\nabla} \Phi^{\prime}_{0 \alpha}\right|^2 dV \bigg) \,,
\end{align}
which has the expected form, that of the ratio of the dissipation rate to the total kinetic energy.

\bibliography{spheroid_paper_1}{}

\begin{thebibliography}{}

\bibitem[\protect\citeauthoryear{Abramowitz \& Stegun}{Abramowitz \&
  Stegun}{1965}]{AbramowitzStegun}
Abramowitz M.,  Stegun I.,  1965, Handbook of Mathematical Functions.
Dover

\bibitem[\protect\citeauthoryear{{Albrecht}, {Winn}, {Johnson}, {Howard},
  {Marcy}, {Butler}, {Arriagada}, {Crane}, {Shectman}, {Thompson}, {Hirano},
  {Bakos} \& {Hartman}}{{Albrecht} et~al.}{2012}]{Albrecht2012}
{Albrecht} S.,  {Winn} J.~N.,  {Johnson} J.~A.,  {Howard} A.~W.,  {Marcy}
  G.~W.,  {Butler} R.~P.,  {Arriagada} P.,  {Crane} J.~D.,  {Shectman} S.~A.,
  {Thompson} I.~B.,  {Hirano} T.,  {Bakos} G.,    {Hartman} J.~D.,  2012, \apj,
  757, 18

\bibitem[\protect\citeauthoryear{Binney \& Tremaine}{Binney \&
  Tremaine}{2008}]{binney2008galactic}
Binney J.,  Tremaine S.,  2008, Galactic Dynamics: (Second Edition).
Princeton Series in Astrophysics, Princeton University Press

\bibitem[\protect\citeauthoryear{{Bryan}}{{Bryan}}{1889}]{Bryan1889}
{Bryan} G.~H.,  1889, Royal Society of London Philosophical Transactions Series
  A, 180, 187

\bibitem[\protect\citeauthoryear{{Chandrasekhar}}{{Chandrasekhar}}{1987}]{Chandrasekhar1987}
{Chandrasekhar} S.,  1987, {Ellipsoidal figures of equilibrium}.
Dover

\bibitem[\protect\citeauthoryear{{Goldreich}}{{Goldreich}}{1963}]{Goldreich1963}
{Goldreich} P.,  1963, \mnras, 126, 257

\bibitem[\protect\citeauthoryear{{Goldreich} \& {Nicholson}}{{Goldreich} \&
  {Nicholson}}{1989}]{GN1989}
{Goldreich} P.,  {Nicholson} P.~D.,  1989, \apj, 342, 1079

\bibitem[\protect\citeauthoryear{{Goodman} \& {Dickson}}{{Goodman} \&
  {Dickson}}{1998}]{GD1998}
{Goodman} J.,  {Dickson} E.~S.,  1998, \apj, 507, 938

\bibitem[\protect\citeauthoryear{{Hansen}}{{Hansen}}{2010}]{Hansen2010}
{Hansen} B.~M.~S.,  2010, \apj, 723, 285

\bibitem[\protect\citeauthoryear{{Hellier}, {Anderson}, {Collier Cameron} \&
  {et al.}}{{Hellier} et~al.}{2009}]{Hellieretal2009}
{Hellier} C.,  {Anderson} D.~R.,  {Collier Cameron} A.,    {et al.} 2009, \nat,
  460, 1098

\bibitem[\protect\citeauthoryear{{Husnoo}, {Pont}, {Mazeh}, {Fabrycky},
  {H{\'e}brard}, {Bouchy} \& {Shporer}}{{Husnoo} et~al.}{2012}]{Husnoo2012}
{Husnoo} N.,  {Pont} F.,  {Mazeh} T.,  {Fabrycky} D.,  {H{\'e}brard} G.,
  {Bouchy} F.,    {Shporer} A.,  2012, \mnras, 422, 3151

\bibitem[\protect\citeauthoryear{{Jackson}, {Greenberg} \& {Barnes}}{{Jackson}
  et~al.}{2008}]{JGB2008}
{Jackson} B.,  {Greenberg} R.,    {Barnes} R.,  2008, \apj, 678, 1396

\bibitem[\protect\citeauthoryear{{Kong}, {Zhang} \& {Schubert}}{{Kong}
  et~al.}{2012}]{Kong2012}
{Kong} D.,  {Zhang} K.,    {Schubert} G.,  2012, \apj, 748, 143

\bibitem[\protect\citeauthoryear{{Kudlick}}{{Kudlick}}{1966}]{Kudlick1966}
{Kudlick} M.~D.,  1966, PhD thesis, MIT

\bibitem[\protect\citeauthoryear{{Lainey}, {Arlot}, {Karatekin} \& {van
  Hoolst}}{{Lainey} et~al.}{2009}]{Lainey2009}
{Lainey} V.,  {Arlot} J.-E.,  {Karatekin} {\"O}.,    {van Hoolst} T.,  2009,
  \nat, 459, 957

\bibitem[\protect\citeauthoryear{{Lindblom} \& {Ipser}}{{Lindblom} \&
  {Ipser}}{1999}]{LI1999}
{Lindblom} L.,  {Ipser} J.~R.,  1999, Physics Review D, 59, 044009

\bibitem[\protect\citeauthoryear{{Meibom} \& {Mathieu}}{{Meibom} \&
  {Mathieu}}{2005}]{MeibomMathieu2005}
{Meibom} S.,  {Mathieu} R.~D.,  2005, \apj, 620, 970

\bibitem[\protect\citeauthoryear{{Ogilvie}}{{Ogilvie}}{2005}]{Ogilvie2005}
{Ogilvie} G.~I.,  2005, {Journal of Fluid Mechanics}, 543, 19

\bibitem[\protect\citeauthoryear{{Ogilvie}}{{Ogilvie}}{2013}]{Ogilvie2013}
{Ogilvie} G.~I.,  2013, \mnras, 429, 613

\bibitem[\protect\citeauthoryear{{Ogilvie} \& {Lin}}{{Ogilvie} \&
  {Lin}}{2007}]{OgilvieLin2007}
{Ogilvie} G.~I.,  {Lin} D.~N.~C.,  2007, \apj, 661, 1180

\bibitem[\protect\citeauthoryear{{Papaloizou} \& {Pringle}}{{Papaloizou} \&
  {Pringle}}{1981}]{Papaloizou1981}
{Papaloizou} J.,  {Pringle} J.~E.,  1981, \mnras, 195, 743

\bibitem[\protect\citeauthoryear{{Rieutord} \& {Valdettaro}}{{Rieutord} \&
  {Valdettaro}}{1997}]{RV1997}
{Rieutord} M.,  {Valdettaro} L.,  1997, {Journal of Fluid Mechanics}, 341, 77

\bibitem[\protect\citeauthoryear{{Terquem}, {Papaloizou}, {Nelson} \&
  {Lin}}{{Terquem} et~al.}{1998}]{Terquem1998}
{Terquem} C.,  {Papaloizou} J.~C.~B.,  {Nelson} R.~P.,    {Lin} D.~N.~C.,
  1998, \apj, 502, 788

\bibitem[\protect\citeauthoryear{Thomson}{Thomson}{1863}]{Thomson1863}
Thomson W.,  1863, Philosophical Transactions of the Royal Society of London,
  153, pp. 583

\bibitem[\protect\citeauthoryear{{Wright}, {Fakhouri}, {Marcy}, {Han}, {Feng},
  {Johnson}, {Howard}, {Fischer}, {Valenti}, {Anderson} \& {Piskunov}}{{Wright}
  et~al.}{2011}]{exoplanets}
{Wright} J.~T.,  {Fakhouri} O.,  {Marcy} G.~W.,  {Han} E.,  {Feng} Y.,
  {Johnson} J.~A.,  {Howard} A.~W.,  {Fischer} D.~A.,  {Valenti} J.~A.,
  {Anderson} J.,    {Piskunov} N.,  2011, \pasp, 123, 412

\bibitem[\protect\citeauthoryear{{Wu}}{{Wu}}{2005a}]{Wu2005a}
{Wu} Y.,  2005a, \apj, 635, 674

\bibitem[\protect\citeauthoryear{{Wu}}{{Wu}}{2005b}]{Wu2005b}
{Wu} Y.,  2005b, \apj, 635, 688

\bibitem[\protect\citeauthoryear{{Yoder} \& {Peale}}{{Yoder} \&
  {Peale}}{1981}]{YP1981}
{Yoder} C.~F.,  {Peale} S.~J.,  1981, \icarus, 47, 1

\bibitem[\protect\citeauthoryear{{Zahn}}{{Zahn}}{1977}]{Zahn1977}
{Zahn} J.-P.,  1977, \aap, 57, 383

\bibitem[\protect\citeauthoryear{Zhang, Liao \& Earnshaw}{Zhang
  et~al.}{2004}]{Zhang2004}
Zhang K.,  Liao X.,    Earnshaw P.,  2004, J. Fluid. Mech., 504, 1

\end{thebibliography}
\bibliographystyle{mn2e}

\end{document}